\documentclass[letterpaper,english,preprint, aps, prd, nofootinbib,showpacs]{revtex4-1}
\usepackage[T1]{fontenc}
\usepackage[latin9]{inputenc}
\usepackage{babel}
\usepackage{verbatim}
\usepackage{textcomp}
\usepackage{amsmath}
\usepackage{amssymb}
\usepackage{graphicx}
\usepackage{makecell}
\usepackage{color}
\usepackage[unicode=true,pdfusetitle,
 bookmarks=true,bookmarksnumbered=false,bookmarksopen=false,
 breaklinks=false,pdfborder={0 0 1},backref=false,colorlinks=false]
 {hyperref}
\makeatletter

\pdfpageheight\paperheight
\pdfpagewidth\paperwidth

\providecommand{\tabularnewline}{\\}

\usepackage{chngcntr}
\counterwithout{figure}{section}

\makeatother

\begin{document}

\title{Quark spins and Anomalous Ward Identity}

\author{\texorpdfstring{Jian Liang$^{1,}$\footnote{jian.liang@uky.edu}, Yi-Bo Yang$^{2,}$\footnote{yangyibo@pa.msu.edu}, Terrence Draper$^{1}$,  Ming Gong$^{3}$, and Keh-Fei Liu$^{1,}$\footnote{liu@g.uky.edu}}}

\affiliation{$^{1}$\mbox{Department of Physics and Astronomy, University of Kentucky, Lexington, KY 40506, USA} 
$^{2}$\mbox{Department of Physics and Astronomy, Michigan State University, East Lansing, MI 48824, USA}
$^{3}$\mbox{Institute of High Energy Physics, Chinese Academy of Sciences, Beijing 100049, China}
\\~\\
\includegraphics[scale=0.12]{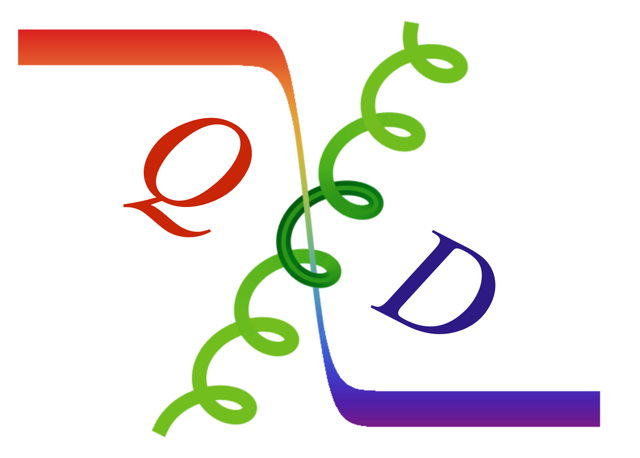}}

\collaboration{$\chi$QCD Collaboration}\noaffiliation
\begin{abstract}
{\normalsize{}We calculate the intrinsic quark spin contribution to
the total proton spin using overlap valence quarks on three ensembles of $2+1$-flavor RBC/UKQCD
domain-wall configurations with different lattice spacings. The lowest pion mass of the ensembles is around
171 MeV which is close to the physical point. 
%
With overlap fermions and topological charge derived from the overlap operator, 
we verify the anomalous Ward identity between nucleon states with momentum transfer.
Both the connected and disconnected insertions of the axial-vector current are calculated.
For the disconnected-insertion
part, the cluster-decomposition error reduction (CDER) technique is utilized
for the lattice with the largest volume and the error can be reduced
by $10\%\sim40\%$. 
Nonperturbative renormalization is carried out and the final results
are all reported in the $\overline{{\rm MS}}$ scheme at 2 GeV. We
determine the total quark spin contribution to the nucleon spin to be $\Delta\Sigma=0.405(25)(37)$,
which is consistent with the recent global fitting result of experimental
data. The isovector axial coupling we obtain in this study is $g_A^3=1.254(16)(30)$,
which agrees well with the experimental value of 1.2723(23).}{\normalsize\par}
\end{abstract}
\preprint{MSUHEP-18-008}

\maketitle

\section{Introduction}

The decomposition of the proton spin into its quark and glue constituents
has long been a puzzle ever since the first deep inelastic scattering
(DIS) experiment around three decades ago \citep{Ashman:1987hv,Ashman:1989ig}
revealed that not all the proton spin originates from the quark intrinsic
spin as depicted in the naive quark model, leading to the so-called ``proton spin crisis.'' Now we understand
that the proton spin, consisting of quark spin, quark orbital angular momentum,
glue spin and glue orbital angular momentum, is the result of complicated
QCD dynamics which cannot be described by the quark model.
However, the precise proportion of the total proton spin carried by
these components remains unclear. On the experimental side, since
the integration of the spin-dependent parton distributions over the
momentum fraction $x$ gives the fraction of the proton spin which
is carried by the corresponding flavor, that is,
\begin{equation}
\Delta q(\mu^2)=\int_{0}^{1}dx\Delta q(x,\mu^2)
\end{equation}
where $\mu$ is the $\overline{\rm{MS}}$ scale,
the global fit to the experimental data of DIS or Drell-Yan processes
for extracting the parton distributions will provide us knowledge
about the quark spin contribution to the proton spin. 
Three recent experimental results from D. de Florian {\it et al.} \cite{deFlorian:2009vb}, the NNPDF collaboration  \cite{Nocera:2014gqa} and the COMPASS collaboration  \cite{Adolph:2015saz}
determine the total quark intrinsic spin contribution $\Delta\Sigma$ to be $0.366^{+0.042}_{-0.062}$, ${0.25(10)}$ and $[0.26, 0.36]$, respectively.
%
On the lattice side, 
a recent calculation \citep{Alexandrou:2017oeh}
is carried out with physical pion mass, but with only one single
lattice ensemble of $N_{f}=2$ clover-improved
twisted mass fermions. 
More careful lattice studies
with ensembles of different lattice spacings and different lattice volumes are imperative to push the results to
the physical limit and to control the corresponding systematic uncertainties.

In this work, we use overlap fermions on three domain-wall ensembles to
calculate the quark spin contribution to the nucleon spin. Since for
each quark flavor the intrinsic spin is actually half of the corresponding
axial coupling of nucleon, we need to calculate the axial coupling for
the flavor-diagonal case. Thus, both the connected-insertions and
disconnected-insertions of the correlation functions need to be included.
The anomalous Ward identity is carefully checked to see if any normalization
due to lattice artifacts needs to be applied to the axial-vector current
to make the identity hold. We actually find that the same normalization
constant for the local axial-vector current as used in the isovector case to satisfy the chiral Ward identity also satisfies 
the anomalous Ward identity. This is not true in general for non-chiral fermions.
For the disconnected-insertion part, the cluster-decomposition error reduction (CDER) technique \citep{Liu:2017man}
is utilized for the lattice with the largest
volume to reduce the statistical error. For the connected-insertion
part, an improved axial-vector current is employed such that the finite
lattice spacing effect can be reduced. All of our results are matched
to the $\overline{{\rm MS}}$ scheme at 2 GeV using nonperturbative
renormalization. We propose a new renormalization pattern where we
separate the connected-insertion part and the disconnected-insertion
part from the beginning which is more natural for the lattice calculation and
offers more information than the conventional flavor irreducible representation approach.

This paper is organized as follows. The formalism of quark spin and anomalous Ward identity are discussed in Sec. \ref{sec:quark_spin}. 
In Sec. \ref{sec:Numerical-details}
we describe all the numerical details of our simulation. Then in Sec.
\ref{sec:Check-of-Anomalous}, we check the axial Ward identity to
address the normalization issue. The bare results of the disconnected
contribution are shown in Sec. \ref{sec:Disconnected-insertion-contribut}.
The detailed results of the connected contribution come in Sec. \ref{sec:Connected-insertion-contribution}.
We discuss the renormalization in Sec. \ref{sec:Renormalization}
and make global fits to get the final results in Sec. \ref{sec:Global-fitting-and}.
A short summary is given in Sec. \ref{sec:Summary}.

\section{Formalism of Quark Spin and Anomalous Ward Identity } \label{sec:quark_spin}

The quark spin contribution to the nucleon spin is associated  from the nucleon matrix element of the flavor-singlet axial-vector current,
 \begin{equation}  \label{g_A_def}
 g_A^0\, s_{\mu}= \frac{\langle p,s| A_{\mu}^0|p,s\rangle}{\langle p,s|p,s\rangle},
 \end{equation}
where $A_{\mu}^0$ is the flavor-singlet axial-vector current
\begin{equation}
A_{\mu}^0 =  \overline{\psi}_{u} i \gamma_{\mu} \gamma_5 \psi_u + \overline{\psi}_{d} i \gamma_{\mu} \gamma_5 \psi_d
+ \overline{\psi}_{s} i \gamma_{\mu} \gamma_5 \psi_s.
\end{equation}
The flavor $u,d$ and $s$ contributions to $g_A^0$ are denoted as $ \Delta u, \Delta d$ and $\Delta s$ in Eq.~(\ref{g_A_def}), so that
\begin{equation}
g_A^0 = \Delta u + \Delta d +\Delta s.
\end{equation}
  A special property of the flavor-singlet axial-vector current is that it satisfies the anomalous Ward identity (AWI) where the Adler-Bell-Jackiw anomaly 
appears from the Jacobian factor of the fermion determinant due to the $U(1)$ chiral transformation~\cite{Fujikawa:1979ay}
\begin{equation}  \label{AWI}
\partial_{\mu} A_{\mu}^0 = \sum_{f = u,d,s} 2m_f P_f -   2i N_f q
 \end{equation}
where the pseudoscalar density $P_f$ and the topological charge density operator $q$ representing the anomaly are
 \begin{eqnarray}
P_f =  \overline{\psi}_{f} i \gamma_5 \psi_f , \hspace{1cm}
q = \frac{1}{16\pi^2} G_{\mu\nu}^a \tilde{G}_{\mu\nu}^a,
\end{eqnarray}
where $G_{\mu\nu}^a$ is the gauge field strength tensor and $\tilde{G}_{\mu\nu}^a=\epsilon_{\mu\nu\rho\sigma}G_{\rho\sigma}^a$.
Note, notations are in Euclidean space and the coupling constant $g$ is absorbed in the definition of the gauge potential $A_{\mu}^a$.

As far as renormalization is concerned, it is shown that $A_{\mu}^0$ has a two-loop renormalization~\cite{Kodaira:1979ib, Espriu:1982bw} and the topological charge has a one-loop mixture with $\partial_{\mu} A_{\mu}^0$~\cite{Espriu:1982bw} so that the renormalized AWI in the dimensional regularization scheme becomes
\begin{equation}    \label{renor_AWI}
\partial_{\mu} A_{\mu}^0 \left( 1 + \gamma N_f \frac{1}{\epsilon}\right)  = \sum_{f = u,d,s} 2m_f^R P_f^R + \left(- 2i N_f q + \gamma N_f \frac{1}{\epsilon} \partial_{\mu} A_{\mu}^0\right)
\end{equation}
with the anomalous dimension $\gamma = - (\alpha_s/\pi)^2 \frac{3}{8}C_F$. $m^R$ and $P^R$ are renormalized quark mass
and pseudoscalar density. We see that, the $\alpha_s^2$ renormalization term on the left is the same as that on the right from mixing. Thus, 
$m P$ and $\partial_{\mu} A_{\mu}^0 + 2i N_f q$ are renormalization group invariant (the latter to second order at least) and the form of AWI
is the same with or without renormalization. 

On the lattice, the AWI is preserved by the overlap fermion which is chiral and satisfies the Ginsparg-Wilson relation~\cite{Hasenfratz:1998ri}. The $m_f P_f$ is renormalization group invariant since $Z_m Z_P = 1$ for the chiral fermion and the local version of the topological charge $q(x)$ derived from the overlap operator is equal to $\frac{1}{16 \pi^2} {\rm tr}_c G_{\mu\nu} \tilde{G}_{\mu\nu}(x)$ in the 
continuum~\cite{Kikukawa:1998pd, Adams:1998eg, Fujikawa:1998if, Suzuki:1998yz}, i.e. 
\begin{eqnarray}  \label{top-charge}  
\!\!q(x)\! =\!  {\rm Tr} \,\left[\gamma_5 \left( \frac{1}{2}D_{\rm ov}(x,x) \!-\!1\right)\right]
 {}_{\stackrel{\longrightarrow}{a \rightarrow 0}} \frac{1}{16 \pi^2}{\rm tr}_c \, G_{\mu\nu} 
\tilde{G}_{\mu\nu}(x),
\end{eqnarray}
where
$D_{\rm ov}$ is the overlap operator. 
In the overlap case, the chiral axial-vector current is derived \cite{Hasenfratz:2002rp} 
and one can directly proceed to carry out the renormalization of the chiral axial-vector current perturbatively or non-perturbatively. However,  
this chiral axial-vector involves a non-local kernel \mbox{$K_{\mu} = -i \frac{\delta D_{\rm ov} (U_{\mu} e^{i \alpha_{\mu}(x)})}{\delta \alpha_{\mu}(x)}|_{\alpha = 0}$} and is somewhat involved to
implement numerically. We shall use the local current in the present study. As such, it invokes a normalization constant $Z_A^0$ which warrants that the unrenormalized AWI in its `semiclassical' form (Eq.~(\ref{AWI})) is satisfied on the lattice and is itself scale independent. Therefore the normalization and renormalization takes two steps. First, one needs to
find the normalization $Z_A^0$ for the local axial-vector current which satisfies the unrenormalized AWI
\begin{equation}   \label{AWI_lattice}
\partial_{\mu} Z_A^0 A_{\mu}^0  = \sum_{f = u,d,s} 2m_f P_f - 2i N_f q
\end{equation}
where $A_{\mu}^0 = \sum_{f = u,d,s} \overline{\psi}_f i \gamma_{\mu} \gamma_5 \hat{\psi}_f$ and 
$P_f = \overline{\psi}_f i \gamma_5 \hat{\psi}_f$ are the local axial-vector current we use on the lattice 
and $\hat{\psi} =( 1 - \frac{1}{2} D_{\rm ov}) \psi$ is for giving rise to the effective quark propagator which conforms to the form in the continuum.

After the normalization constant $Z_A^0$ is determined, one then takes on the renormalization procedure. We shall discuss the 
determination of $Z_A^0$ in Section \ref{sec:Check-of-Anomalous}  after we give the numerical details of the calculation and will carry out the renormalization in Section \ref{sec:Renormalization}. 

Before we check the AWI on the lattice, we shall first give some numerical details of the lattice calculation.

\section{Numerical details\label{sec:Numerical-details}}

We use overlap fermions \citep{Neuberger:1997fp} as valence quarks
to perform our calculation. Since the overlap action preserves chiral
symmetry at finite lattice spacing via the Ginsparg-Wilson relation
\citep{Ginsparg:1981bj}, there is no additive renormalization for
the quark mass. The effective quark propagator of the massive overlap
fermion is the inverse of operator $D_{c}+m$ \citep{Chiu:1998eu,Liu:2002qu}
where $D_{c}$ satisfying \mbox{$\left\{ D_{c},\gamma_{5}\right\} =0$} is
exactly chiral and can be defined from the original overlap operator
$D_{{\rm ov}}$ as $D_{c}=\frac{\rho D_{{\rm ov}}}{1-D_{{\rm ov}}/2}$.
The overlap operator can be expressed as $D_{{\rm ov}}=1+\gamma_{5}\epsilon(\gamma_{5}D_{{\rm w}}(\rho))$
where $\epsilon$ is the matrix sign function and $D_{{\rm w}}$ is
the Wilson kernel with $\kappa=0.2$ (corresponding to parameter $\rho=1.5$).
As discussed above, another great feature of the overlap operator is that the local version of the topological
charge of the gauge field can be defined as $q(x)={\rm Tr}[\gamma_{5}(\frac{1}{2\rho}D_{{\rm ov}}(x,x)-1)]$~\cite{Kikukawa:1998pd, Adams:1998eg, Fujikawa:1998if, Suzuki:1998yz}.
The Atiyah-Singer index theorem \citep{Atiyah:1971rm} is satisfied which relates the
total topological charge to the index of zero modes of the overlap
operator so no multiplicative renormalization is needed for this definition
of $q$. These two features help us to feasibly check the AWI which we can use as a normalization condition in the
disconnected-insertion case. We use multiple partially-quenched valence
quark masses to cover a wide range of pion mass using the multi-mass algorithm.
More details regarding the calculation of the overlap operator and
eigenmodes deflation in the inversion of the fermion matrix can be
found in \citep{Li:2010pw}.

The three lattice ensembles we use for the calculation are 2+1-flavor
domain-wall fermion (DWF) ensembles generated by the RBC/UKQCD collaboration
\citep{Aoki:2010dy,Blum:2014tka}. They are labeled as 24I, 32I and
32ID and the detailed parameters of the ensembles can be found in
Table \ref{parameters_of_ensembles}. We have three different lattice
spacings and the lowest pion mass at $171$ MeV is close to the physical one.

\begin{table}[ht]
\caption{The parameters of the 2+1-flavor RBC/UKQCD configurations: label, spatial/temporal
size, lattice spacing, the degenerate light sea quark mass, strange
sea quark mass, the corresponding pion mass and the number of configurations
used in this work. \label{parameters_of_ensembles}}
\centering{}%
\begin{tabular}{ccccccc}
label & $L^{3}\times T$ & $a^{-1}$ (GeV) & $m_{l}^{(s)}a$ & $m_{s}^{(s)}a$ & $m_{\pi}$ (MeV) & $N_{{\rm cfg}}$\tabularnewline
\hline 
32I & $32^{3}\times64$ & 2.3833(86) & 0.004 & 0.03 & 302 & 309\tabularnewline
\hline 
24I & $24^{3}\times64$ & 1.7848(50) & 0.005 & 0.04 & 337 & 203\tabularnewline
\hline 
32ID & $32^{3}\times64$ & 1.3784(68) & 0.001 & 0.045 & 171 & 200\tabularnewline
\hline 
\end{tabular}
\end{table}

\begin{figure}[tbp]
\begin{centering}
\includegraphics[scale=0.3]{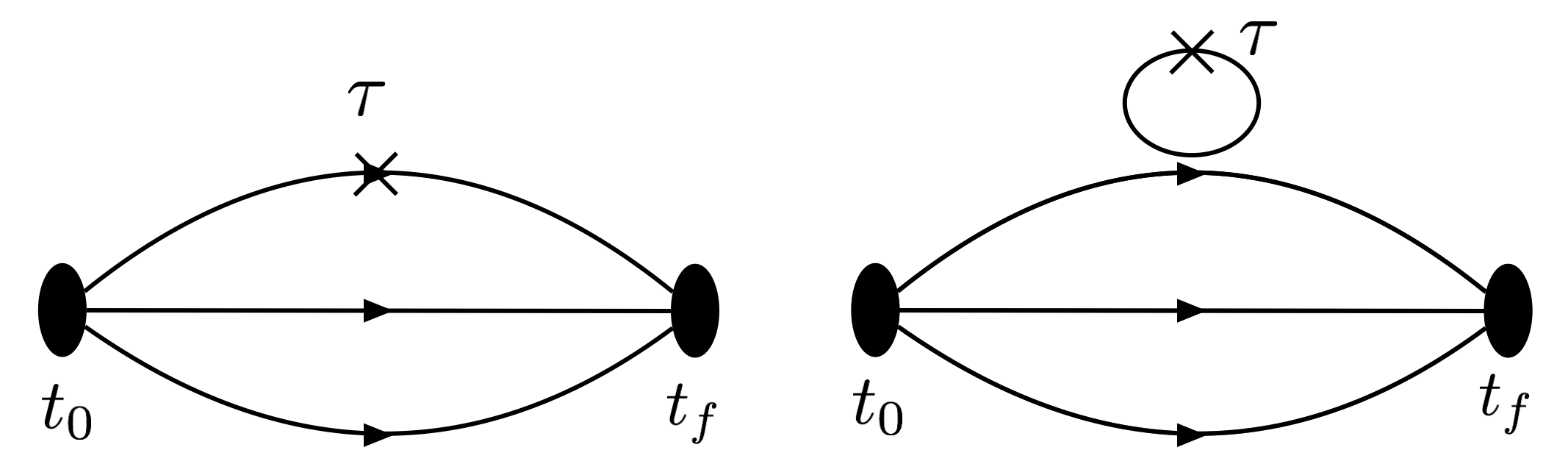}
\par\end{centering}
\caption{The illustration of the connected-insertion (left) and disconnected-insertion (right).\label{fg:CIDI}}
\end{figure}

To calculate the quark spin or, in practice, to calculate the axial
coupling, we need to construct 3-point correlation functions
\begin{equation}
C_{3,\mu}(t_{f},\tau)=\sum_{\vec{x},\vec{y}}\langle\chi(t_{f},\vec{y})A_{\mu}(\tau,\vec{x})\bar{\chi}(0,\mathcal{G})\rangle
\end{equation}
where $\chi$ is the nucleon interpolation field, $\mathcal{G}$ denotes the
source grid and $A_{\mu}=\bar{\psi}i\gamma_{\mu}\gamma_{5}\hat{\psi}$
is the local axial-vector current with $\hat{\psi}=(1-\frac{1}{2}D_{{\rm ov}})\psi$
for giving rise to the effective quark propagator $\left(D_{c}+m\right)^{-1}$.
The correlation function can have two kinds of current insertions,
i.e., the connected-insertion (CI) and the disconnected-insertion
(DI), corresponding to two ways of Wick contractions. 
They are depicted in Figure~\ref{fg:CIDI}.    \\

For the CI calculation,
we use the stochastic sandwich method (SSM) with low-mode substitution
(LMS) \citep{Yang:2015zja} to better control the statistical uncertainty.
We use $Z_{3}$ noise grid sources with Gaussian (24I and 32I) or
block smearing (32ID) \citep{Liang:2016fgy} coherently at $t_{{\rm {src}}}=0$
and $t_{{\rm {src}}}=32$ in one inversion. The sinks are block smeared and located
at different positions with different separations in time from the source.
Setups regarding the valence simulation of the CI case are listed
in Table \ref{tb:valence_setup_CI}. Technical details regarding the
LMS of random $Z_{3}$ grid source with mixed momenta and the SSM
with LMS for constructing 3-point functions can be found in references
\citep{Gong:2013vja,Yang:2015zja,Liang:2016fgy}.
Due to the fact that multi-mass inversion algorithm is uniquely applicable to
the overlap fermion with eigenvector deflation, we calculate $5\sim6$ valence masses each for the three lattices. 

\begin{table}[h]
\caption{The details of the overlap simulation in the valence sector for the
CI case, including the name of the lattice, the grid type of source
$\mathcal{G}_{{\rm src}}$ (the notations such as 12-12-12 denote
the intervals of the grid in the three spatial directions; see
reference \citep{Yang:2015zja} for more details), the number of source
grids $N_{{\rm src}}$, the positions of sources $t_{{\rm src}}$,
the grid type of sink $\mathcal{G}_{{\rm sink}}$, the number of the noises for the sink
grids $N_{{\rm sink}}$, the source-sink separations $(t_{{\rm sink}}-t_{{\rm src}})$
and the bare valence quark masses $m_{q}^{v}a$. \label{tb:valence_setup_CI}}
\centering{}%
\begin{tabular}{r@{\extracolsep{0pt}.}lr@{\extracolsep{0pt}.}lr@{\extracolsep{0pt}.}lr@{\extracolsep{0pt}.}lr@{\extracolsep{0pt}.}lccc}
\multicolumn{2}{c}{Lattice} & \multicolumn{2}{c}{$\mathcal{G}_{{\rm src}}$} & \multicolumn{2}{c}{$N_{{\rm src}}$} & \multicolumn{2}{c}{$t_{{\rm src}}$} & \multicolumn{2}{c}{$\ensuremath{\mathcal{G}_{{\rm sink}}}$} & $N_{{\rm sink}}$ & $(t_{{\rm sink}}-t_{{\rm src}})$ & $m_{q}^{v}a$\tabularnewline
\hline 
\multicolumn{2}{c}{} & \multicolumn{2}{c}{} & \multicolumn{2}{c}{} & \multicolumn{2}{c}{} & \multicolumn{2}{c}{} & 5 & 0.88 fm & \tabularnewline
\cline{11-12} 
\multicolumn{2}{c}{24I} & \multicolumn{2}{c}{12-12-12} & \multicolumn{2}{c}{1} & \multicolumn{2}{c}{(0, 32)} & \multicolumn{2}{c}{2-2-2} & 3 & 1.11 fm & (0.00809, 0.0102, 0.0135, 0.0160, 0.0203) \tabularnewline
\cline{11-12} 
\multicolumn{2}{c}{} & \multicolumn{2}{c}{} & \multicolumn{2}{c}{} & \multicolumn{2}{c}{} & \multicolumn{2}{c}{} & 5 & 1.22 fm & \tabularnewline
\cline{11-12} 
\multicolumn{2}{c}{} & \multicolumn{2}{c}{} & \multicolumn{2}{c}{} & \multicolumn{2}{c}{} & \multicolumn{2}{c}{} & 5 & 1.33 fm & \tabularnewline
\hline 
\multicolumn{2}{c}{} & \multicolumn{2}{c}{} & \multicolumn{2}{c}{} & \multicolumn{2}{c}{} & \multicolumn{2}{c}{} & 3 & 0.99 fm & \tabularnewline
\cline{11-12} 
\multicolumn{2}{c}{32I} & \multicolumn{2}{c}{16-16-16} & \multicolumn{2}{c}{1} & \multicolumn{2}{c}{(0, 32)} & \multicolumn{2}{c}{1-1-1} & 3 & 1.16 fm & (0.00585, 0.00765, 0.00885, 0.0112, 0.0152)\tabularnewline
\cline{11-12} 
\multicolumn{2}{c}{} & \multicolumn{2}{c}{} & \multicolumn{2}{c}{} & \multicolumn{2}{c}{} & \multicolumn{2}{c}{} & 3 & 1.24 fm & \tabularnewline
\hline 
\multicolumn{2}{c}{} & \multicolumn{2}{c}{} & \multicolumn{2}{c}{} & \multicolumn{2}{c}{} & \multicolumn{2}{c}{} & 2 & 1.00 fm & \tabularnewline
\cline{11-12} 
\multicolumn{2}{c}{} & \multicolumn{2}{c}{} & \multicolumn{2}{c}{} & \multicolumn{2}{c}{} & \multicolumn{2}{c}{} & 3 & 1.15 fm & \tabularnewline
\cline{11-12} 
\multicolumn{2}{c}{32ID} & \multicolumn{2}{c}{16-16-16} & \multicolumn{2}{c}{6} & \multicolumn{2}{c}{(0, 32)} & \multicolumn{2}{c}{1-1-1} & 4 & 1.29 fm & (0.0042, 0.0060, 0.011, 0.014, 0.017, 0.022)\tabularnewline
\cline{11-12} 
\multicolumn{2}{c}{} & \multicolumn{2}{c}{} & \multicolumn{2}{c}{} & \multicolumn{2}{c}{} & \multicolumn{2}{c}{} & 5 & 1.43 fm & \tabularnewline
\cline{11-12} 
\multicolumn{2}{c}{} & \multicolumn{2}{c}{} & \multicolumn{2}{c}{} & \multicolumn{2}{c}{} & \multicolumn{2}{c}{} & 12 & 1.57 fm & \tabularnewline
\cline{11-12} 
\end{tabular}
\end{table}

For disconnected-insertion calculations, we use the low-mode average
(LMA) technique to calculate the quark loops which improves the signal-to-noise
ratio particularly for the pseudoscalar and scalar currents. 
The low-mode part is calculated exactly while the high-mode part is estimated
with 8 sets of $Z_{4}$-noise on a 4-4-4-2 space-time grid with even-odd dilution
and additional time shift. The same $Z_{3}$-noise grid source with
smearing as in the CI case is used in the production of the nucleon
propagators. We make multiple measurements by shifting the source
time-slice to improve statistics; the spatial position of the
center of the grid is randomly chosen for each source time-slice to
reduce autocorrelation. References \citep{Gong:2013vja,Gong:2015iir,Yang:2015uis}
contain more details regarding the DI calculation. When constructing
quark loops, we include more valence quark masses to cover the strange
region. The bare strange quark mass is determined by the global-fit
value at  2 GeV in the $\overline{{\rm MS}}$ scale calculated in our previous study
\citep{Yang:2014sea} and the nonperturbative mass renormalization
constant calculated in \citep{Liu:2013yxz}.

To obtain the axial coupling, we construct a ratio of the 3-point correlation
function to the nucleon 2-point function
\begin{equation}
R(t_{f},\tau)=f_k\frac{{\rm Tr}\left[\Gamma_{p}C_{3}(t_{f},\tau)\right]}{{\rm Tr}\left[\Gamma_{e}C_{2}(t_{f})\right]}
\end{equation}
where $f_k$ is a kinematic factor which is related to the Lorentz index of the current, $\Gamma_{p}$ is the polarized projector of the nucleon spin,
$\Gamma_{e}$ is the non-polarized projector and $C_{2}(t_{f})=\sum_{\vec{x}}\langle\chi(t_{f},\vec{x})\bar{\chi}(0,\mathcal{G})\rangle$.
The matrix element $g_{A}$ can then be obtained asymptotically $g_{A}=R(t_{f}\gg\tau,\tau\gg0)$.
However at finite $t_{f}$ and $\tau$, the excited states will contribute
to the ratio and we need to extract $g_{A}$ by fitting the ratio
to more complicated function forms. A commonly used form with two-state
fit reads
\begin{equation}
R(t_{f},\tau)=g_{A}+c_{1}e^{-\delta m\left(t_{f}-\tau\right)}+c_{2}e^{-\delta m\tau}+c_{3}e^{-\delta mt_{f}}\label{eq:two-state-fit}
\end{equation}
which assumes only the first excited state has effects and $\delta m$
is the energy difference between the ground state and the first excited
state. In practice, higher excited-states' contribution can alter
the value of $\delta m$, making it a free parameter, accounting for
an effective energy difference.

\section{Anomalous Ward Identity on the lattice} \label{sec:Check-of-Anomalous}

    To verify the AWI in Eq.~(\ref{AWI_lattice}), we note that there is no flavor-mixing in this unrenormalized form. Thus, one
can check it for individual flavors and, furthermore, since the lattice calculations of matrix elements are separated in the CI and DI cases as shown in Figure~\ref{fg:CIDI}, 
one can separately check the chiral Ward identity for the connected matrix elements
\begin{equation}   \label{WI_CI}
\langle p'|\partial_{\mu} Z_A({\rm CI}) A_{\mu}|p\rangle ({\rm CI})  = \langle p'|2 m_q P |p\rangle ({\rm CI}).
\end{equation}
In this case, the matrix elements are for the $u$ or $d$ valence quark with quark mass $m_q$. 
Here, the normalization constant $Z_A({\rm CI})$ is due to the fact that we use the local current in this calculation.

Similarly, the AWI for the matrix elements in the DI case is
\begin{equation}  \label{AWI-DI}
\langle p'|\partial_{\mu} Z_A({\rm DI}) A_{\mu}|p\rangle ({\rm DI})  = \langle p'|2 m_q P - 2i q |p\rangle ({\rm DI})
\end{equation}
In principle, the normalization constants $Z_A(\rm{CI})$ and  $Z_A(\rm{DI})$ can be different, especially when non-chiral fermions are used
in the lattice calculation and also when the topological charge $q$ is not calculated from the overlap operator $D_{\rm ov}$ as in Eq.~(\ref{top-charge}).
We shall check them in the following.

\subsection{Disconnected insertion (DI) case}

We shall check the disconnected insertion (DI) case first which has been investigated in our previous study~\citep{Gong:2015iir}. 
The anomalous Ward identity (AWI) in the DI case in Eq.~(\ref{AWI-DI}) 
relating the nucleon matrix element of the divergence of the axial-vector current $A_{\mu}$ to those of 
the product of the quark mass $m_{q}$ and the pseudo-scalar current
$P$ and also to the topological charge term $q$ is an important check
for lattice spin calculations involving the flavor-diagonal matrix
elements (MEs) of the axial-vector current. This is especially true for the strange quark as it only contributes in the DI.
Only properly extracted MEs plus correct lattice normalization will make this identity hold.
Our previous work~\citep{Gong:2015iir} utilized the AWI via the form factors 
which is actually an extended form of the Goldberger-Treiman
relation for the flavor-singlet case at finite momentum transfer $q^2$ and is expressed as
\begin{equation}\label{eq:AWI_FF}
g_{A}(q^{2})+\frac{q^{2}}{2m_{N}}h_{A}(q^{2})=\frac{2m_{q}}{2m_N}g_{P}(q^{2})+2g_{Q}(q^{2})
\end{equation}
where $g_{A}$ and $h_{A}$ are the axial and induced pseudoscalar
form factors respectively from the nucleon matrix element of the axial-vector current 
\begin{equation}    \label{g_AFF}
\langle p', s| A_{\mu}| p, s\rangle = \bar{u} (p', s) [ i \gamma_{\mu} \gamma_5 g_A(q^2)  -i q_{\mu} \gamma_5 h_A (q^2)] u (p, s),
\end{equation}
and $g_{P}(q^2)$ and $g_Q(q^2)$ are the pseudoscalar and anomaly form factors defined in
\begin{eqnarray}
&&\langle p', s| P| p, s\rangle = \bar{u} (p', s) i \gamma_5\,  u (p, s) g_P (q^2),  \label{PFF}\\
&&\langle p', s|-iq| p, s\rangle = \bar{u} (p', s) i \gamma_5\,  u (p, s) m_N g_Q (q^2). \label{AFF}
\end{eqnarray}
Eqs.~(\ref{g_AFF}) and (\ref{PFF}) have the same form separately for the CI and DI, while Eq. (\ref{AFF}) for the topological form factor
only appears in the DI.
Equation (\ref{eq:AWI_FF}) can be derived
by inserting the currents between nucleon states with momenta $\vec{p}$ and $\vec{p'}$ and applying the divergence to the nucleon states
\begin{equation}
\partial_{\mu}\langle p'|A_{\mu}|p\rangle=(E'-E)\langle p'|A_{4}|p\rangle+iq_{i}\langle p'|A_{i}|p\rangle\label{eq:partial_ME}
\end{equation}
where $E$ and $E'$ are the energies of the source and sink nucleons
and $q_{i}$ is the momentum transfer $\vec{q}=\vec{p'}-\vec{p}$ in the $i{\rm th}$ direction.
In the earlier study \citep{Gong:2015iir} we found that a normalization
factor of $\kappa_{A}\sim1.4$ on the axial-vector side is needed
in order to satisfy the identity which is much larger than the isovector
normalization constant\footnote{We choose to call it normalization constant rather than renormalization
constant since it is a finite renormalization which has no scale dependence and deviates from unity only
because of finite lattice spacing effects.} $Z_{A}^{3}({\rm 24I})=1.111(6)$ computed by using the chiral Ward
identity in the pion 2-point function case \citep{Liu:2013yxz}. Since
on the right hand side of Eq. (\ref{eq:AWI_FF}) we have $Z_{m}Z_{P}=1$
and there is no multiplicative renormalization of the topological
charge defined by the overlap operator, and, as shown above, the
renormalized\footnote{When we say renormalization, we mean there is a non-zero anomalous dimension
and therefore is scale dependent.} AWI is the same as the un-renormalized one 
at two-loop level \citep{Gong:2015iir}, the factor $\kappa_{A}$
was believed to be a necessary normalization factor in the DI case
for compensating the violation of the AWI induced by lattice artifacts
and was used to normalize the DI axial-vector MEs. In this study,
we shall have a critical reexamination of this issue. We also make a similar check for the light quarks case of 24I and
32I and for the new 32ID lattice by calculating the following ratio
\begin{equation}
R_{1}(\tau,t_{f},q^{2})=\frac{\frac{2m_{q}}{2m_N}g_{P}(\tau,t_{f},q^{2})+2g_{Q}(\tau,t_{f},q^{2})}{g_{A}(\tau,t_{f},q^{2})+\frac{q^{2}}{2m_{N}}h_{A}(\tau,t_{f},q^{2})}.\label{eq:check_1}
\end{equation}
Note that we keep the dependence of the sink time $t_{f}$
and the current time $\tau$ for all the form factors and, therefore,
the excited-state effects are not handled until we fit the final ratio.

We also check the AWI more carefully at the ME level, in other words,
treating $\partial_\mu A_\mu$ as an operator insertion between the nucleon states $p$ and $p'$.
The lattice
version of the AWI reads

\begin{equation}
\sum_{\vec{x}}\left(\sum_{i}\left\langle A_{i}(\tau,\vec{x})\text{\textminus}A_{i}(\tau,\vec{x}-\hat{i})\right\rangle +\left\langle A_{4}(\tau,\vec{x})\text{\textminus}A_{4}(\tau-1,\vec{x})\right\rangle \right)e^{-i\vec{q}\cdot\vec{x}}=\frac{2m_{q}}{2m_N}\left\langle P(\tau,\vec{q})\right\rangle -2i\left\langle q(\tau,\vec{q})\right\rangle ,
\end{equation}
where $\left\langle A_{i}(\tau,\vec{x})\right\rangle $ is an abbreviated
form of the ME $\langle p'|A_{i}(\tau,\vec{x})|p\rangle$, $\hat{i}$
is the unit vector along the $i{\rm th}$ direction and the continuum
partial derivative is replaced by the backward-difference on the lattice.
This equation cannot be checked directly since the momentum projection
is always done before we can take the spatial difference $A_{i}(\tau,\vec{x})\text{\textminus}A_{i}(\tau,\vec{x}-\hat{i})$
in the 3-point function calculation. However,
\begin{equation}
\sum_{\vec{x}}\sum_{i}\left\langle A_{i}(\tau,\vec{x})\text{\textminus}A_{i}(\tau,\vec{x}-\hat{i})\right\rangle e^{-i\vec{q}\cdot\vec{x}}=\left(1-e^{-iq_{i}}\right)\sum_{\vec{x}}\left\langle A_{i}(\tau,\vec{x})\right\rangle e^{-i\vec{q}\cdot\vec{x}}\sim iq_{i}\left\langle A_{i}(\tau,\vec{q})\right\rangle 
\end{equation}
is a good approximation if $q_{i}$ is small enough. So we have a
simplified form
\begin{equation}
iq_{i}\left\langle A_{i}(\tau,\vec{q})\right\rangle +\left\langle A_{4}(\tau,\vec{q})\text{\textminus}A_{4}(\tau-1,\vec{q})\right\rangle =\frac{2m_{q}}{2m_N}\left\langle P(\tau,\vec{q})\right\rangle -2i\left\langle q(\tau,\vec{q})\right\rangle ,\label{eq:AWI_check_1}
\end{equation}
which can be checked numerically. A second ratio $R_{2}$ is thus
defined as 
\begin{equation}
R_{2}=\frac{\frac{2m_{q}}{2m_N}\left\langle P(\tau,\vec{q})\right\rangle -2i\left\langle q(\tau,\vec{q})\right\rangle }{iq_{i}\left\langle A_{i}(\tau,\vec{q})\right\rangle +\left\langle A_{4}(\tau,\vec{q})\text{\textminus}A_{4}(\tau-1,\vec{q})\right\rangle }.\label{eq:check_2}
\end{equation}
Furthermore, for the temporal part, by inserting complete sets of
intermediate states and using the time evolution operator, the time
dependence of the ME can be formulated as 
\begin{equation}
\left\langle A_{4}(\tau,\vec{x})\right\rangle =\left\langle A_{4}(0,\vec{q})\right\rangle e^{+\Delta E\tau}\label{eq:time_depen_A4}
\end{equation}
up to exponential suppression of the excited-states contamination,
where $\Delta E=E'-E$ is the energy difference between the sink and
source nucleon states, such that 
\begin{equation}
\left\langle A_{4}(\tau,\vec{x})-A_{4}(\tau-1,\vec{x})\right\rangle \sim\Delta E\left\langle A_{4}(\tau,\vec{x})\right\rangle ,
\end{equation}
leading to 
\begin{equation}
iq_{i}\left\langle A_{i}(\tau,\vec{q})\right\rangle +\Delta E\left\langle A_{4}(\tau,\vec{q})\right\rangle =2m_{q}\left\langle P(\tau,\vec{q})\right\rangle -2i\left\langle q(\tau,\vec{q})\right\rangle \label{eq:AWI_check_2}
\end{equation}
if the excited-states contamination can be ignored or completely
removed by fit. This is actually the counterpart of Eq. (\ref{eq:partial_ME})
and we now have the third ratio to check the AWI
\begin{equation}
R_{3}=\frac{\frac{2m_{q}}{2m_N}\left\langle P(\tau,\vec{q})\right\rangle-2i\left\langle q(\tau,\vec{q})\right\rangle }{iq_{i}\left\langle A_{i}(\tau,\vec{q})\right\rangle +\Delta E\left\langle A_{4}(\tau,\vec{q})\right\rangle }.\label{eq:check_3}
\end{equation}

\begin{figure}[tbp]
\begin{centering}
\includegraphics[scale=0.5,page=1]{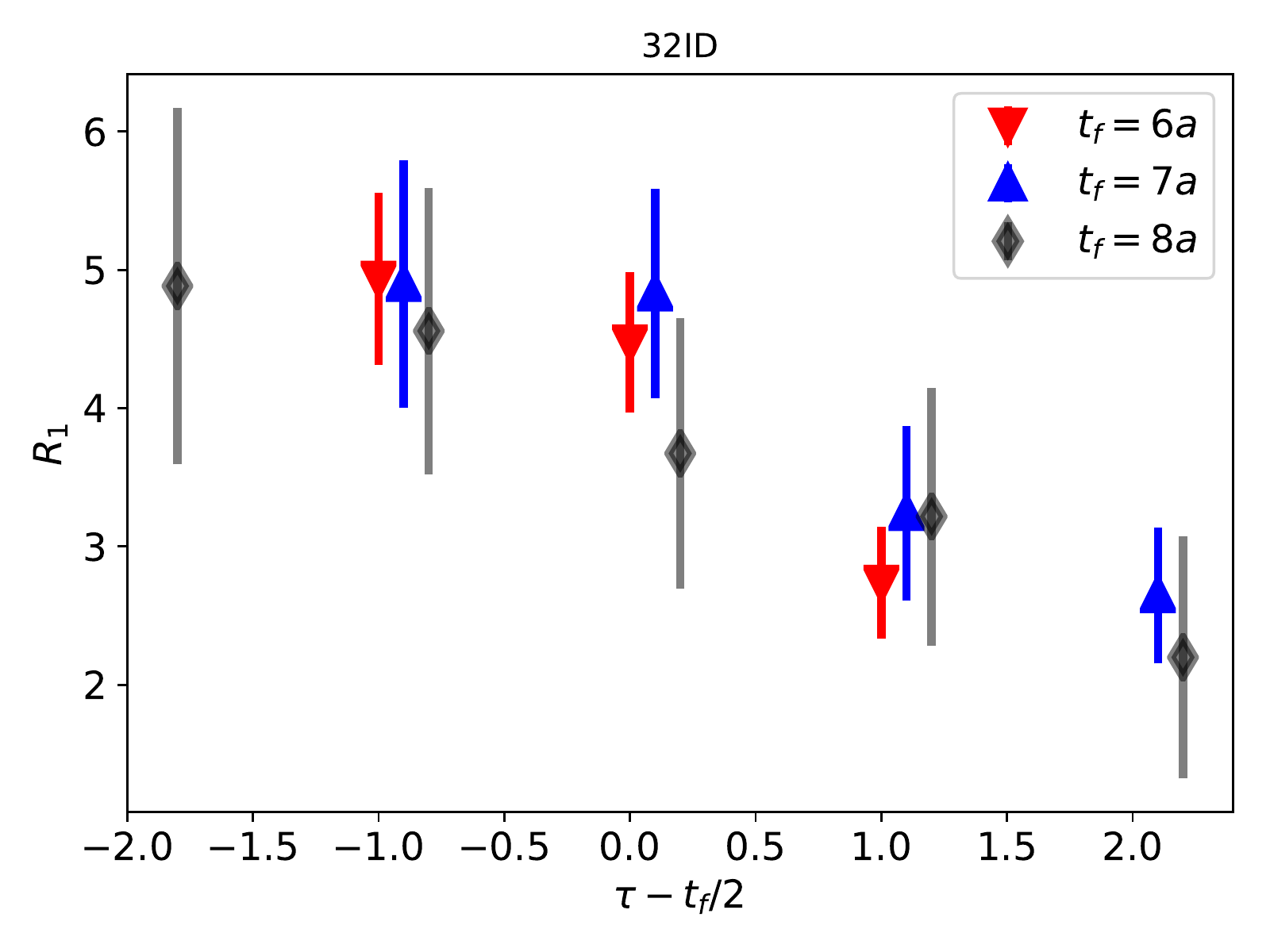}\includegraphics[scale=0.5,page=1]{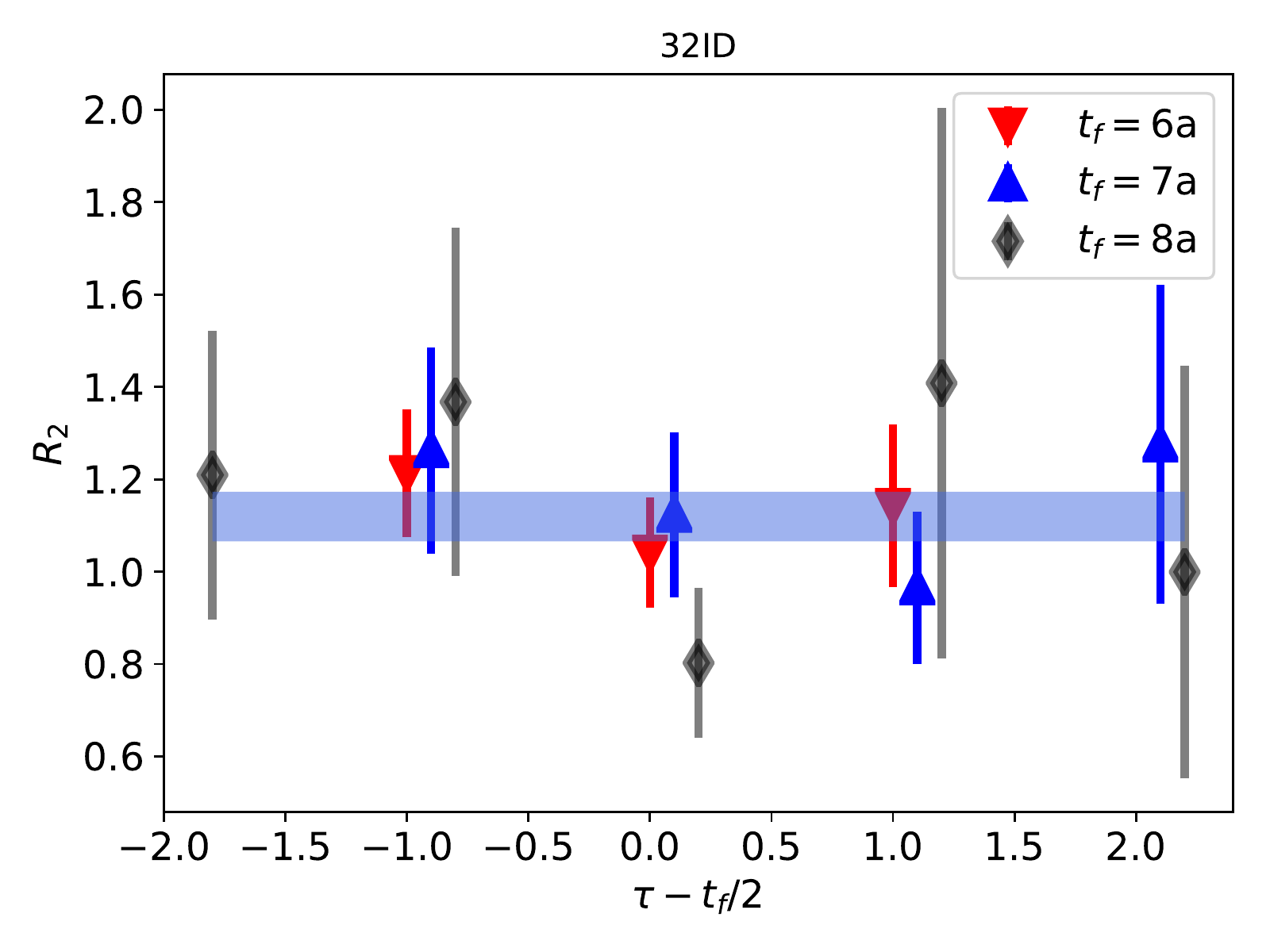}
\par\end{centering}
\begin{centering}
\includegraphics[scale=0.5,page=1]{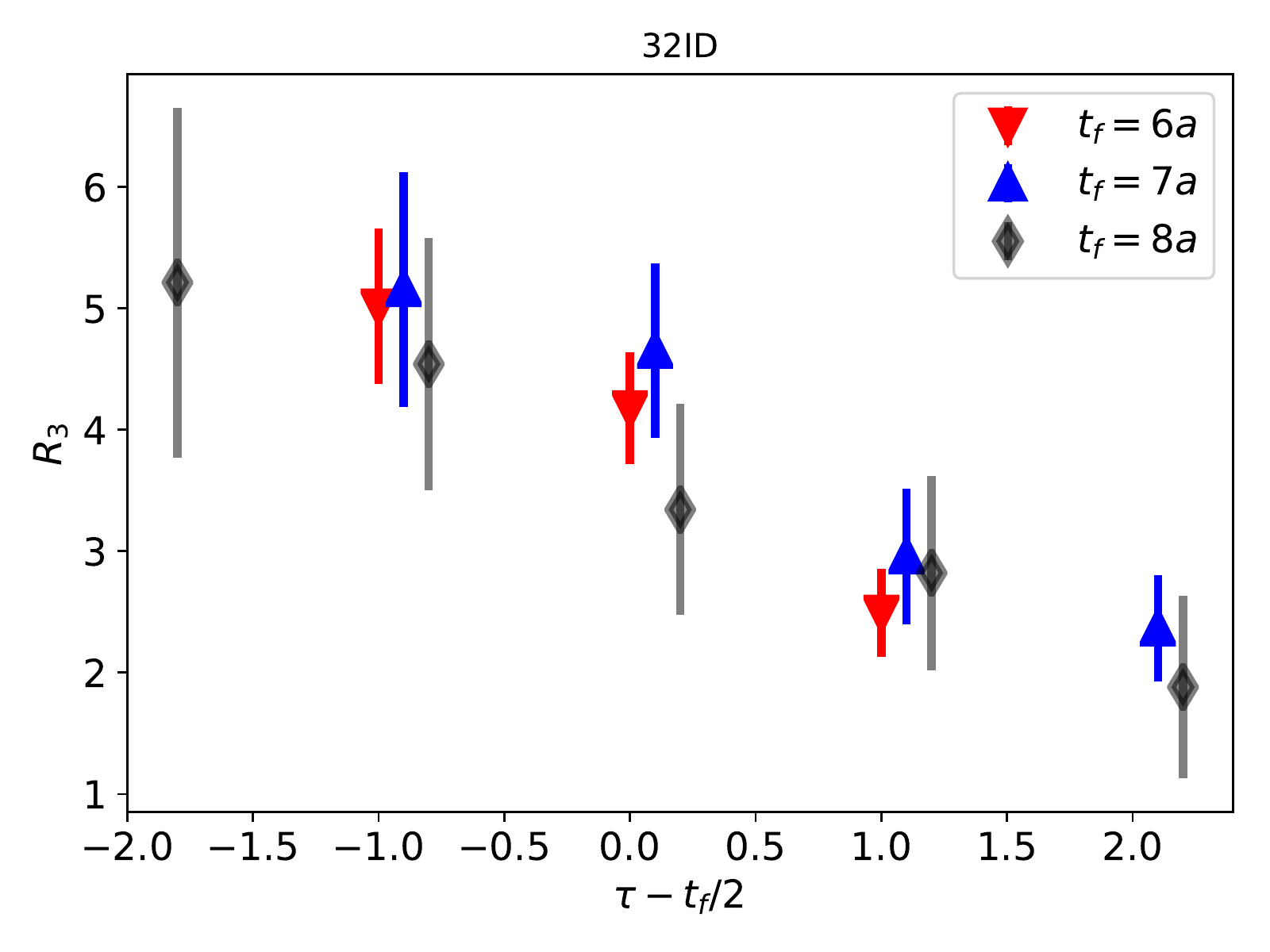}\includegraphics[scale=0.5,page=1]{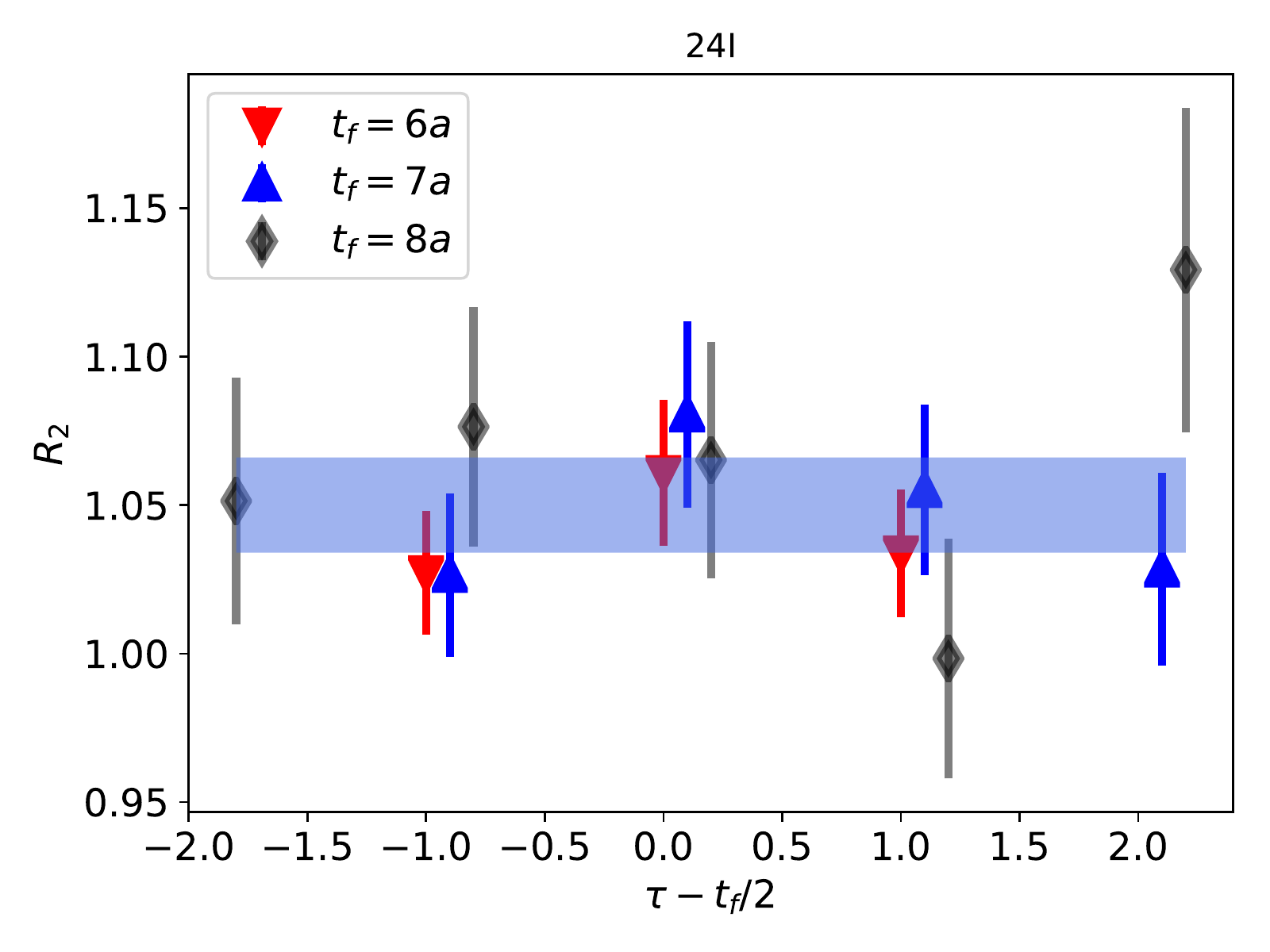}
\par\end{centering}
\caption{Ratios $R_{1}$, $R_{2}$ and $R_{3}$ on the 32ID lattice and the
ratio $R_{2}$ on the 24I lattice around the unitary points with momentum
transfer $|\vec{q}|=\frac{2\pi}{L}$ are plotted as a function of $\tau-\frac{t_f}{2}$. Three source-sink separations $t_f$
are included. The blue bands show the constant fit results of $R_{2}$.
Points of different $t_f$ are shifted slightly to enhance the legibility and the transparency of the points with $t_f=8a$ is increased for the same purpose.
\label{fg:AWI_check}}
\end{figure}

The numerical results of ratios $R_{1}$, $R_{2}$ and $R_{3}$ on
the 32ID lattice around the unitary point with momentum transfer $|\vec{q}|=\frac{2\pi}{L}$
are plotted in Figure \ref{fg:AWI_check} as a function of $t$. Three
$t_{f}$ are included so the $t_{f}$ dependence can be checked. It
can be seen that $R_{3}$ (lower left panel) is merely slightly different from $R_{1}$  (upper left panel) 
and they agree with each other quite well within errors, meaning that
there would be no difference regardless of whether we check the AWI
on the form factor level or on the ME level with the partial derivatives
replaced by energy and momentum transfer in the latter case. This
also serves as a sanity check of our calculation. The values of $R_{1}$
or $R_{3}$ are far away from $1$ and are not flat versus $t$, making
it very hard to have a reliable fit. But the situation of $R_{2}$ in the upper right panel
is much different. The points are more regular and a value of
$1.091(76)$ can be easily extracted by a constant fit which is quite
consistent with the isovector normalization constant $Z_{A}^{3}({\rm 32ID})=1.141(1)$
computed using pion 2-point functions as in Reference \citep{Liu:2013yxz}.
The ratio of $R_{2}$ on the 24I lattice is also calculated and shown
in the lower right panel of Figure \ref{fg:AWI_check}. Unlike the normalization factor $k_{A}\sim1.4$
obtained in the previous study through the ratio $R_1$ by using form factors, $R_{2}$ is also  consistent with the isovector
normalization constant $Z_{A}^{3}({\rm 24I})=1.111(6)$ where the
value from a constant fit is $1.074(24)$.

\begin{figure}[tbp]
\begin{centering}
\includegraphics[scale=0.5,page=1]{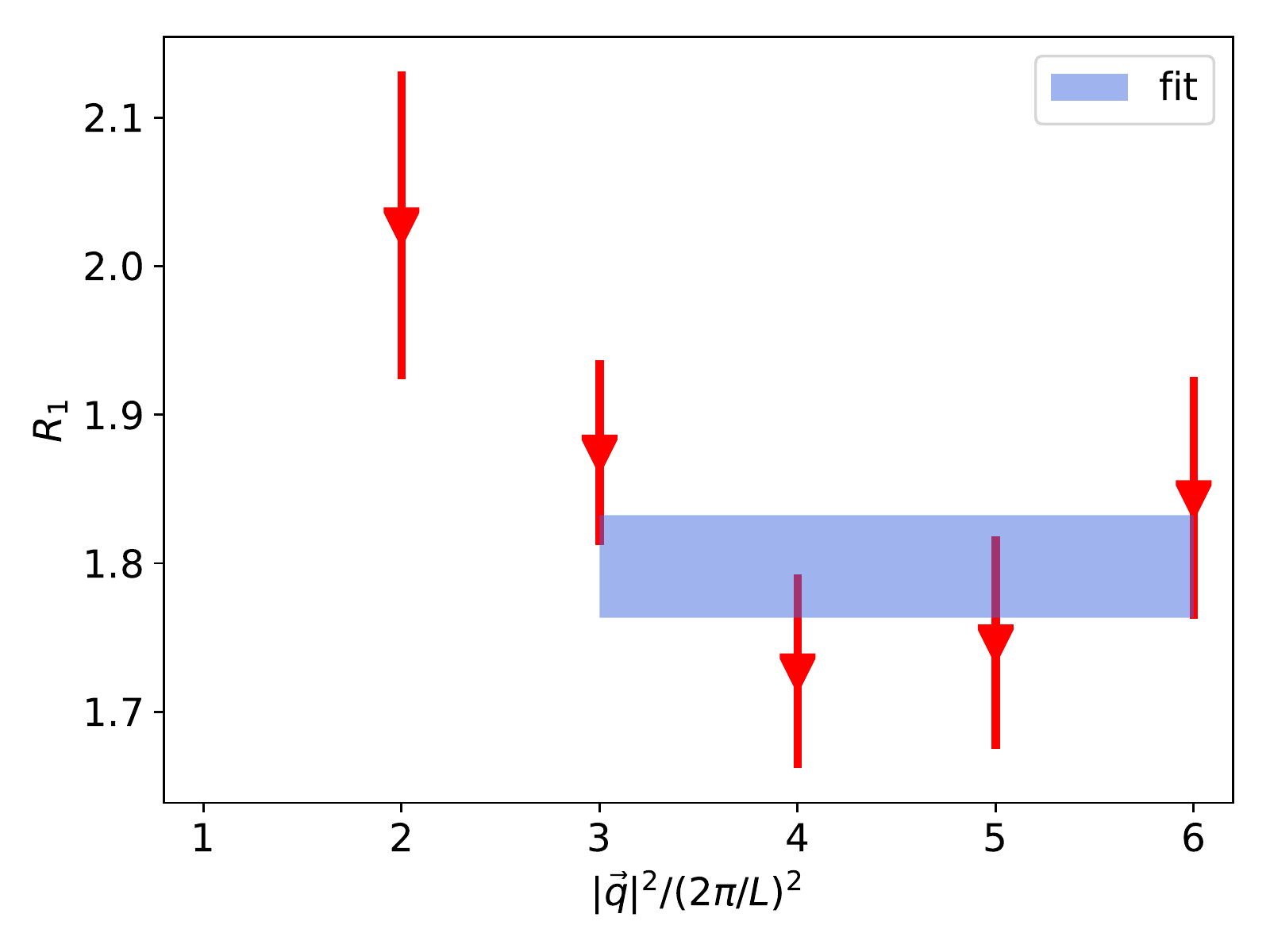}\includegraphics[scale=0.5,page=1]{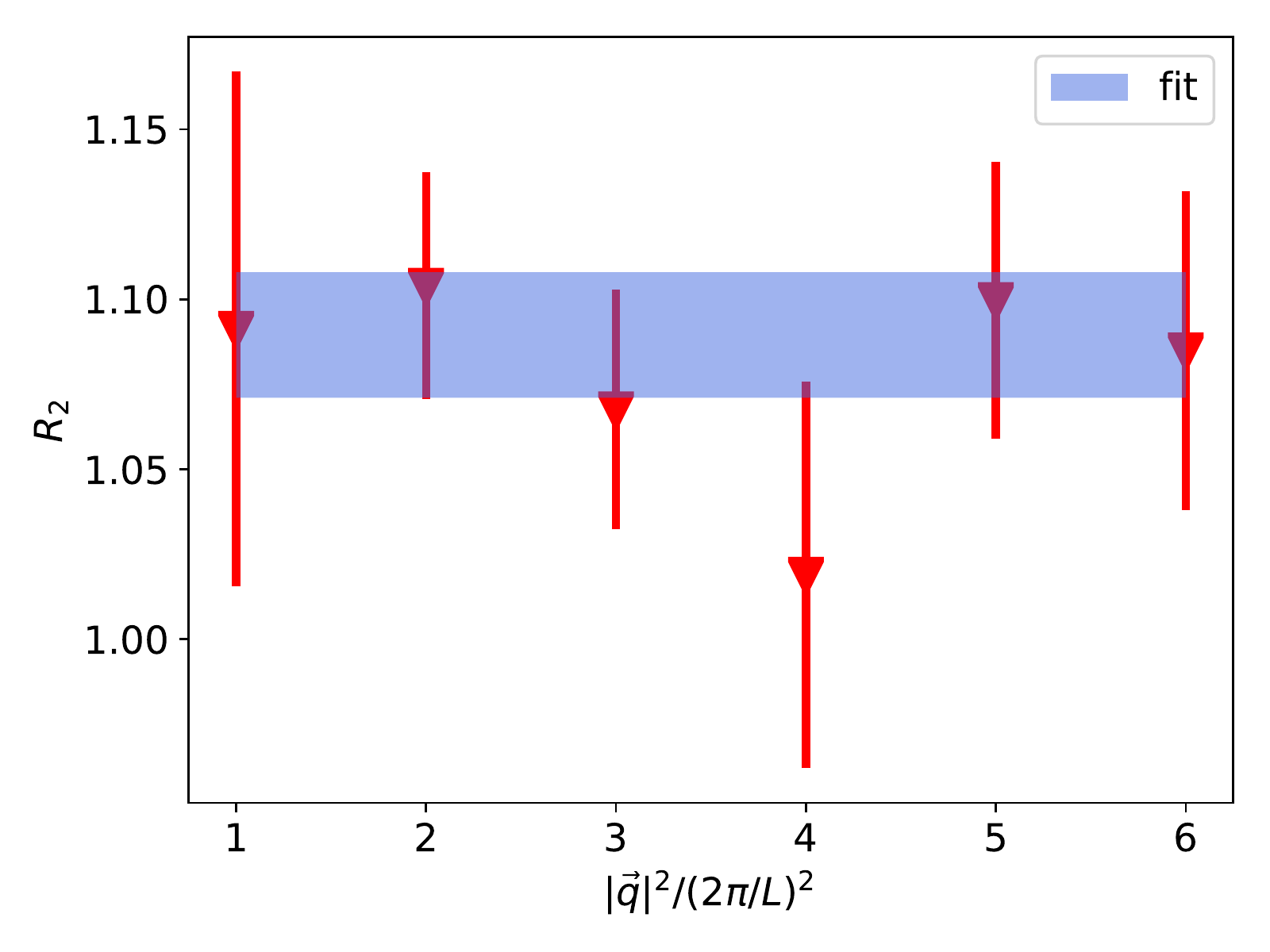}
\par\end{centering}
\caption{The momentum transfer dependence of $R_{1}$ and $R_{2}$ on the 32ID
lattice at the unitary point. The blue bands are the constant fit
results. Each point comes from a fit combining different $t_f$. The point at the first momentum transfer in the left panel
is missing because the corresponding two-state fit fails. \label{fg:AWI_vs_mom}}
\end{figure}

The momentum transfer dependence of $R_{1}$ and $R_{2}$ are plotted
in Figure \ref{fg:AWI_vs_mom}. Each point on the plot of $R_{1}$
comes from a two-state fit while the points on the plot of $R_{2}$
come from constant fits. In the $R_{1}$ plot, except for the first
two $|\vec{q}|^{2}$ (the point of the first $|\vec{q}|^{2}$ is not shown in the figure since the corresponding two-state fit fails), 
the values are basically flat within errors and the fitted
value of a constant fit is $1.798(35)$. If we believe that
the ratio $R_{1}$ is a proper check of the AWI, this value should
be used as a normalization factor. In the $R_{2}$ case, all the points
lie on a constant line within errors and a constant fit excluding
the 4th point gives $1.096(15)$, quite consistent with $Z_{A}^{3}({\rm 24I})=1.111(6)$. The problem now is to understand
the discrepancy and to determine which one is correct.

\begin{figure}[tbp]
\begin{centering}
\includegraphics[scale=0.5,page=1]{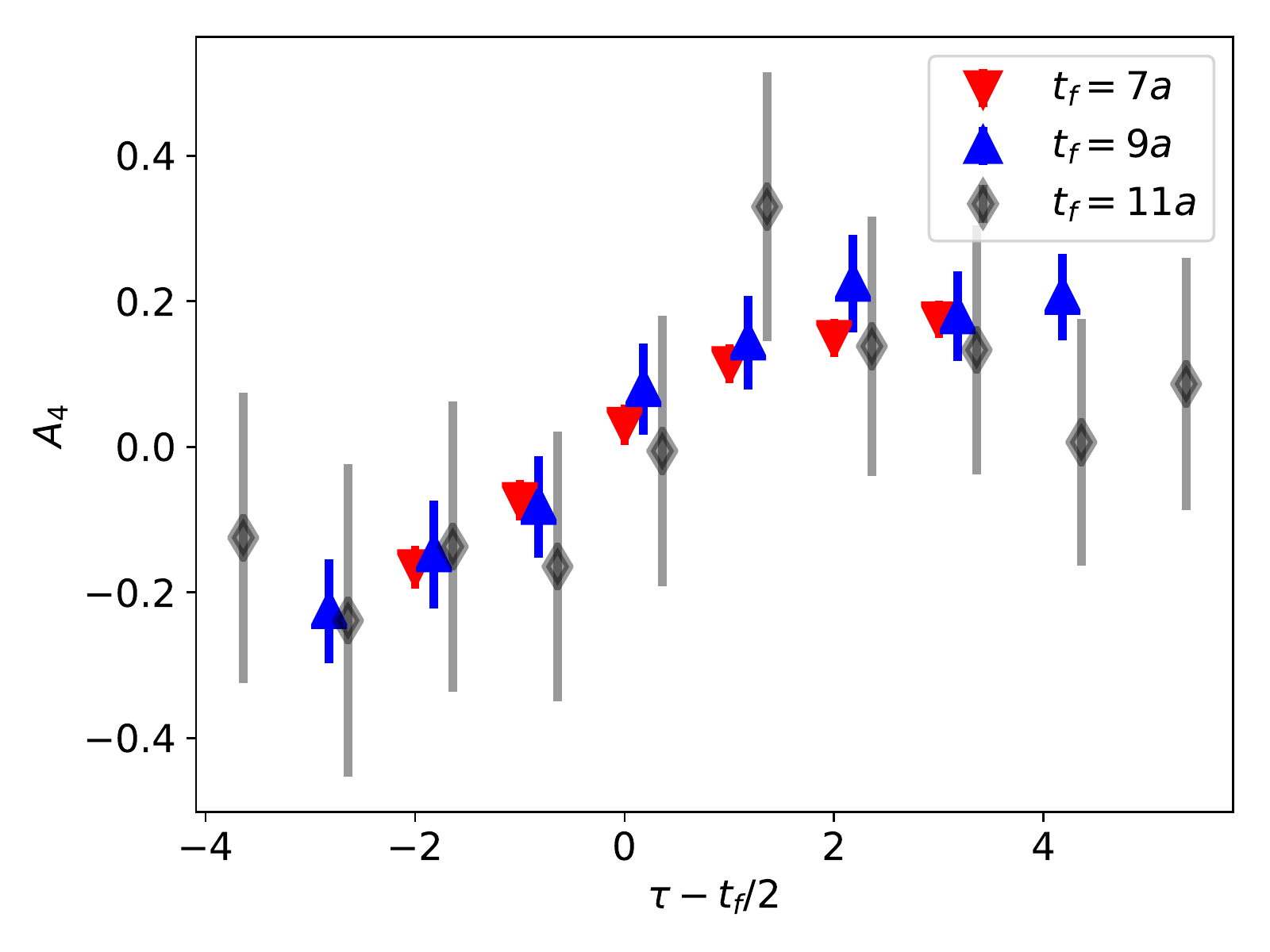}\includegraphics[scale=0.5,page=1]{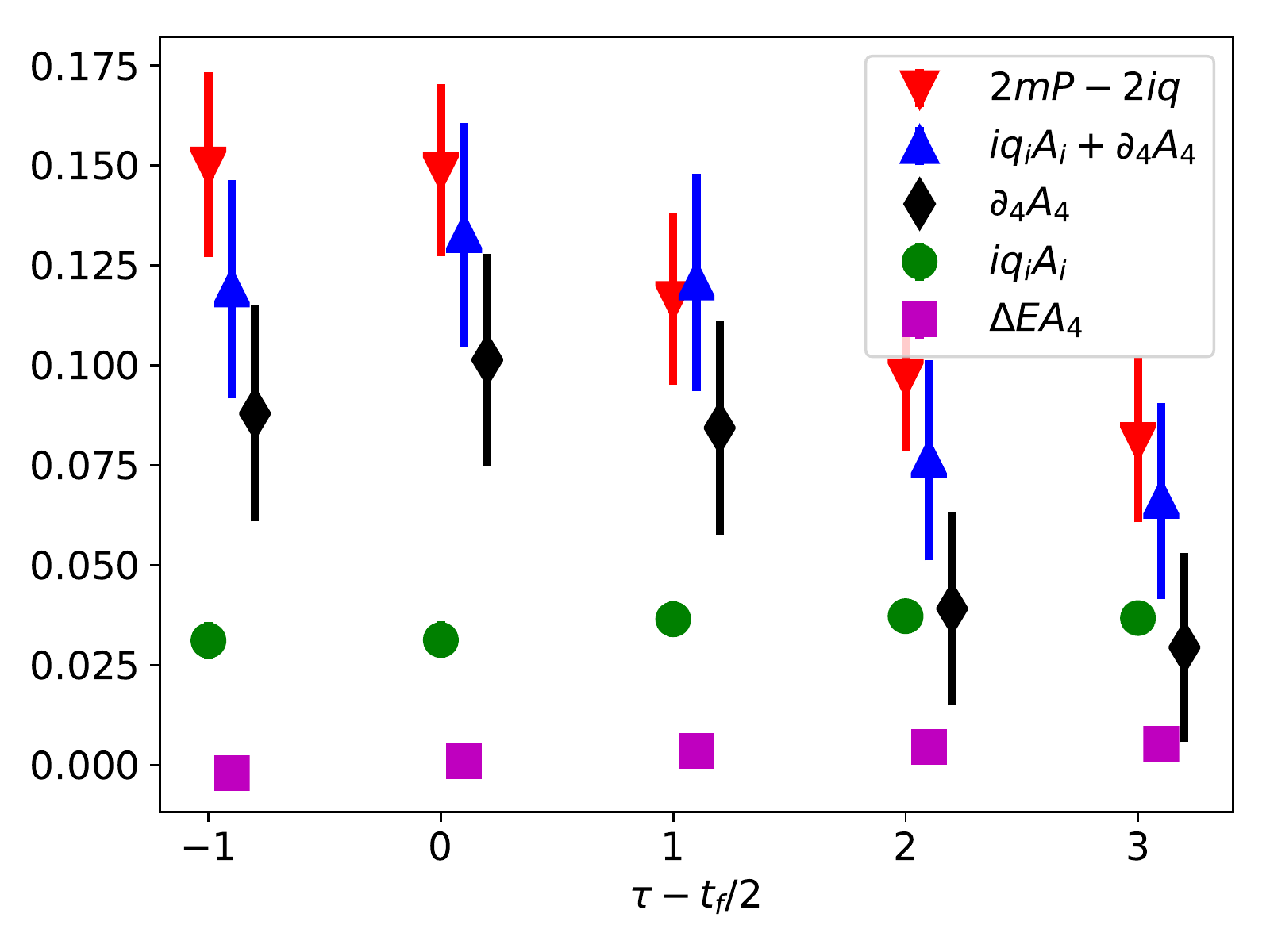}
\par\end{centering}
\caption{The behavior of $\left\langle A_{4}\right\rangle $ with respect to
$\tau$ and $t_{f}$ (left panel) and the components of the AWI at $t_f=7a$ (right
panel). The results are from the 32ID lattice at the unitary points
with momentum transfer $|\vec{q}|=\frac{2\pi}{L}$. The legend $\partial_{4}A_{4}$
stands for the $\left\langle A_{4}(\tau,\vec{q})\text{\textminus}A_{4}(\tau-1,\vec{q})\right\rangle $
term. Points of different $t_f$ are shifted slightly to enhance the legibility and the transparency of the points with $t_f=11a$ in the left panel is increased for the same purpose.
\label{fg:AWI_compare}}
\end{figure}

Since $R_1$ and $R_3$ in Figure \ref{fg:AWI_check} are quite similar, we shall only compare $R_3$ and $R_2$.
It is easy to see that the only difference between $R_{2}$ and $R_{3}$
is to use $\left\langle A_{4}(\tau,\vec{q})\text{\textminus}A_{4}(\tau-1,\vec{q})\right\rangle $
or $\Delta E\left\langle A_{4}(\tau,\vec{q})\right\rangle $. We have
proven that they are exactly the same if there are no excited-state
effects. So it is useful to see what $\left\langle A_{4}(\tau,\vec{q})\right\rangle $
looks like. The left panel of Figure \ref{fg:AWI_compare} shows the
behavior of $\left\langle A_{4}(\tau,|\vec{q}|=\frac{2\pi}{L})\right\rangle $ as a function of $\tau$;
no obvious plateau is discernible even we go to relatively larger $t_{f}$,
which means that the excited-states contamination are large and, in this
case, even a two-state fit cannot extract the ME reliably. To be more
specific, all the components of the AWI are plotted in the right panel
of Figure \ref{fg:AWI_compare}. In this $|\vec{q}|=\frac{2\pi}{L}$
case $\Delta Ea\sim0.03$, so the values of $\Delta E\left\langle A_{4}(\tau,\vec{q})\right\rangle $
are very close to $0$; however the $\left\langle \partial_{4}A_{4}\right\rangle $
values are of order $0.1$ which pinpoints the problem. 
One can ask why the $\left\langle \partial_{4}A_{4}\right\rangle $
case is so different since there should also be some excited-states
contamination. The answer is that the AWI is actually a relation
of the current operators and it should hold regardless of whether the currents
are inserted between two nucleon ground states or excited states.
The only problem is that when we use $\left\langle A_{4}(\tau,\vec{q})\text{\textminus}A_{4}(\tau-1,\vec{q})\right\rangle =\Delta E\left\langle A_{4}(\tau,\vec{q})\right\rangle $,
we are assuming that the ME is the ground-state ME and the $\Delta E$
is the energy difference between two ground states which is, apparently,
not the case. We can thus conclude that if the conditions $\tau\gg0$
and $t_{f}\gg \tau$ are satisfied, the three ratios will be the same;
for finite $\tau$ and $t_{f}$, the ratio of $R_{2}$ is the preferable
check of the AWI. The results of $R_{2}$ show that the AWI is well
satisfied in our case and we do not need any extra normalization factor
in addition to the isovector one $Z_A^3$ to make the AWI hold for the DI calculations
for all the three lattices and all the quark masses. In other words, we have $Z_A^0=Z_A({\rm CI}) = Z_A({\rm DI})=Z_A^3$.

\subsection{CI case}

\begin{figure}[tbp]
\begin{centering}
\includegraphics[scale=0.5,page=1]{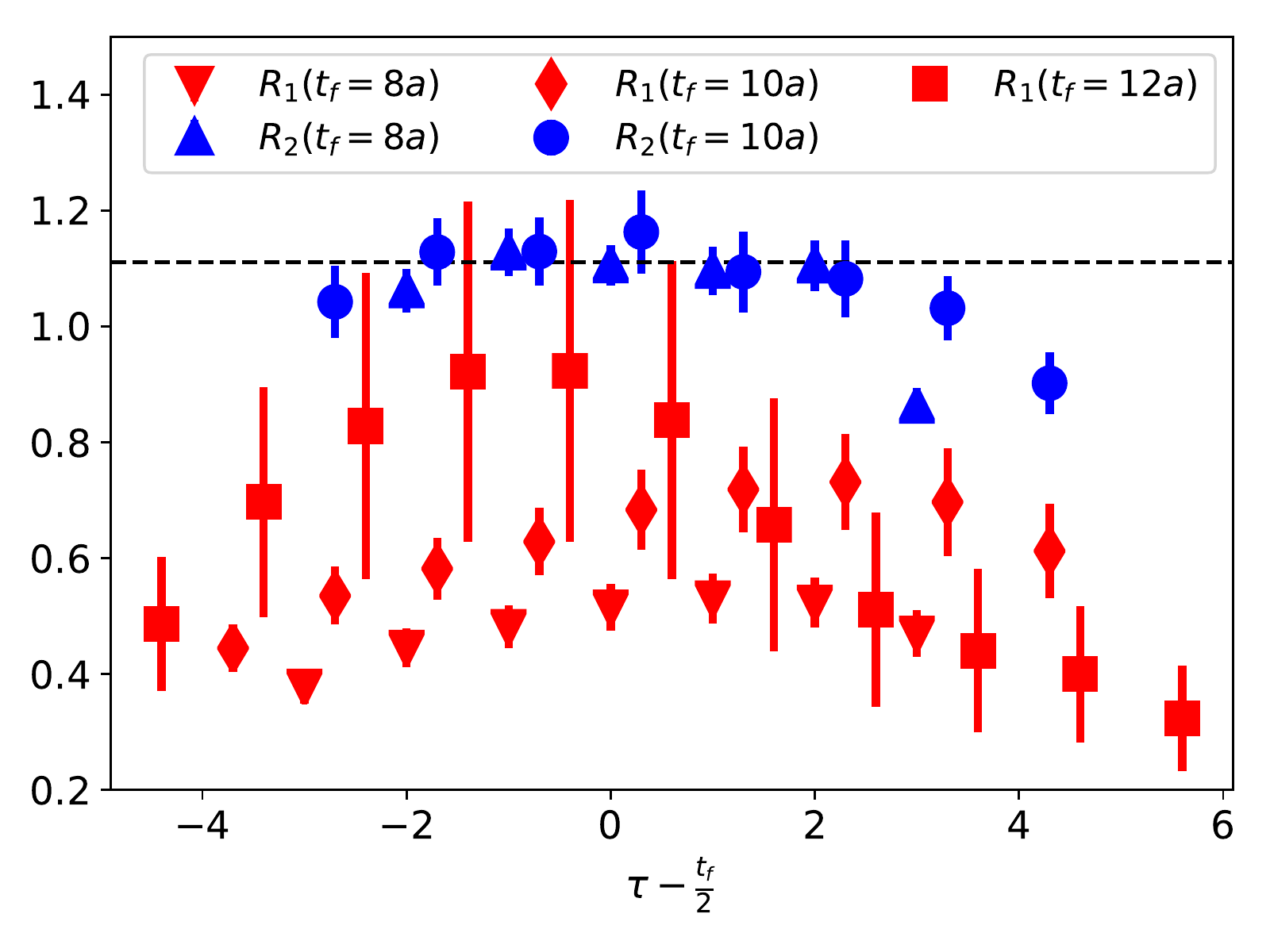}
\par\end{centering}
\caption{The ratios of $R_{1}$ and $R_{2}$ in the CI case at different $t_{f}$.
The results are from the 24I lattice at the unitary point with momentum
transfer $|\vec{q}|=\frac{2\pi}{L}$. The horizontal line indicating
$Z_{A}^{3}=1.111(6)$ shows that the values of $R_{2}$ are consistent
with this normalization factor except for the boundary points. $R_{1}$
shows large discrepancy and approaches the horizontal line with large
$t_{f}$.  Points of different $t_f$ are shifted slightly to enhance the legibility.\label{fg:AWI_CI_24}}
\end{figure}

We also check the chiral Ward identity in the CI case. In fact, the
violation of the chiral Ward identity in terms of form factors at
small momentum transfers in the CI case is also observed
and reported in \citep{Rajan:2017lxk}, where their formula is equivalent to checking
the ratio $R_{1}$. In the CI case, the definitions of $R_{1}$ and
$R_{2}$ are the same as those in the DI case but without the topological
charge term. The results of $R_{1}$ and $R_{2}$ at different $t_{f}$
of the 24I lattice are plotted in Figure \ref{fg:AWI_CI_24} as an
example. The horizontal line in the figure indicating $Z_{A}^{3}=1.111(6)$
shows where the points of the ratios should be if the chiral Ward identity
holds. Again, the points of $R_{1}$ show obvious discrepancy. But
a trend that the points are approaching the horizontal line with larger
$t_{f}$ can be observed. As a contrast, the points of $R_{2}$ at
both $t_{f}=8a$ and $t_{f}=10a$ do lie on the target line except
for the boundary points, showing valid Ward identity.

The results of the CI case are similar and the conclusion is the
same. The ratio $R_{2}$ shows well established chiral Ward identity
while $R_{1}$ shows violation. The difference of $R_{1}$ and $R_{2}$
reflects how we treat the $A_{4}$ term. Even if the form factors $g_A(q^2)$ and $h_A(q^2)$ are calculated
using $A_{i}$ only, the ratio $R_{1}$ is still problematic since
Eq. (\ref{eq:partial_ME}) is used in the derivation and it assumes
that the ME of $A_{4}$ gives the same form factors without excited-states
contamination. A cure to this problem is to go to large enough $t_{f}$
where the excited-states contamination can be ignored.
Unfortunately, this requires much larger statistics. We will test this in the future.

\section{Disconnected-insertion contribution\label{sec:Disconnected-insertion-contribut}}

\begin{figure}[tbp]
\centering{}
\includegraphics[scale=0.5,page=1]{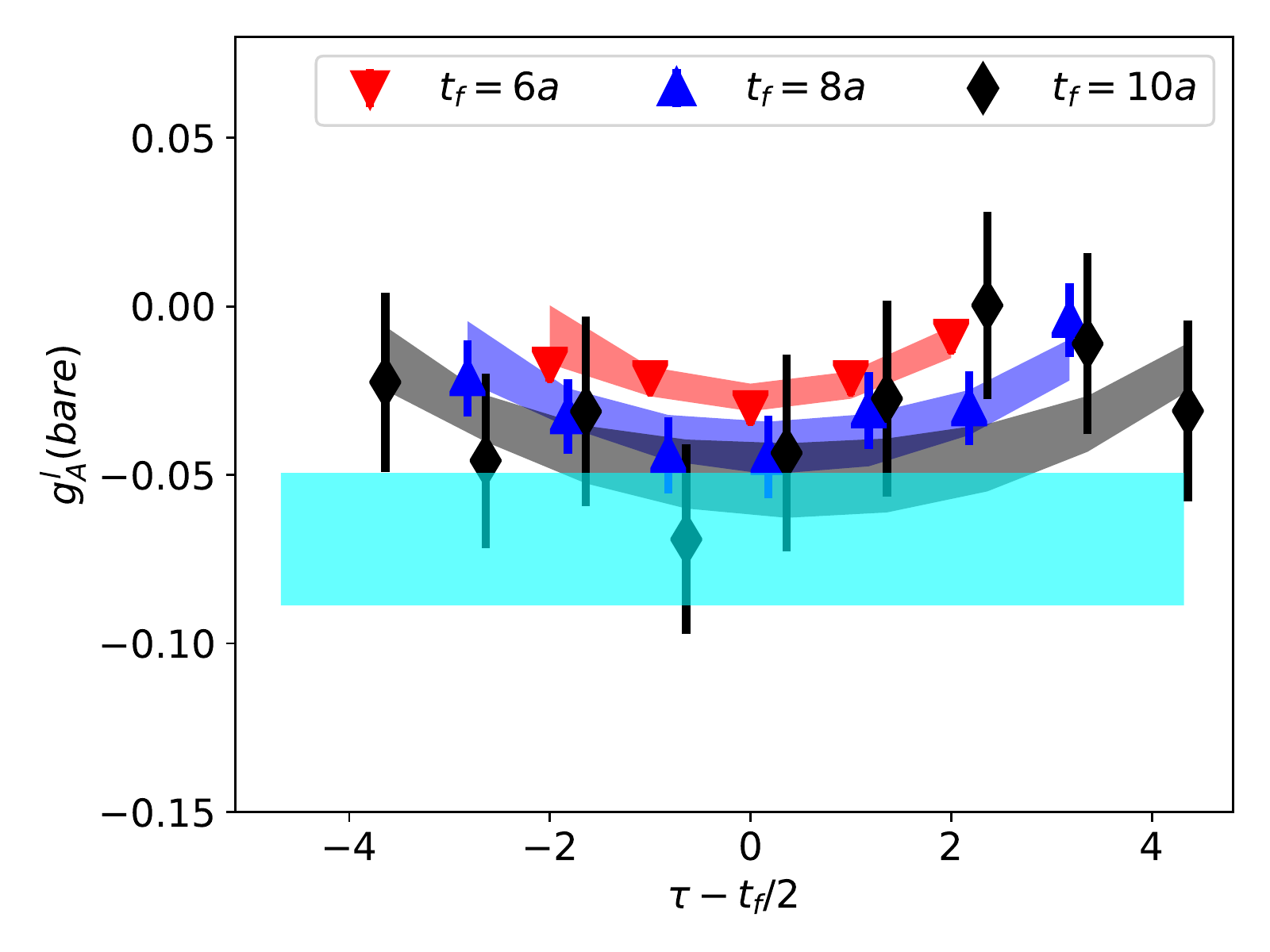}
\includegraphics[scale=0.5,page=1]{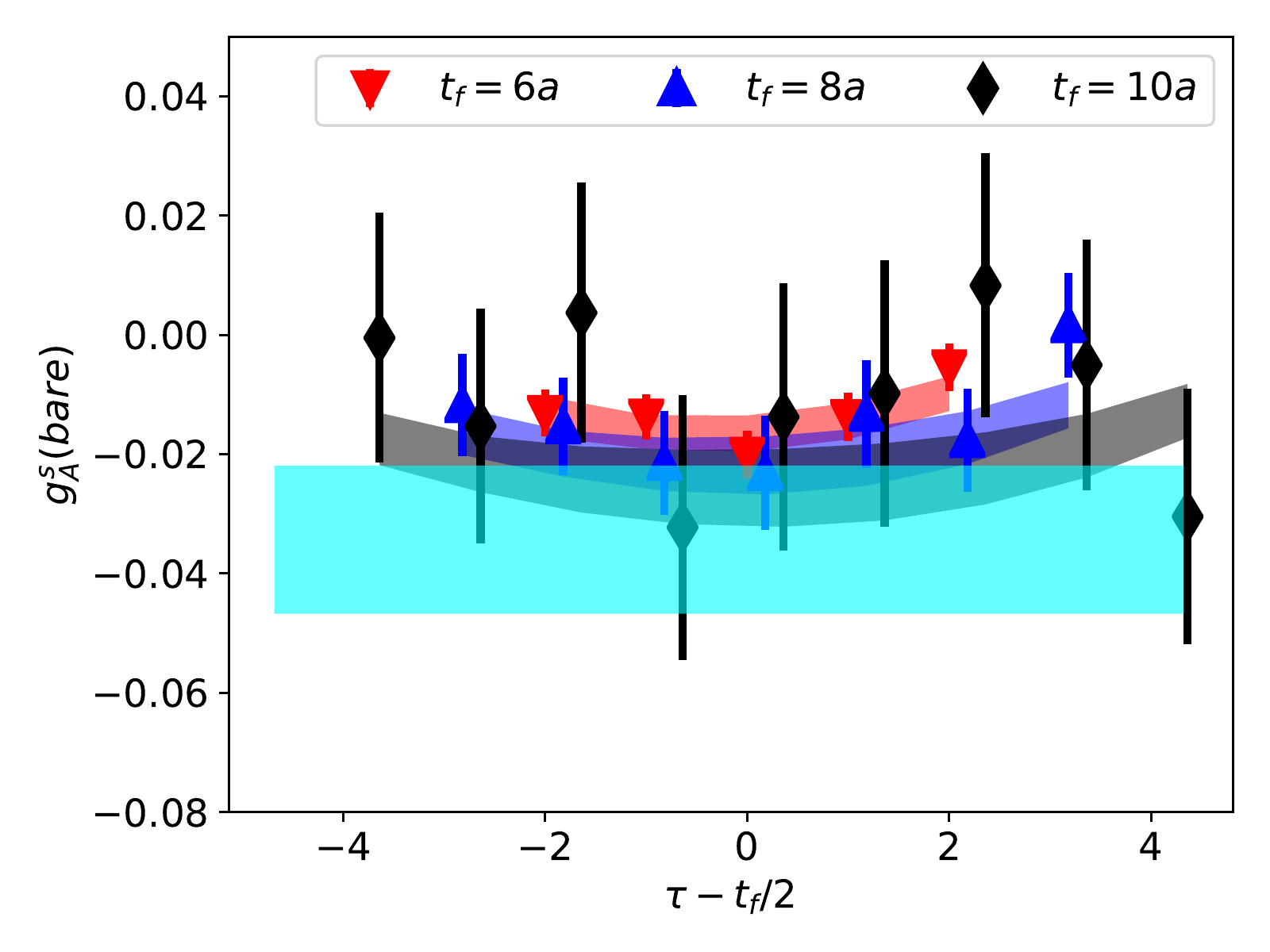}\caption{Examples of two-state fits on the 32ID lattice at the unitary point.
The light quark result is on the left while the strange quark result
is on the right. Cyan bands are the fitting results from $t_f=6a$, $7a$ and $8a$ which are consistent
with the points of large separations. Points of different $t_f$ are shifted slightly in the horizontal direction to enhance the legibility. \label{fig:two-state-fits}}
\end{figure}

As is mentioned above, we use the two-state fit to extract the MEs.
Examples of the fitting on the 32ID lattice at the unitary point can
be found in Figure \ref{fig:two-state-fits}; both the light and strange
quark results are included. We use three source-sink separations $t_{f}=6a$,
$7a$ and $8a$, which correspond to 0.86 fm, 1.00 fm and 1.15 fm respectively,
to carry out the fit. The fitting results are shown by the cyan bands,
which are consistent with the data points of largest separation within errors.
The contact points on either the source
or the sink side are always excluded and more points may be excluded
to have better $\chi^{2}$. A prior value of $\delta m$ is used to
stabilize the fits. A criterion used to choose the prior value and
width is that the fitting result of $\delta m$ should have statistical
significance and the final result of $g_{A}$ should not be too far
away from the data points of large separations. 
The difference of the fitting results due to the choice of prior values is included in the systematic uncertainty.
For the
24I and 32I lattice, the two-state fits are done similarly. A table
listing all the fitting setups is given below (Table \ref{tab:2s_fitting_DI}).

\begin{table}
\caption{Setups of the two-state fits in the DI case. The source-sink separations
used in the fits, the number of points dropped on the source side,
the number of points dropped on the sink side and the prior value
and width of $\delta m$ are listed for each lattice and for both light and strange flavors. \label{tab:2s_fitting_DI}}
\begin{centering}
\begin{tabular}{ccccc}
\cline{1-5}
\multicolumn{1}{c|}{lattice/flavor} &\multicolumn{1}{c|}{separations (a)}&\multicolumn{1}{c|}{source drop}
&\multicolumn{1}{c|}{sink drop}&\multicolumn{1}{c}{prior $\delta ma$}\tabularnewline
\hline 
32ID/$l$ & 6, 7, 8 & 2 & 2 & 0.4(0.1)\tabularnewline
\hline 
32ID/$s$ & 6, 7, 8 & 1 & 1 & 0.3(0.1)\tabularnewline
\hline 
24I/$l$ & 8, 9, 10, 11 & 2 & 2 & 0.4(0.1)\tabularnewline
\hline 
24I/$s$ & 8, 9, 10, 11 & 2 & 1 & 0.3(0.2)\tabularnewline
\hline 
32I/$l$ & 9,10,11,12 & 3 & 2 & 0.4(0.1)\tabularnewline
\hline 
32I/$s$ & 9,10,11,12 & 3 & 2 & 0.3(0.2)\tabularnewline
\hline 
\end{tabular}
\par\end{centering}
\end{table}

\begin{figure}[tbp]
\begin{centering}
\includegraphics[scale=0.5,page=1]{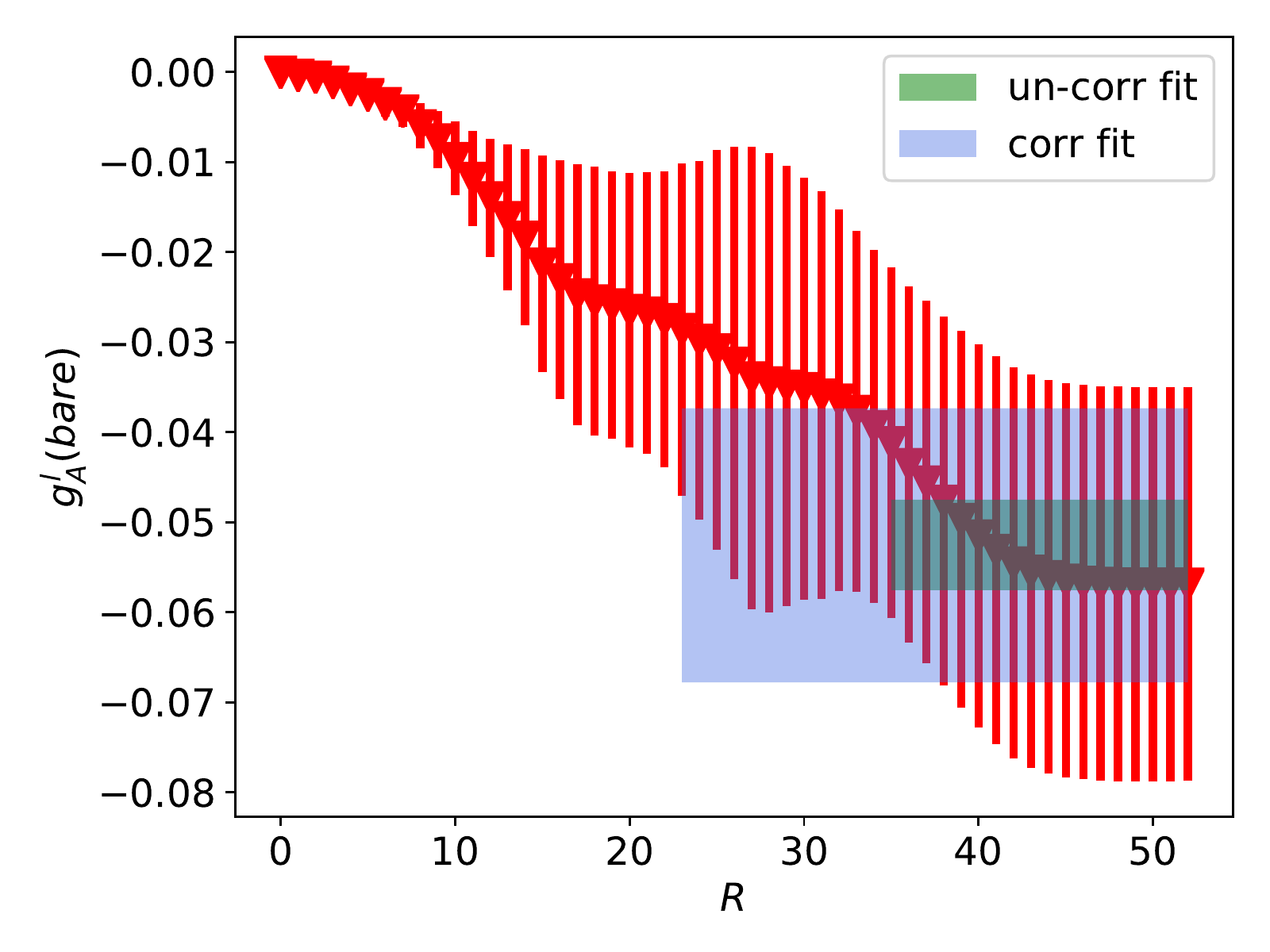}\includegraphics[scale=0.5,page=1]{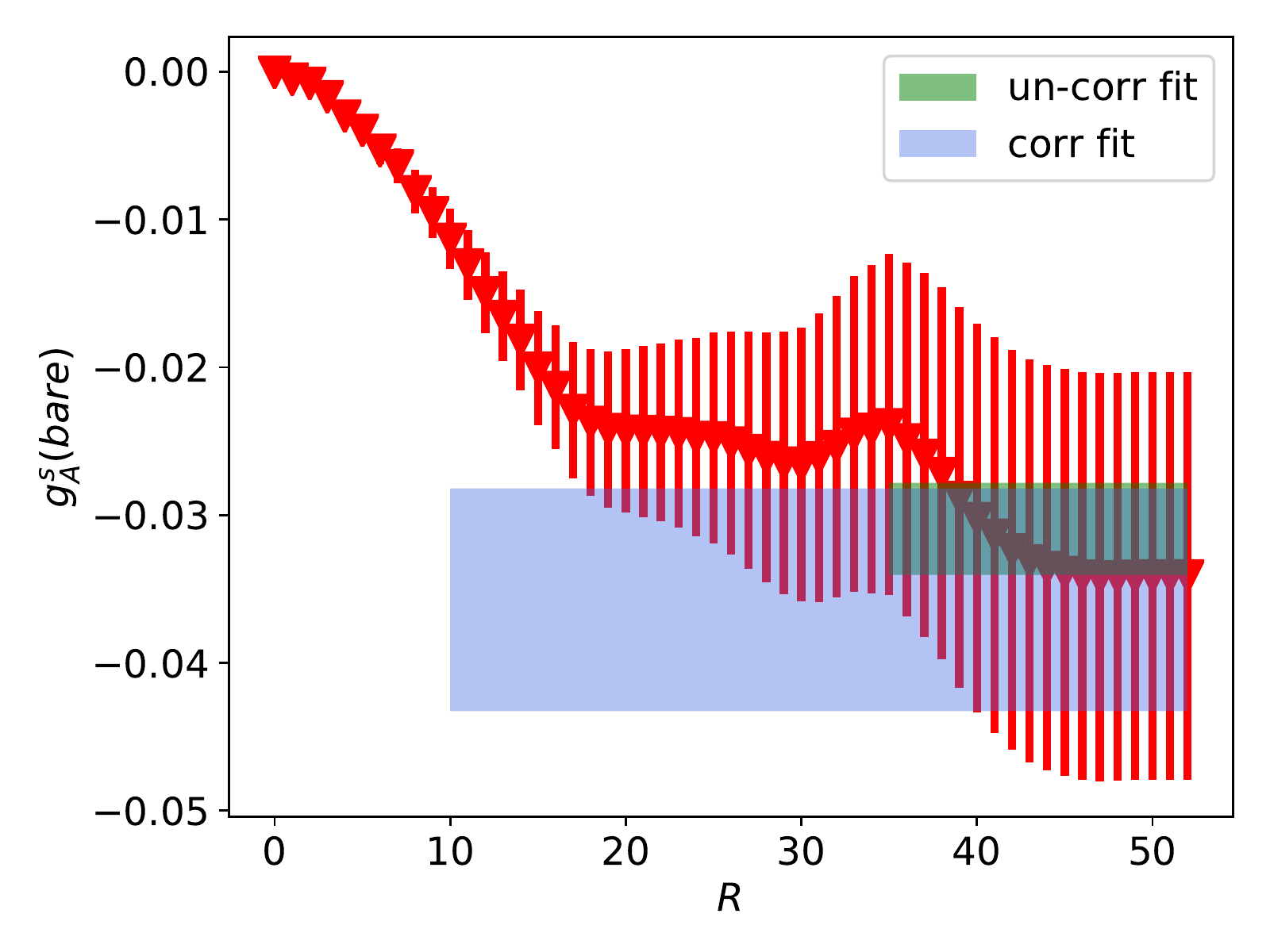}
\par\end{centering}
\caption{The $R$ dependence of the axial coupling of the 32ID lattice at the
unitary point. The left panel is for the light quark case and the
right panel is for the strange quark case. Blue bands and green bands
are for correlated fit and uncorrelated fit, respectively. \label{fig:cut-of-R}}
\end{figure}

The cluster-decomposition error reduction (CDER) technique \citep{Liu:2017man}
is used in order to better control the statistical uncertainties for
the 32ID lattice where the CDER technique may improve the signal more
significantly since the size of this lattice $L\sim4.6$ fm is relatively
large, while we do not use this technique for the 24I and 32I lattice
due to their small sizes ($L\sim2.7$ fm and $L\sim2.6$ fm respectively).
In order to use the CDER technique, the 3-point functions can be rewritten
as 
\begin{equation}
\label{C_3_R}
C_{3,\mu}(t_{f},\tau,R)=\sum_{\vec{x},|\vec{r}|<R}\langle\chi(t_{f},\vec{x})A_{\mu}(\tau,\vec{x}+\vec{r})\bar{\chi}(0,\mathcal{G})\rangle
\end{equation}
where we put a cutoff $R$ to the distance between the quark loop
and the sink of the nucleon propagator and we can vary $R$ to obtain
different 3-point functions. As demonstrated in \citep{Liu:2017man},
the signal will saturate after $R$ is larger than the corresponding correlation
length but the noise will keep growing due to the the fact that the variances of the two disconnected operators
in their vacuum expectation values are independent of each other.
Therefore an optimal cutoff $R$ can be found if the lattice size
is larger than the correlation length between the operators whereupon
the signal-to-noise ratio is improved. However, as is shown in Figure
\ref{fig:cut-of-R} where the $g_A({\rm DI})$ from Eq.~(\ref{C_3_R}) is plotted as a function of $R$, 
no clear plateau shows until at very large $R$, especially for
the light quark case, which is probably because the correlation length
is not much smaller than the lattice size. So we cannot find an optimal
$R$ in this case. Nevertheless a correlated fit using the following
asymptotic form \citep{Liu:2017man}
\begin{equation}
C_{3}(R)=C_{3}(\infty)+k\sqrt{R}\frac{e^{-MR}}{M}
\end{equation}
with $k$ and $M$ being free parameters helps in extracting $C_{3}(\infty)$
properly. The blue bands in the figure show the results of the correlated
fit while the green bands show the results of an uncorrelated fit
in contrast. The reason why we need this comparison is because the
data points of different $R$ are strongly correlated and an uncorrelated
fit will underestimate the error very much. To keep the correlation
of different $R$, we cannot do single two-state fits respectively
for each $R$. Instead, a simultaneous two-state fit combining
all the $R$ to keep the whole correlation matrix is carried out. The error of the
correlated fit, which is not much smaller than the error of the data
points with large cutoff, is believed to be a reasonable estimation.
In this way, the final statistical uncertainties of the MEs on the
32ID lattice can be reduced by $10\%\sim40\%$ for different quark masses. 

\section{Connected-insertion contribution\label{sec:Connected-insertion-contribution}}

\begin{figure}[tbp]
\begin{centering}
\includegraphics[scale=0.5,page=1]{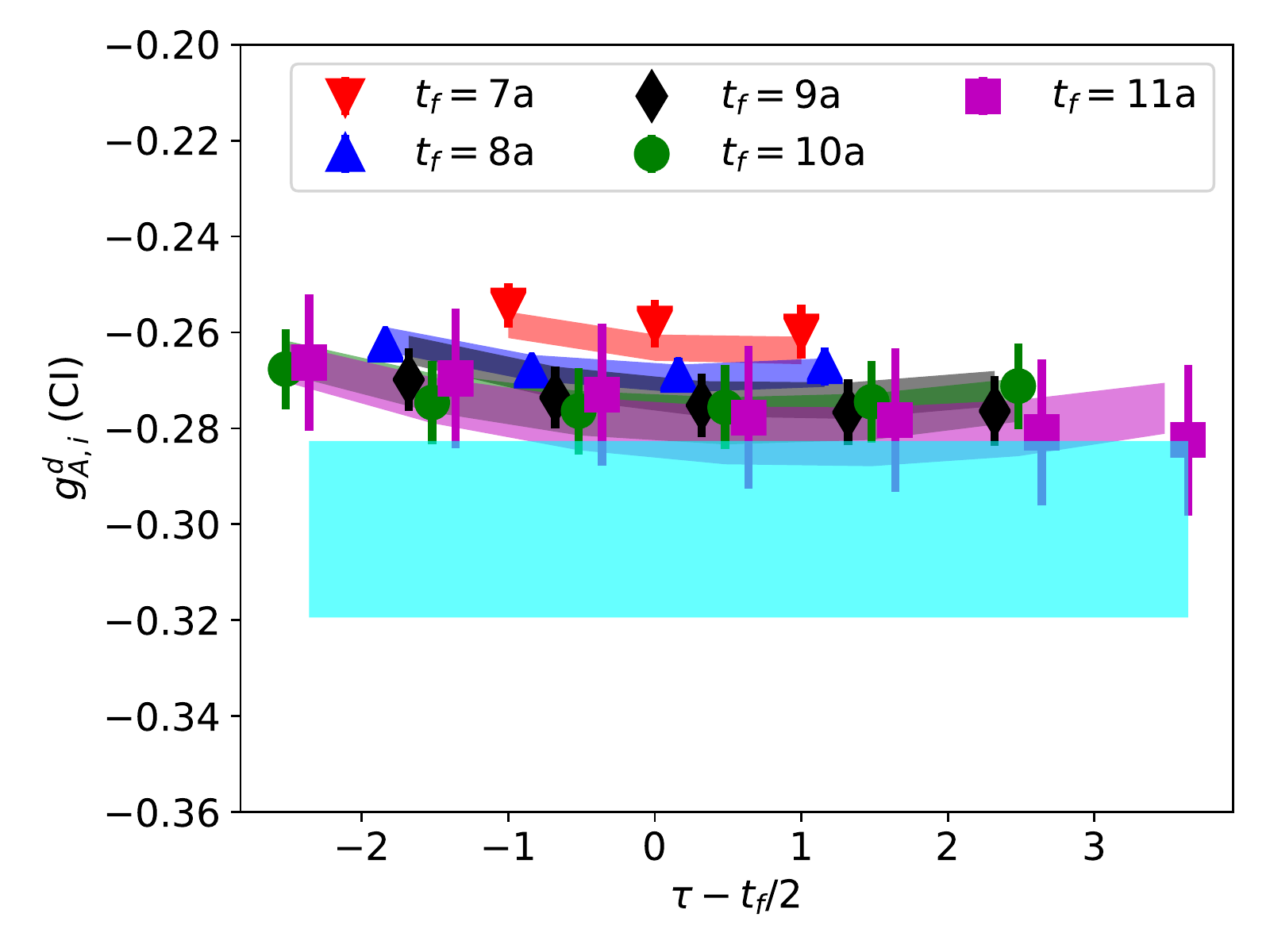}\includegraphics[scale=0.5,page=1]{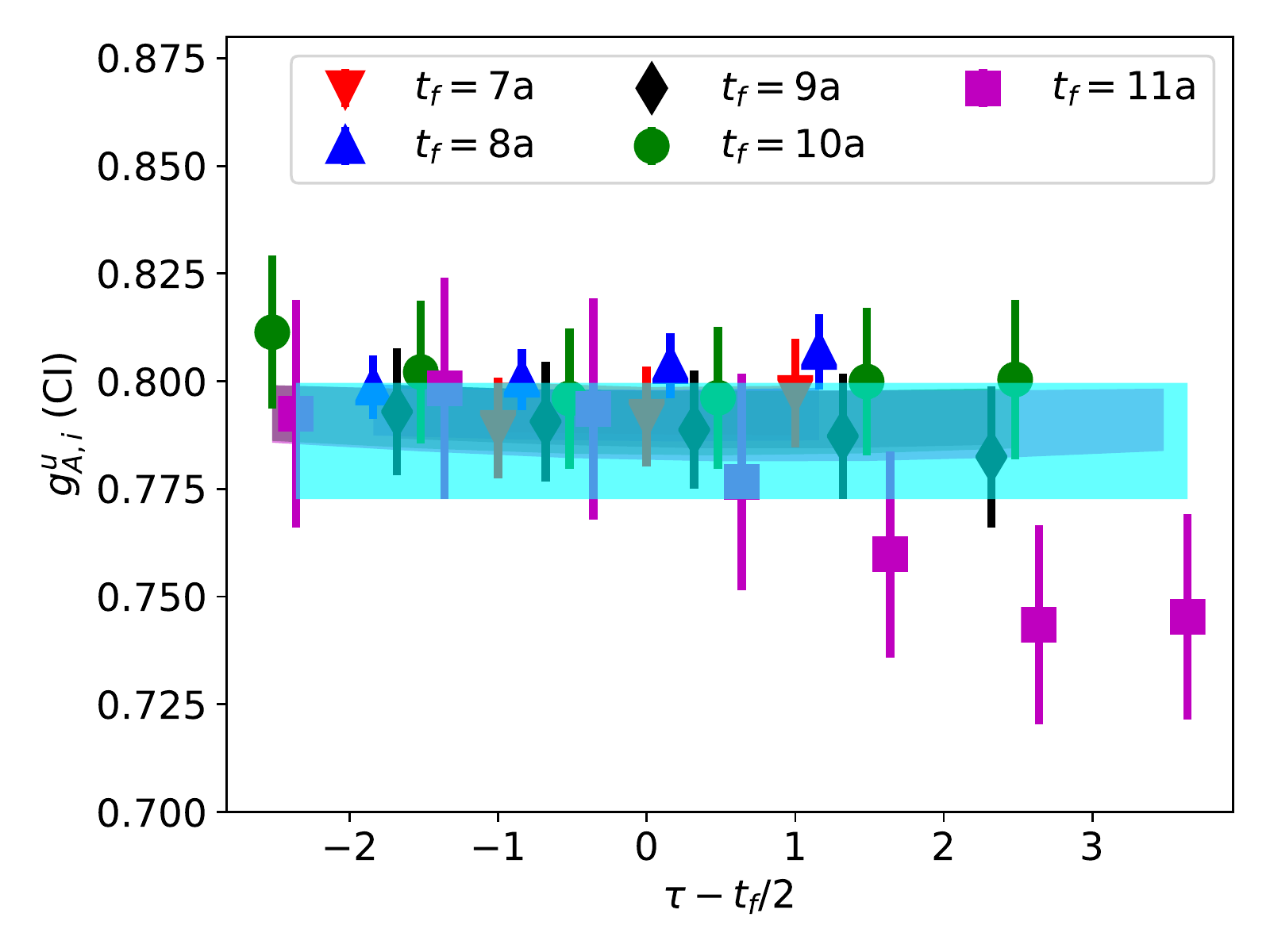}
\par\end{centering}
\caption{Two-state fit examples for $d$ quark and $u$ quark respectively
on the 32ID lattice. The label $g_{A,i}^{d}$ and $g_{A,i}^{u}$
denote the axial couplings for $d$ and $u$ quarks calculated from
current $A_{i}=\bar{\psi}i\gamma_{i}\gamma_{5}\hat{\psi}$. 
Points of different $t_f$ are shifted slightly in the horizontal direction to enhance the legibility. \label{fig:2st_fit_CI}}
\end{figure}

\begin{figure}[tbp]
\begin{centering}
\includegraphics[scale=0.5,page=1]{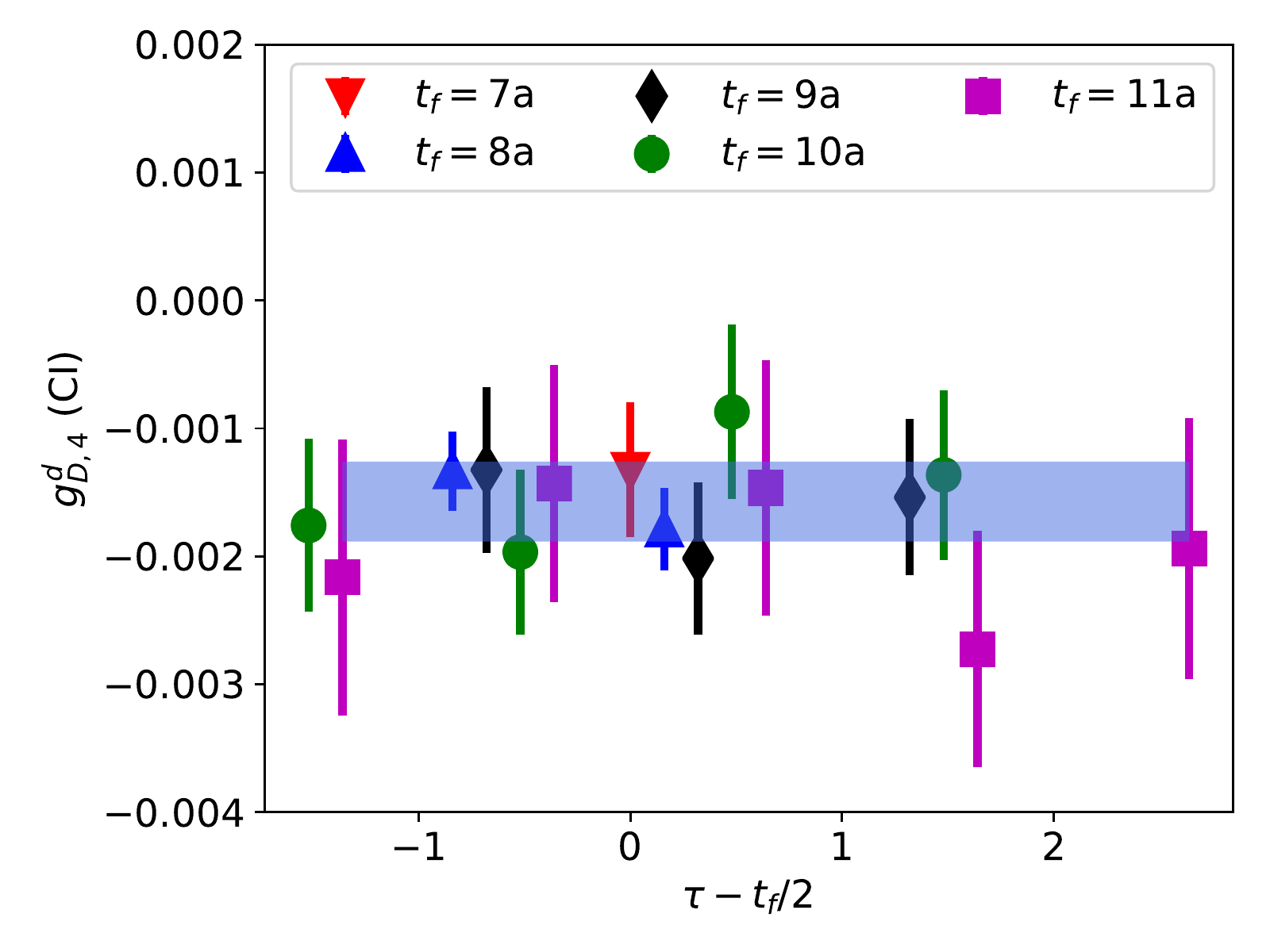}
\par\end{centering}
\caption{Constant fit example for the $d$ quark case of current $D_{4}$. Points of different $t_f$ are shifted slightly in the horizontal direction to enhance the legibility. 
\label{fig:2st_fit_CI_D4}}
\end{figure}

As for the CI case, we use the improved axial-vector current following
our previous work on the 24I and 32I lattice \citep{Liang:2016fgy} to reduce the discretization
errors on the 32ID lattice as well. For the 24I and 32I lattice, we reanalyze
the data with $u$ quark and $d$ quark separately. Two-state fits
are also applied to the MEs of $A_{i}=\bar{\psi}i\gamma_{i}\gamma_{5}\hat{\psi}$,
and the fitting setups are also listed in a similar table (Table \ref{tab:2s_fitting_CI}).
We plot the fitting results of the
32ID lattice at the unitary point in Figure \ref{fig:2st_fit_CI} 
as an example. To implement the improvement, we also need to fit for
the MEs of three more currents: $A_{4}=\bar{\psi}i\gamma_{5}\gamma_{5}\hat{\psi}$,
$D_{i}=\bar{\psi}i\sigma_{i\mu}\overleftrightarrow{D}_{\mu}\gamma_{5}\hat{\psi}$
and $D_{4}=\bar{\psi}i\sigma_{4i}\overleftrightarrow{D}_{i}\gamma_{5}\hat{\psi}$.
For these currents, the signal-to-noise ratio is not as good as that
for the $A_{i}$ case and no obvious excited-state contribution can
be observed; we are only able to make a constant fit combining several
separations. A example of $D_{4}$ is plotted in Figure \ref{fig:2st_fit_CI_D4}, note that we drop three points on each of the source and sink sides.

\begin{table}
\caption{Setups of the two-state fits in the CI case for the current $A_{i}=\bar{\psi}\gamma_{5}\gamma_{i}\hat{\psi}$.
The source-sink separations used in the fits, the number of points
dropped on the source side, the number of points dropped on the sink
side and the prior value and width of $\delta m$ are listed for each lattice and for both $u$ and $d$ quarks. \label{tab:2s_fitting_CI}}
\centering{}%
\begin{tabular}{ccccc}
\cline{1-5}
\multicolumn{1}{c|}{lattice/flavor} &\multicolumn{1}{c|}{separations (a)}&\multicolumn{1}{c|}{source drop}
&\multicolumn{1}{c|}{sink drop}&\multicolumn{1}{c}{prior $\delta ma$}\tabularnewline
\hline 
32ID/$u$ & 7, 8, 9, 10, 11 & 3 & 2 & 0.35(0.1)\tabularnewline
\hline 
32ID/$d$ & 7, 8, 9, 10, 11 & 2 & 2 & 0.35(0.1)\tabularnewline
\hline 
24I/$u$ & 8, 10, 11, 12 & 2 & 1 & 0.3(0.1)\tabularnewline
\hline 
24I/$d$ & 8, 10, 11, 12 & 3 & 1 & 0.3(0.1)\tabularnewline
\hline 
32I/$u$ & 12, 14, 15 & 2 & 2 & 0.2(0.1)\tabularnewline
\hline 
32I/$d$ & 12, 14, 15 & 4 & 3 & -\tabularnewline
\hline 
\end{tabular}
\end{table}

\begin{figure}[tbp]
\begin{centering}
\includegraphics[scale=0.7,page=1]{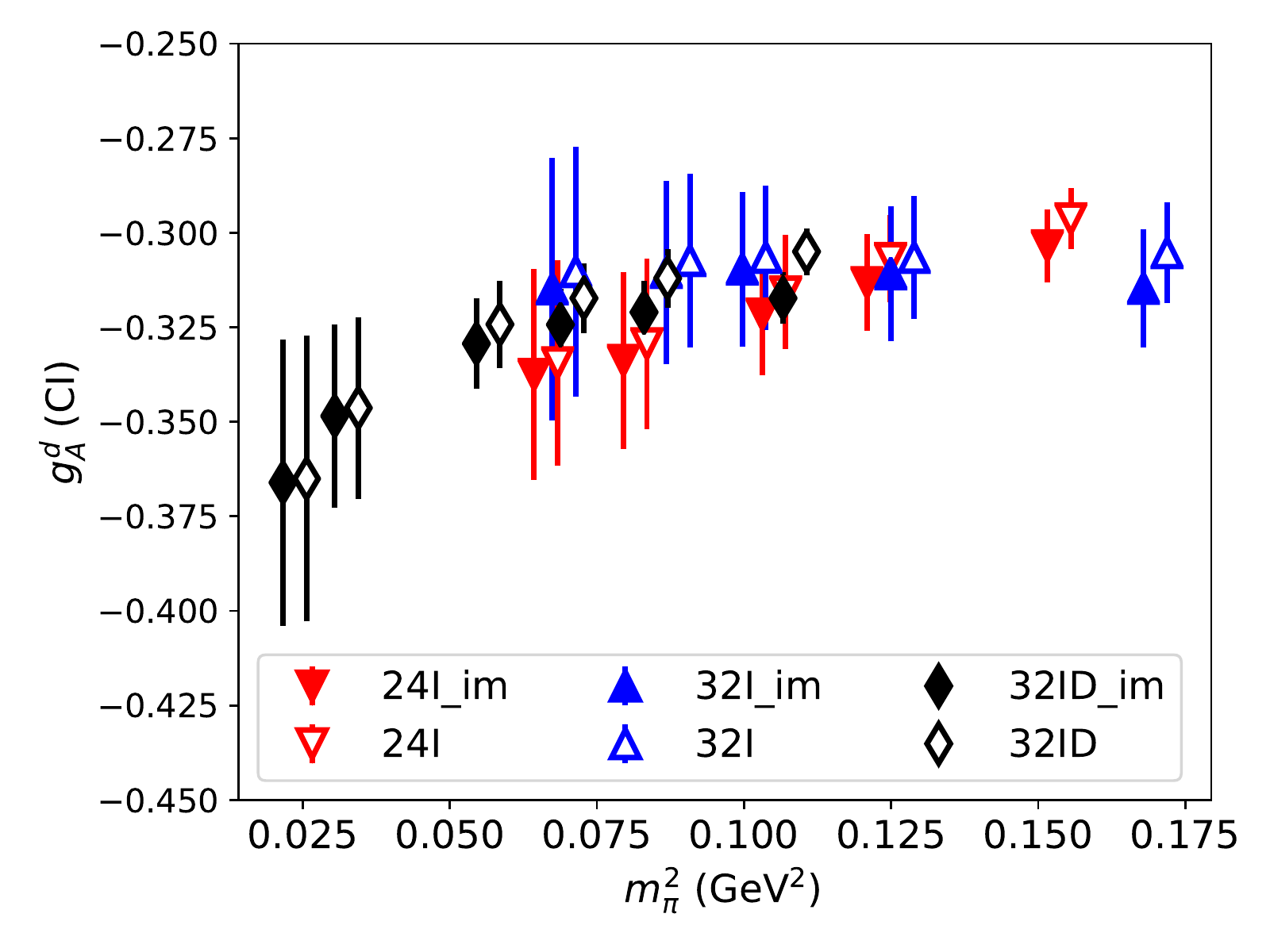}\\
\includegraphics[scale=0.7,page=1]{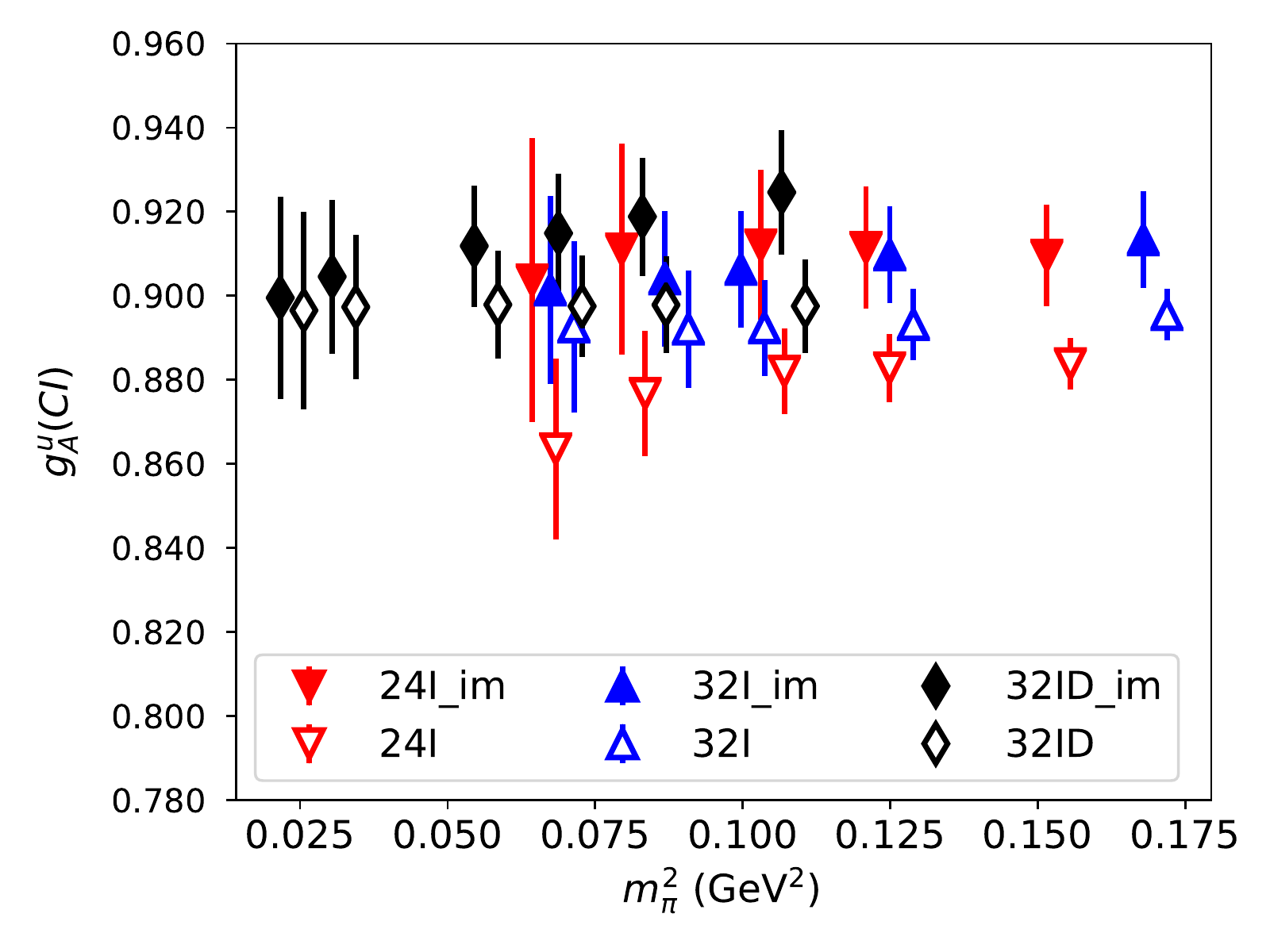}
\par\end{centering}
\caption{The comparison of the MEs before and after the improvement as a function
of pion mass squared. The left panel is for $d$ quark and the right
panel is for $u$ quark. 
Points of the unimproved results are shifted slightly in the horizontal direction to enhance the legibility. \label{fig:improvement}}
\end{figure}

The spatial and temporal components of the improved axial-vector current are defined as $A_{i}^{\rm im}=A_{i}+gD_{i}$ and $A_{4}^{\rm im}=A_{4}+gD_{4}$,
where the factor $g$ is determined by assuming the final $g_A$ calculated
from the two components of the improved current are identical \citep{Liang:2016fgy}.
Although the results of the currents with derivative are noisy and
the constant fit may not be a perfect choice, it is enough for this
calculation since the improvement itself is only around $3\%$ or less. 
Plots of the improvement are shown in Figure \ref{fig:improvement}.
For the $d$ quark case, the improvement has no effect basically, while
for the $u$ quark case, especially for the 24I lattice, the improvement
is at the $2\sigma$ level and the improved data points are closer
to the points of the other two lattices around similar pion mass,
manifesting smaller lattice spacing effects.

\section{Renormalization\label{sec:Renormalization}}

The renormalization of the axial-vector current is indispensable for
comparing our result with experiment and phenomenology. The scale-independent 
isovector normalization constant $Z_{A}(\rm{CI})$ can be calculated
by imposing the chiral Ward identity in CI as in Eq.~(\ref{WI_CI}) or between the vacuum and a
pion state \citep{Liu:2013yxz}. There is no difference between the $u$ and $d$ quark in this 
case, as can be seen in the RI/MOM non-perturbative procedure. Hence, $Z_A(\rm{CI}) = Z_A^3$, 
the isovector normalization constant. Since we adopt the mass-independent renormalization scheme, it is also
the same as the octet renormalization $Z_A^8$. We shall define $Z_A \equiv Z_A^3 = Z_A^8$ as conventionally used
in the literature. After checking the AWI in the DI, we concluded in Sec.~\ref{sec:Check-of-Anomalous} that 
axial-vector current with the normalization of $Z_A({\rm CI})$ satisfies the AWI, thus there is no additional normalization
factor for the AWI and $Z_A({\rm DI}) = Z_A({\rm CI})=Z_A$ is the only normalization constant as far as tree-level AWI is concerned.
Through the chiral Ward identity in the CI, we can determine $Z_{A}$ to a high precision. Since we have calculated MEs of both CI and DI, 
the disconnected part of vertex functions also needs to be computed and the corresponding
renormalization can be obtained by the lattice nonperturbative approach
in the RI/MOM scheme \citep{Martinelli:1994ty}. This part contains
a scale-dependent DI piece and also mixing effects and is referred to as renormalization, to be
distinguished from the normalization, discussed so far, in upholding the AWI at the tree level.

\subsection{Formalisms}

The axial-vector coupling has conventionally been classified as the isovector $g_A^3 = \Delta u - \Delta d$, the 
octet $g_A^8= \Delta u + \Delta d - 2 \Delta s$ through the diagonal $SU(3)$ chiral transformation and the singlet 
$g_A^0 = \Delta u + \Delta d + \Delta s$ through the U(1) transformation and their renormalization follows.
One can obtain the renormalized $\Delta u, \Delta d$ and $\Delta s$ in term of their unrenormalized counterparts 
through these flavor-irreducible representations and the details are given in Ref.~\cite{Green:2017keo}.
On the other hand, the lattice calculations are carried out in terms of flavors and MEs in the CI  and DI.
It is natural to use them as the basis in renormalization. As we shall see, this has the advantage of preserving the
the CI piece which is scale independent and can be compared in different lattice calculations. Moreover, it is physical and
can be extracted from the global fitting of the polarized PDF. 

In the RI/MOM renormalization scheme, the renormalized quantities are related to the unrenormalized ones through the vertex
and the field renormalization. The most general form from the lattice classification is the following

\begin{equation}
\left(\begin{array}{c}
\Delta u ({\rm CI})\\
\Delta d ({\rm CI})\\
\Delta u ({\rm DI})\\
\Delta d ({\rm DI})\\
\Delta s ({\rm DI})
\end{array}\right)=\left(\begin{array}{ccccc}
\Sigma_{C} & 0 & 0 & 0 & 0\\
0 & \Sigma_{C} & 0 & 0 & 0\\
\Sigma_{D} & \Sigma_{D} & \Sigma_{C}+\Sigma_{D} & \Sigma_{D} & \Sigma_{D}\\
\Sigma_{D} & \Sigma_{D} & \Sigma_{D} & \Sigma_{C}+\Sigma_{D} & \Sigma_{D}\\
\Sigma_{D} & \Sigma_{D} & \Sigma_{C} & \Sigma_{C} & \Sigma_{C}+\Sigma_{D}
\end{array}\right)\left(\begin{array}{c}
\Delta u^{{\rm RI}}({\rm CI})\\
\Delta d^{{\rm RI}}({\rm CI})\\
\Delta u^{{\rm RI}} ({\rm DI})\\
\Delta d^{{\rm RI}} ({\rm DI})\\
\Delta s^{{\rm RI}} ({\rm DI})
\end{array}\right).\label{eq:renorm_1}
\end{equation}
where $\Delta f (f=u,d,s)$ is the bare axial-vector current matrix element
for a particular flavor $f$ and $\Delta f ^{{\rm RI}}$ is the corresponding
renormalized one in the RI scheme. $\Sigma_{C}$ or $\Sigma_{D}$ in the matrix is defined
by the following trace indicating the renormalization condition.
\begin{equation}
\Sigma_{C/D}=Z_{q}^{-1}\frac{1}{12}{\rm Tr}\left[\Lambda_{C/D}(p)\Lambda^{{\rm tree}}(p)^{-1}\right]\label{eq:sigmas}
\end{equation}
where $Z_{q}$ is the quark field renormalization constant, $\Lambda_{C/D}(p)$
is the connected or disconnected part of the vertex function and $\Lambda^{{\rm tree}}(p)$
is the tree-level vertex. The vertex function $\Lambda_{C/D}(p)$
is the following amputated Green\textquoteright s function 
\begin{equation}
\Lambda_{C/D}(p)=S^{-1}(p)G_{A,C/D}(p)S^{-1}(p)
\end{equation}
where $S^{-1}(p)$ is a quark propagator in the momentum space in the Landau gauge and
$G_{A,C/D}(p)$ is the connected piece or the disconnected piece of
the forward Green\textquoteright s function $G_{A}(p)=\sum_{x,y}e^{-ip\cdot(x-y)}\langle\psi(x)\Gamma_A\bar{\psi}(y)\rangle$
with $\Gamma_A = \gamma_{\mu}\gamma_5$. To be more specific,
the two ways of Wick contraction of $G_{A}(p)$ lead to two kinds 
of the vertex function which are the connected part $\Lambda_{C}$,
where the quark fields in the bilinear operator contract with the
other two external quark fields, and the disconnected one $\Lambda_{D}$, where the
quark fields in the bilinear operator contract with each other. Since
only in $\Lambda_{D}$ can the flavor of the bilinear operator be
different from that of the external legs, the off-diagonal entries
of the matrix in Eq.~(\ref{eq:renorm_1}) which represent the flavor
mixing effect contain $\text{\ensuremath{\Sigma_{D}}}$ alone. We should stress that the entries of zero reflect
the fact that the CI MEs do not receive mixing from the DIs.
On the other hand, the DI MEs receive contributions from both CI and DI. 
These equations are defined in the quark massless limit so that the RI-MOM
is a mass-independent renormalization scheme. In practice, we do calculations
at finite quark masses and then extrapolate to the chiral limit. 
In principle, $Z_{q}$ can be determined by considering the derivative
of the quark propagator with respect to the discretized momenta. However, $Z_q$ so determined is known to
have large discretization error. We shall use $Z_A$ from the chiral Ward identity as an input; therefore, 
we have $\Sigma_C = \frac{1}{Z_A}$ and $Z_{q}$ is determined via Eq.~(\ref{eq:sigmas}) instead as employed in Ref.~\cite{Liu:2013yxz}. 

The renormalization constants come from the inverse of the matrix in Eq.~(\ref{eq:renorm_1}). The renormalized quark spins in
the RI scheme are
\begin{equation}
\left(\begin{array}{c}
\Delta u^{{\rm RI}}({\rm CI})\\
\Delta d^{{\rm RI}}({\rm CI})\\
\Delta u^{{\rm RI}} ({\rm DI})\\
\Delta d^{{\rm RI}} ({\rm DI})\\
\Delta s^{{\rm RI}} ({\rm DI})
\end{array}\right)=\left(\begin{array}{ccccc}
Z_A & 0 & 0 & 0 & 0\\
0 & Z_A & 0 & 0 & 0\\
Z_{A}^{{\rm D,RI}} & Z_{A}^{{\rm D,RI}} & Z_A +Z_{A}^{{\rm D,RI}} & Z_{A}^{{\rm D,RI}} & Z_{A}^{{\rm D,RI}}\\
Z_{A}^{{\rm D,RI}} & Z_{A}^{{\rm D,RI}} & Z_{A}^{{\rm D,RI}} & Z_A+Z_{A}^{{\rm D,RI}} & Z_{A}^{{\rm D,RI}}\\
Z_{A}^{{\rm D,RI}} & Z_{A}^{{\rm D,RI}} & Z_{A}^{{\rm D,RI}} & Z_{A}^{{\rm D,RI}} & Z_A +Z_{A}^{{\rm D,RI}}
\end{array}\right)\left(\begin{array}{c}
\Delta u ({\rm CI})\\
\Delta d ({\rm CI})\\
\Delta u ({\rm DI})\\
\Delta d ({\rm DI})\\
\Delta s ({\rm DI})
\end{array}\right). \label{eq:renorm_2}
\end{equation} 
where
\begin{equation}
Z_{A}^{{\rm D,RI}}\equiv-Z_A\left(\frac{\Sigma_{D}}{\Sigma_{C}+N_f\Sigma_{D}}\right),Z_A=\frac{1}{\Sigma_C}
\end{equation}
In the present calculation, $N_f =3$. 

To compare with experiments, we need to match the results from the 
above RI scheme to that of $\overline{\rm MS}$ at 2 GeV. As we see from Sec. \ref{sec:quark_spin}, it 
entails a two-loop perturbative calculation of the axial-vector current in the RI and $\overline{\rm MS}$ 
schemes respectively. On the other hand, it is shown in Eq.~(\ref{renor_AWI}), this is the same renormalization 
constant from the one-loop mixing of the topological charge. We carry out the simpler one-loop mixing calculation of
the topological charge for the matching factor from the RI scheme at momentum $p$ to the $\overline{\rm MS}$ 
scheme at scale $\mu$ based on Package-X \cite{Patel:2015tea,Patel:2016fam} and this matching ratio can be represented as a matrix $R_m$ which needs to be %
\begin{equation}
R_m=\left(\begin{array}{ccccc}
1 & 0 & 0& 0& 0 \\
0 & 1 & 0& 0 & 0 \\
f_m& f_m& 1+f_{m} & f_{m} & f_{m}\\
f_m & f_m& f_{m} & 1+f_{m} & f_{m}\\
f_m & f_m& f_{m} & f_{m} & 1+f_{m}
\end{array}\right)
\end{equation}
where $f_{m}=\left(\frac{\alpha_s}{4\pi}\right)^24C_F\left(-\frac{3}{2}{\rm log}\left(\frac{\mu^2}{p^2}\right)+\frac{7}{2}\right)$ with $C_F = 4/3$.

Thus, after $R_m$ is multiplied to the renormalization matrix in 
Eq.~(\ref{eq:renorm_2}), the renormalized quark spin in the $\overline{\rm MS}$ scheme is 
\begin{equation}
\left(\begin{array}{c}
\Delta u^{N}({\rm CI})\\
\Delta d^{N}({\rm CI})\\
\Delta u^{\overline{\rm MS}}({\rm DI})(\mu)\\
\Delta d^{\overline{\rm MS}}({\rm DI})(\mu)\\
\Delta s^{\overline{\rm MS}}({\rm DI})(\mu)
\end{array}\right)=\left(\begin{array}{ccccc}
Z_A & 0 & 0 & 0 & 0\\
0 & Z_A & 0 & 0 & 0\\
Z_{A}^{{\rm D, \overline{MS}}} & Z_{A}^{{\rm D, \overline{MS}}} & Z_A +Z_{A}^{{\rm D, \overline{MS}}} & Z_{A}^{{\rm D, \overline{MS}}} & 
Z_{A}^{{\rm D, \overline{MS}}}\\
Z_{A}^{{\rm D, \overline{MS}}} & Z_{A}^{{\rm D, \overline{MS}}} & Z_{A}^{{\rm D, \overline{MS}}} & Z_A+Z_{A}^{{\rm D, \overline{MS}}} & 
Z_{A}^{{\rm D, \overline{MS}}}\\
Z_{A}^{{\rm D, \overline{MS}}} & Z_{A}^{{\rm D, \overline{MS}}} & Z_{A}^{{\rm D, \overline{MS}}} & Z_{A}^{{\rm D, \overline{MS}I}} & Z_A 
+Z_{A}^{{\rm D, \overline{MS}}}
\end{array}\right)\left(\begin{array}{c}
\Delta u ({\rm CI})\\
\Delta d ({\rm CI})\\
\Delta u ({\rm DI})\\
\Delta d ({\rm DI})\\
\Delta s ({\rm DI})\end{array}\right) \label{eq:renorm_3}
\end{equation}
where the notations of $\Delta u^{N}({\rm CI})$ and $\Delta d^{N}({\rm CI})$ mean they have normalization only and
\begin{equation} 
Z_{A}^{{\rm D, \overline{MS}}} =Z_{A}^{{\rm D,RI}} + f_m + N_f f_{m}Z_{A}^{{\rm D,RI}}.
\end{equation}

In practice, $Z_{A}^{{\rm D, \overline{MS}}}$ are to be evolved to a given scale such as 2 GeV for each $p^2a^2$ in RI and extrapolated to
$p^2a^2 = 0$. This involves an evolution 
\begin{equation}
\label{evolve}
\mu^2 \frac{d}{d\mu^2} \log \left(Z_{A}^{{\rm D, \overline{MS}}}(\mu)\right) = \gamma (\alpha_s) = - \sum_i \gamma_i \alpha_s^{i+1}.
\end{equation}
For the axial-vector current, the anomalous dimensions are
\begin{equation}
\gamma_0 = 0, \hspace{1cm} \gamma_1 = \frac{1}{(4 \pi)^2} 6 C_F N_f
\end{equation}
It is shown in~\cite{Green:2017keo} that, at two-loop order, the evolution of the flavor-singlet renormalization factor is given by
\begin{equation}  
\frac{Z_A+3Z_{A}^{{\rm D, \overline{MS}}}(\mu)}{Z_A+3Z_{A}^{{\rm D, \overline{MS}}}(\mu_0)} = \left(\frac{\beta_0 + \beta_1 \alpha_s(\mu)}
{\beta_0 + \beta_1 \alpha_s(\mu_0)}\right)^{\gamma_1/\beta_1}
\end{equation}
where $Z_A+3Z_{A}^{{\rm D, \overline{MS}}}(\mu)$ is the renormalization constant for the flavor-singlet case which we will show later,
and the relevant constants are $\beta_0=\frac{1}{4\pi}\left(\frac{11}{3}C_A-\frac{4}{3}T_FN_f\right)=\frac{1}{4\pi}\left(11-\frac{2}{3}N_f\right)$
and $\beta_1=\frac{1}{(4\pi)^2}\left(\frac{34}{3}C^2_A-\frac{20}{3}C_AT_FN_f-4C_FT_FN_f\right)=\frac{1}{(4\pi)^2}\left(102-\frac{38}{3}N_f\right)$
with $C_A=3$ and $T_F=\frac{1}{2}$. The the evolution of $\alpha_s$ at two-loop level is given in \cite{Prosperi:2006hx},
\begin{equation}
\alpha_s(\mu)=-\frac{\beta_0}{\beta_1}\frac{1}{1+W_{-1}(\zeta)}, \hspace{1cm} \zeta=-\frac{\beta_0^2}{e\beta_1}\left(\frac{\Lambda^2}{\mu^2}\right)^{\beta_0^2/\beta_1}
\end{equation}
where $W_{-1}$ is the lower branch of the Lambert function and $\Lambda$ is set to be the PDG value $332(19)$ MeV.

The final results of the renormalized $u/d$ quark spin can be decomposed into the CI part and the DI part for 
each flavor
\begin{equation}  \label{CI/DI-1}
\left(\Delta u/\Delta d\right) ^{\rm \overline{MS}} (\mu) = \left(\Delta u/\Delta d\right)^{N} ({\rm CI}) + \left(\Delta u/\Delta d\right) ^{\rm \overline{MS}} ({\rm DI}) (\mu),
\end{equation}
where the connected insertion part
\begin{equation}  \label{CI/DI-2}
\left(\Delta u/\Delta d\right)^{N} ({\rm CI}) = Z_A \left(\Delta u/\Delta d\right) ({\rm CI})
\end{equation}
is scale independent. We should caution that this is true for the axial-vector case due to the chiral Ward identity. This is not
true in general, such as for the case of the scalar and the energy-momentum tensor matrix elements where the CI parts are also scale dependent. 
On the other hand, the disconnected insertion parts depend on the $\overline{\rm MS}$ scale of $\mu$
\begin{eqnarray}  \label{CI/DI-3}
\left(\Delta u/\Delta d\right)^{\rm \overline{MS}} (\rm DI) (\mu) &=& Z_A\, \left(\Delta u/\Delta d\right) ({\rm DI}) + Z_A^{\rm{D, \overline{MS}}} (\mu)\, \Sigma \nonumber \\
\Delta s^{\rm \overline{MS}} (\mu) &=&  Z_A \Delta s + Z_A^{\rm{D, \overline{MS}}} (\mu)\, \Sigma
\end{eqnarray}
where
\begin{equation}
\Sigma = \Delta u + \Delta d + \Delta s = \Delta u ({\rm CI}) + \Delta d ({\rm CI}) + (\Delta u + \Delta u + \Delta s) ({\rm DI}).
\end{equation}

This decomposition of the quark spin in terms of flavor and CI and DI is common for all renormalization schemes, and is not limited
to the RI or $\overline{\rm MS}$ scheme. When the CI and DI components are added together from Eq.~(\ref{eq:renorm_3}) to get the total matrix elements, 
one arrives at a simpler expression 
\begin{equation} \label{CI+DI}
\Delta\! f^{\rm \overline{MS}} (\mu) = Z_A \, \Delta\! f+ Z_A^{\rm{D, \overline{MS}}} (\mu)\, \Sigma
\end{equation}
where $f = {u,d,s}$. 
In terms of the flavor irreducible representations, they are
\begin{eqnarray}
g_A^{3}&=&\Delta u^{\rm \overline{MS}} - \Delta d^{\rm \overline{MS}} = Z_A (\Delta u - \Delta d) \\ 
g_A^8 &=& \Delta u^{\rm \overline{MS}} + \Delta d ^{\rm \overline{MS}} - 2 \Delta s^{\rm \overline{MS}} = Z_A (\Delta u + \Delta d - 2\Delta s) \\ 
g_A^{0, {\rm \overline{MS}}}(\mu) &=& \Delta u^{\rm \overline{MS}} + \Delta d ^{\rm \overline{MS}} + \Delta s^{\rm \overline{MS}} 
= \left(Z_A + 3 Z_{A}^{{\rm D, \overline{MS}}}(\mu)\right) \Sigma.
\end{eqnarray}

Eq.~(\ref{CI+DI}) can be derived by starting the renormalization from the combined CI and DI matrix elements so that
Eq.~(\ref{eq:renorm_3}) becomes 
\begin{equation}
\left(\begin{array}{c}
\Delta u^{\overline{\rm MS}} (\mu)\\
\Delta d^{\overline{\rm MS}} (\mu)\\
\Delta s^{\overline{\rm MS}} (\mu)
\end{array}\right)=\left(\begin{array}{ccc}
 Z_A +Z_{A}^{{\rm D, \overline{MS}}}(\mu) & Z_{A}^{{\rm D, \overline{MS}}} (\mu)& Z_{A}^{{\rm D, \overline{MS}}} (\mu)\\
Z_{A}^{{\rm D, \overline{MS}}} (\mu)& Z_A+Z_{A}^{{\rm D, \overline{MS}}} (\mu)& Z_{A}^{{\rm D, \overline{MS}}}(\mu)\\
Z_{A}^{{\rm D, \overline{MS}}} (\mu)& Z_{A}^{{\rm D, \overline{MS}I}} (\mu)& Z_A 
+Z_{A}^{{\rm D, \overline{MS}}}(\mu)
\end{array}\right)\left(\begin{array}{c}
\Delta u \\
\Delta d \\
\Delta s 
\end{array}\right). \label{eq:renorm_4}
\end{equation}
Similarly, Eq.~(\ref{CI+DI}) for the renormalized quark spin for each flavor can be derived from the the basis of flavors
irreducible representations $g_A^3, g_A^8$ and $g_A^0$
\begin{equation}
\left(\begin{array}{c}
g_A^{3}\\
g_A^{8}\\
g_A^{0, {\rm \overline{MS}}}(\mu)
\end{array}\right)=\left(\begin{array}{ccc}
Z_{A} & 0 & 0 \\
0 & Z_{A} & 0 \\
0 & 0 & Z_{A}+ N_f Z_A^{\rm{D, \overline{MS}}}(\mu)
\end{array}\right)\left(\begin{array}{c}
\Delta u - \Delta d\\
\Delta u + \Delta d - 2 \Delta s\\
\Delta u + \Delta d + \Delta s
\end{array}\right).  \label{eq:renorm_5}
\end{equation}
This has been worked out in \cite{Green:2017keo} with the same results. 

It is not surprising that one arrives at the same renormalized results in Eq.~(\ref{CI+DI}) irrespective of the starting basis in Eq.~(\ref{eq:renorm_3}),
(\ref{eq:renorm_4}) or (\ref{eq:renorm_5}), since they involve linear equations. Eq.~(\ref{eq:renorm_5}) is the conventional way of presenting
the renormalized results both in experiments and phenomenology. 
However, we should stress that there are advantages of separating them further in terms of CI and DI parts as in Eqs.~(\ref{CI/DI-1}), (\ref{CI/DI-2}) and (\ref{CI/DI-3}) for each flavor. 
First of all, we note that the CI parts are renormalization
group invariant due to the chiral Ward identity and they are easier to calculate on the lattice than those of the DI parts so that they can be readily
compared from lattice calculations involving different systematics owing to different actions and lattice parameters. More importantly, they can
be deduced from experiments. The parton degrees of freedom of the nucleon structure functions in the DIS has been
classified in the Euclidean path-integral formalism of the hadronic tensor~\cite{Liu:1993cv,Liu:1999ak}. 
It is found that there is a connected-sea parton which is in the connected-insertion of the current-current correlator in addition to the disconnected-sea partons in the corresponding disconnected-insertion. 
The former is responsible for the Gottfried sum-rule violation~\cite{Liu:1993cv}. These two sea partons have not been separated in
the global fittings so far. However, it is demonstrated, in one example, that by combining the strange parton distribution from the semi-inclusive
DIS experiment of HERMES, the global fitting result of $\bar{u}(x) + \bar{d}(x)$ and the lattice calculation of the ratio of 
$\frac{\langle x\rangle_{s}}{\langle x\rangle_{u} ({\rm DI})}$, one can separate the connected-sea from the disconnected-sea distribution of 
the $u$ and $d$ partons~\cite{Liu:2012ch}. 
It is shown that in the operator-product-expansion, it is the moments of the combined connected-sea and valence parton 
distributions that correspond to the local matrix elements of the CI in the lattice calculation~\cite{Liu:1999ak}. The parton evolution equations with separate connected- and disconected-sea parton is formulated~\cite{Liu:2017lpe}. Provided that future global fitting take this separation into account when fitting experiments at different $Q^2$, one can obtain the moments of the valence and connected-sea to extract $\Delta u ({\rm CI})$ and $\Delta d ({\rm CI})$ and other moments from the unpolarized and polarized partons and compare directly with the lattice calculation of moments.

\subsection{Numerical results of the renormalization}

\begin{figure}[tbp]
\centering{}\includegraphics[scale=0.5,page=1]{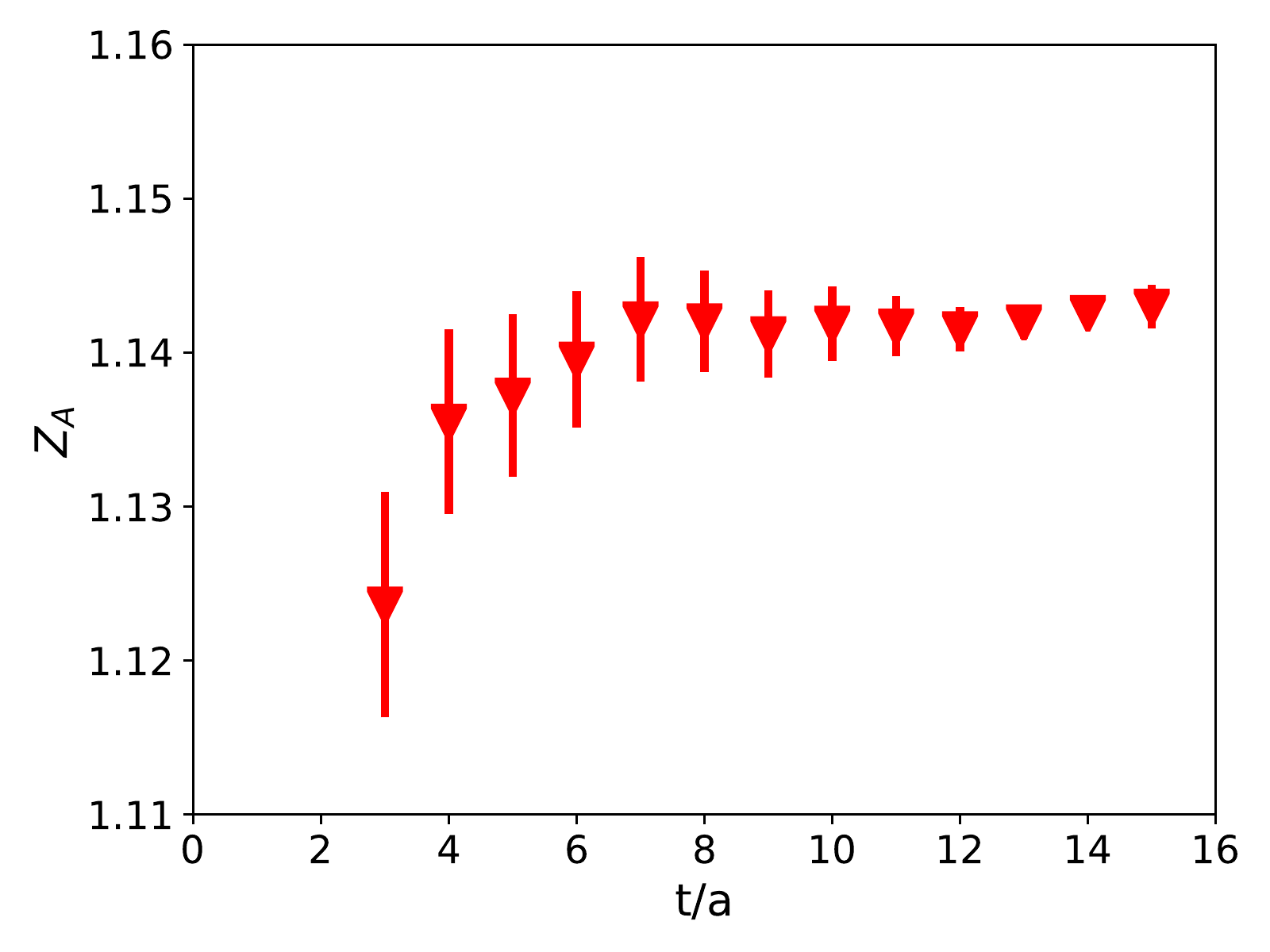}\includegraphics[scale=0.5,page=1]{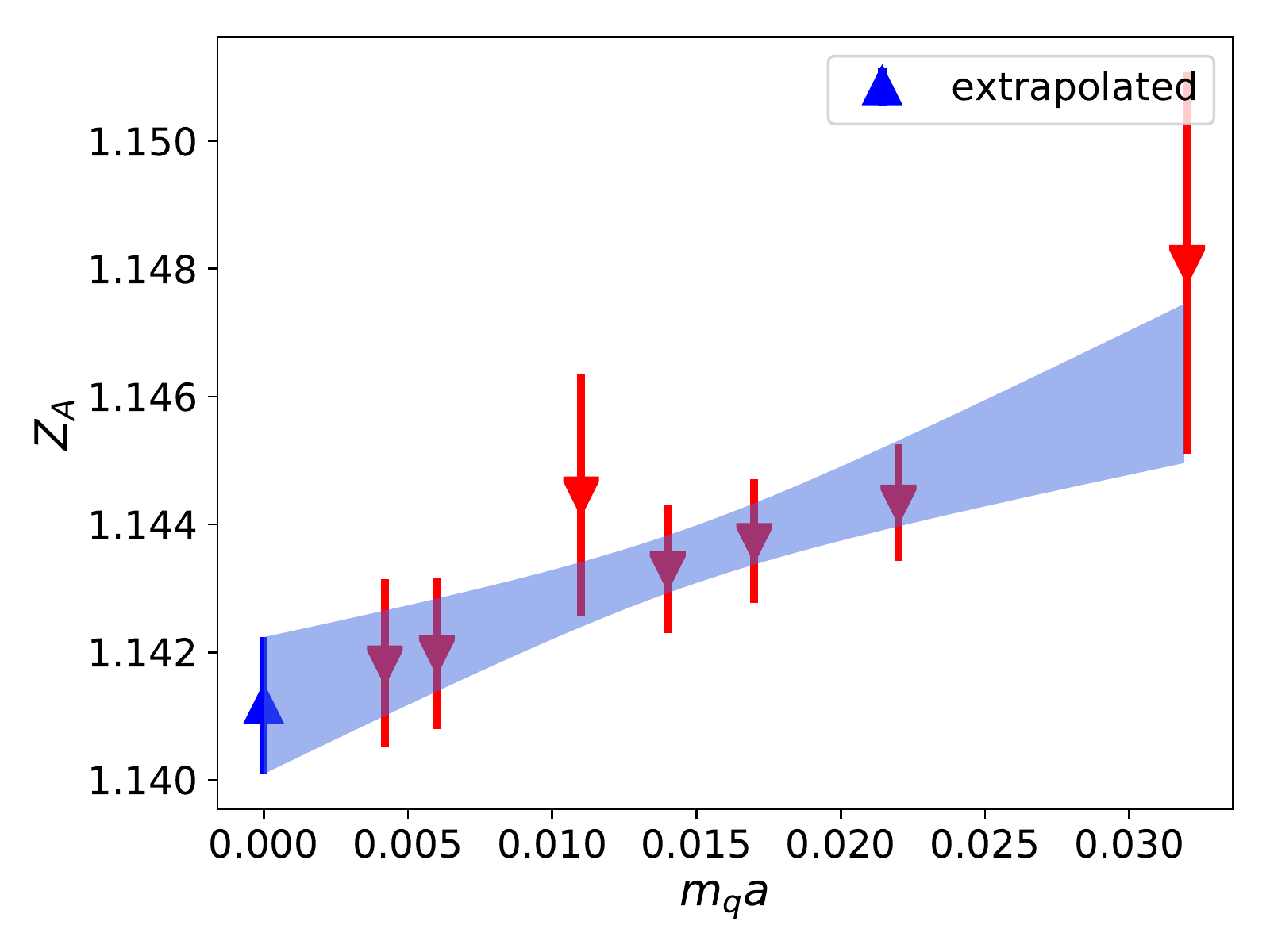}\caption{$Z_{A}$ on the 32ID lattice at the unitary point as a
function of $t$ is shown in the left panel. The corresponding chiral
extrapolation is shown in the right panel. \label{fg:ZA_32ID}}
\end{figure}

The results of $Z_{A}$ on the 24I and 32I lattice have
been obtained in our previous study \citep{Liu:2013yxz} to be $1.111(6)$
and $1.086(2)$ at the massless limit for both valence and sea quarks.
The $Z_{A}$ on the 32ID lattice is calculated in this
study using the same strategy: $Z_{A}=\frac{2m_{q}\langle0|P|\pi\rangle}{m_\pi\langle0|A_{4}|\pi\rangle}$
where $P$ and $A_{4}$ are the pseudo-scalar quark bilinear operator
and the temporal component of the axial-vector operator, respectively.
Pion 2-point correlators are calculated to obtain the corresponding
MEs. Figure \ref{fg:ZA_32ID} shows the ratio of $Z_{A}$
as a function of Euclidean time of the pion correlators at the unitary
point in the left panel and the chiral extrapolation in the right
panel. The final value we get is $Z_{A}({\rm 32ID})=1.141(1)$.

\begin{figure}[tbp]
\centering{}\includegraphics[scale=0.5,page=1]{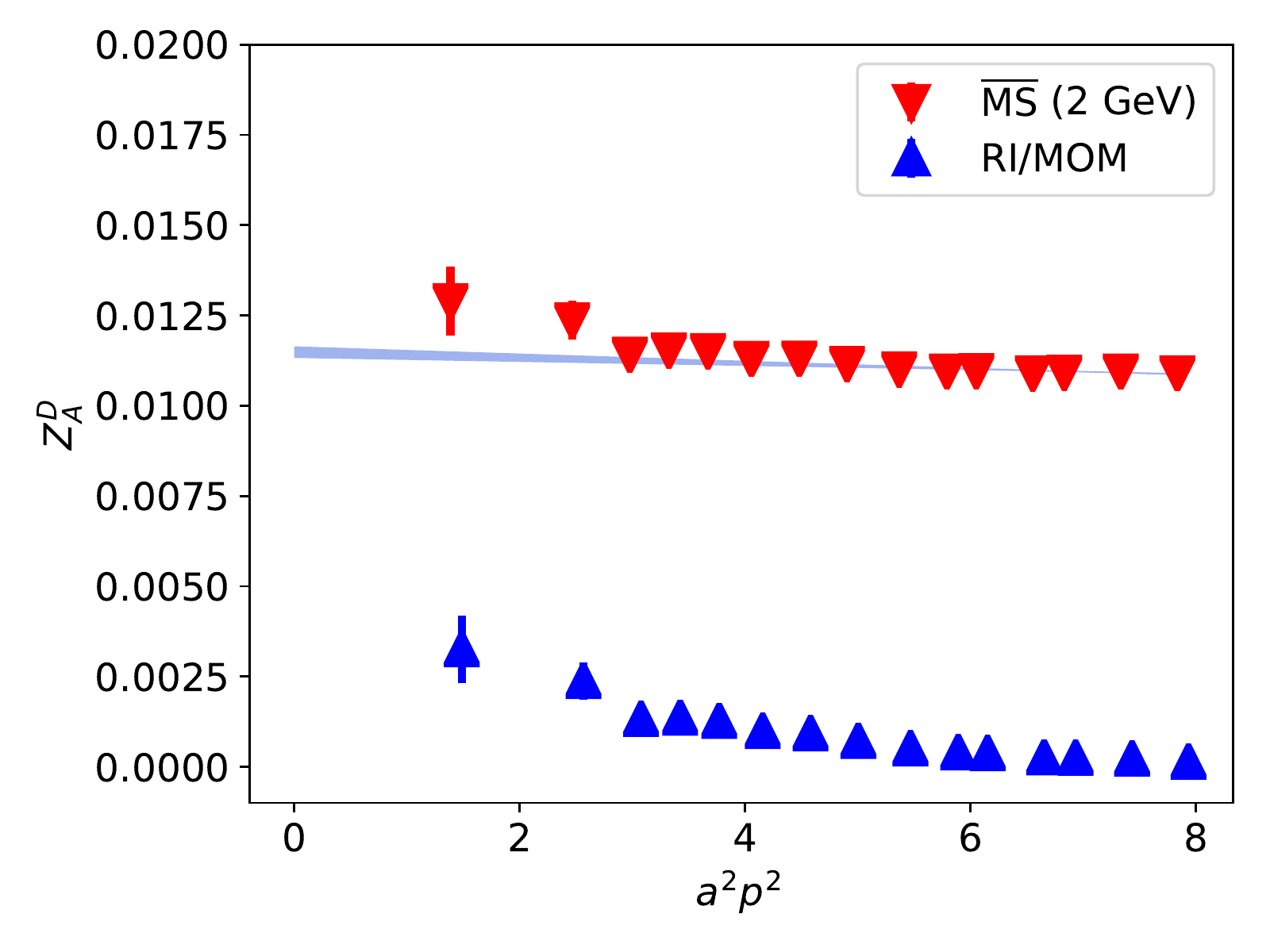}\includegraphics[scale=0.5,page=1]{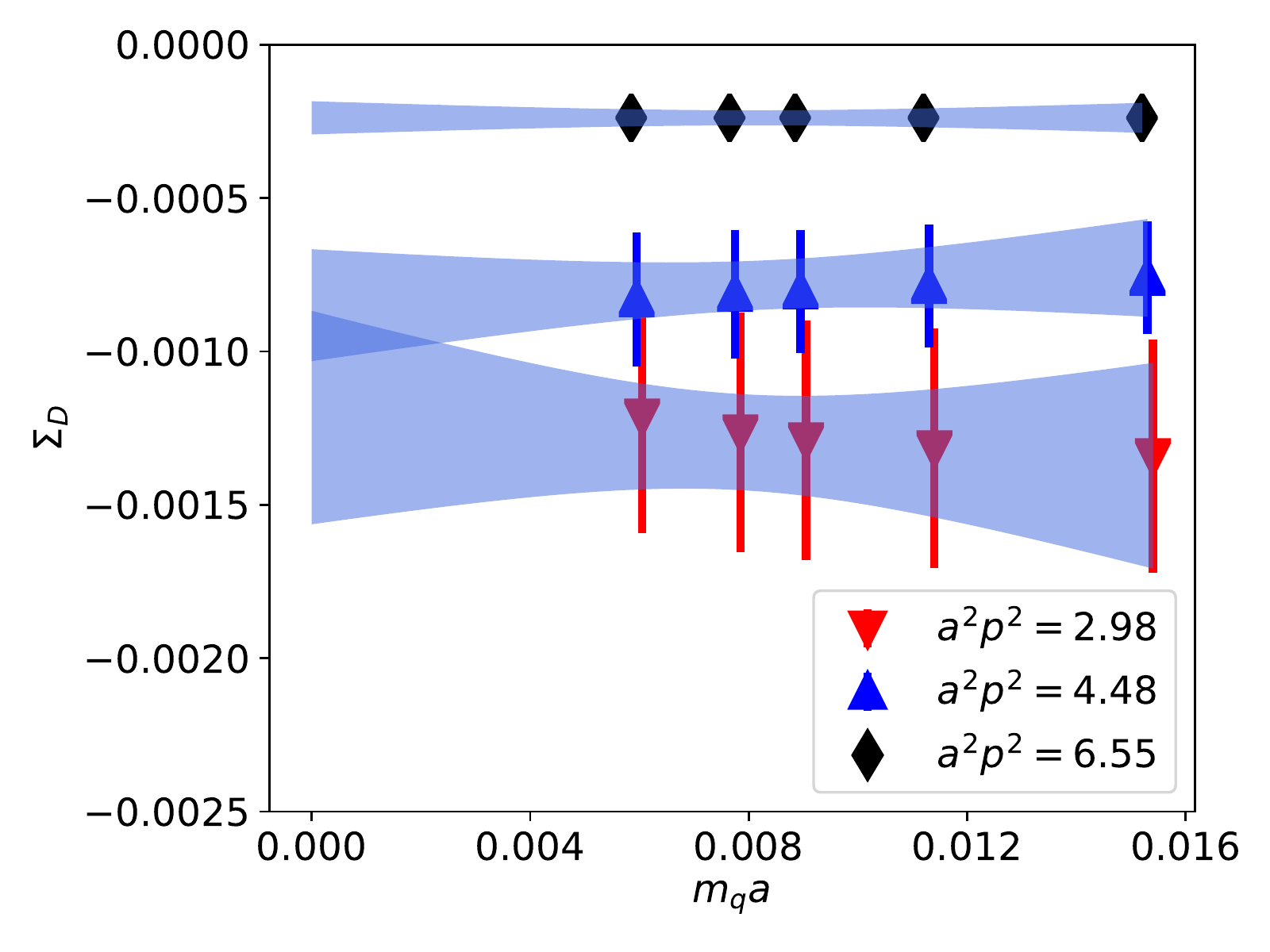}\caption{Renormalization calculation on the 32I lattice. The left panel shows
the $a^{2}p^{2}$ dependence of $Z_{A}^{{\rm D,RI}}$ and also the
remaining $a^{2}p^{2}$ effects of $Z_{A}^{{\rm D,\overline{{\rm MS}}}}$
after matching to the $\overline{{\rm MS}}$ scheme at $\mu=2$ GeV.
The blue band of the left plot shows the linear extrapolation of $Z_{A}^{{\rm D,\overline{{\rm MS}}}}$;
 the first two points are not included. The right panel shows the
$m_{q}a$ dependence and the linear chiral extrapolation of $\Sigma_{D}$
at three typical $a^{2}p^{2}$ values. \label{fg:renorm-32I}}
\end{figure}

\begin{figure}[tbp]
\centering{}\includegraphics[scale=0.5,page=1]{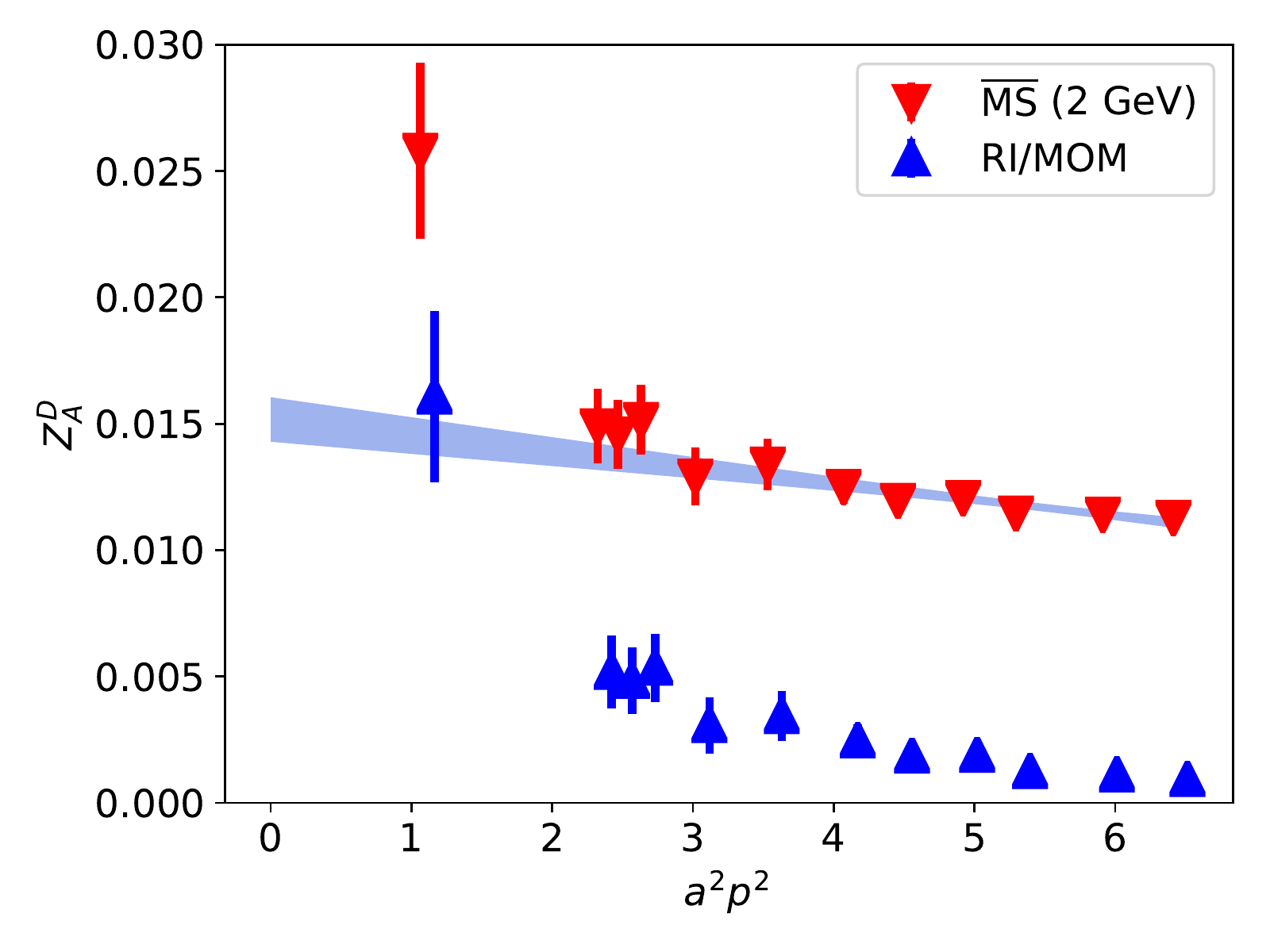}\includegraphics[scale=0.5,page=1]{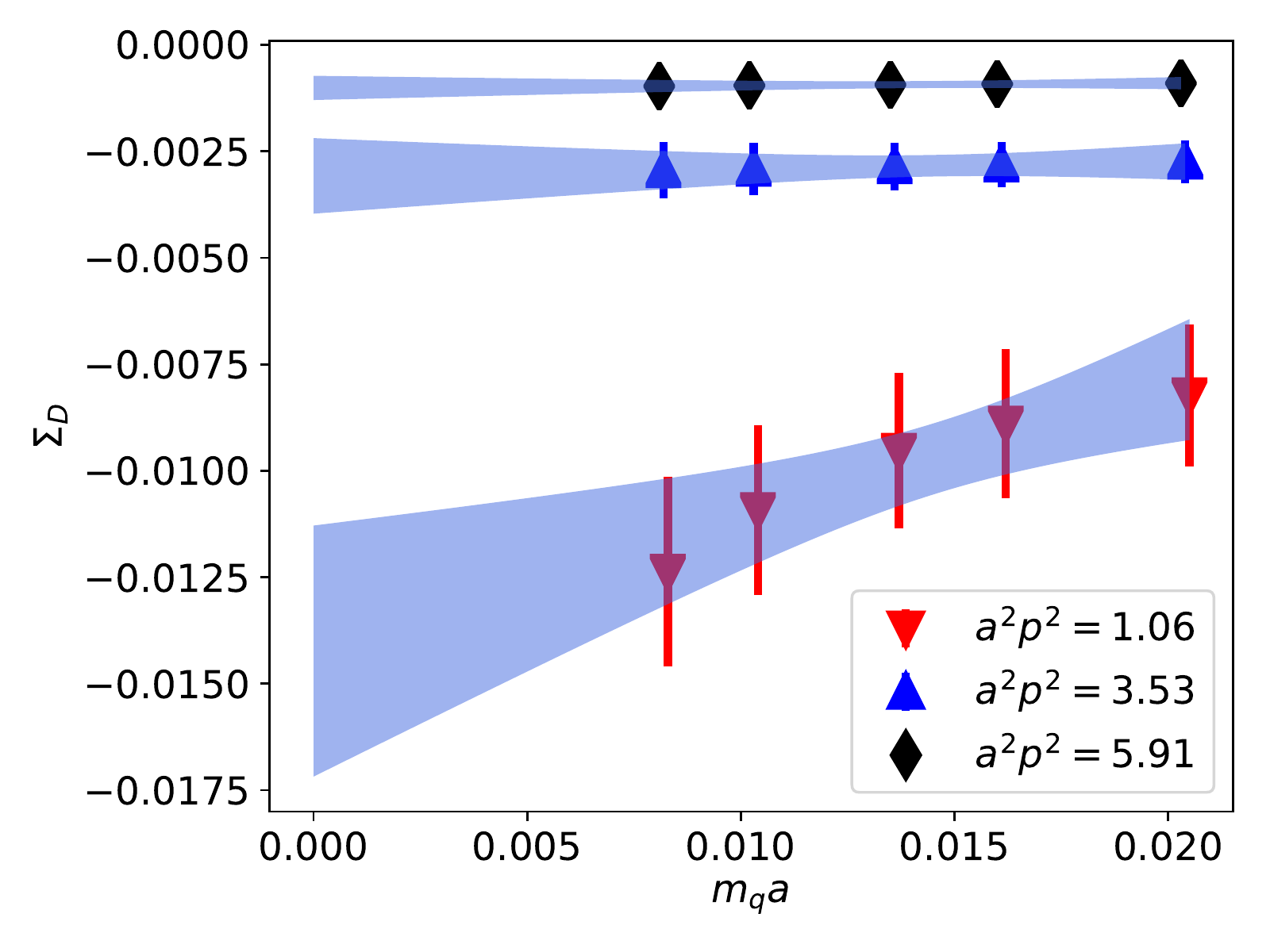}\caption{The same figure as Figure \ref{fg:renorm-32I} but for the 24I lattice.
\label{fg:renorm-24I}}
\end{figure}

\begin{figure}[tbp]
\centering{}\includegraphics[scale=0.5,page=1]{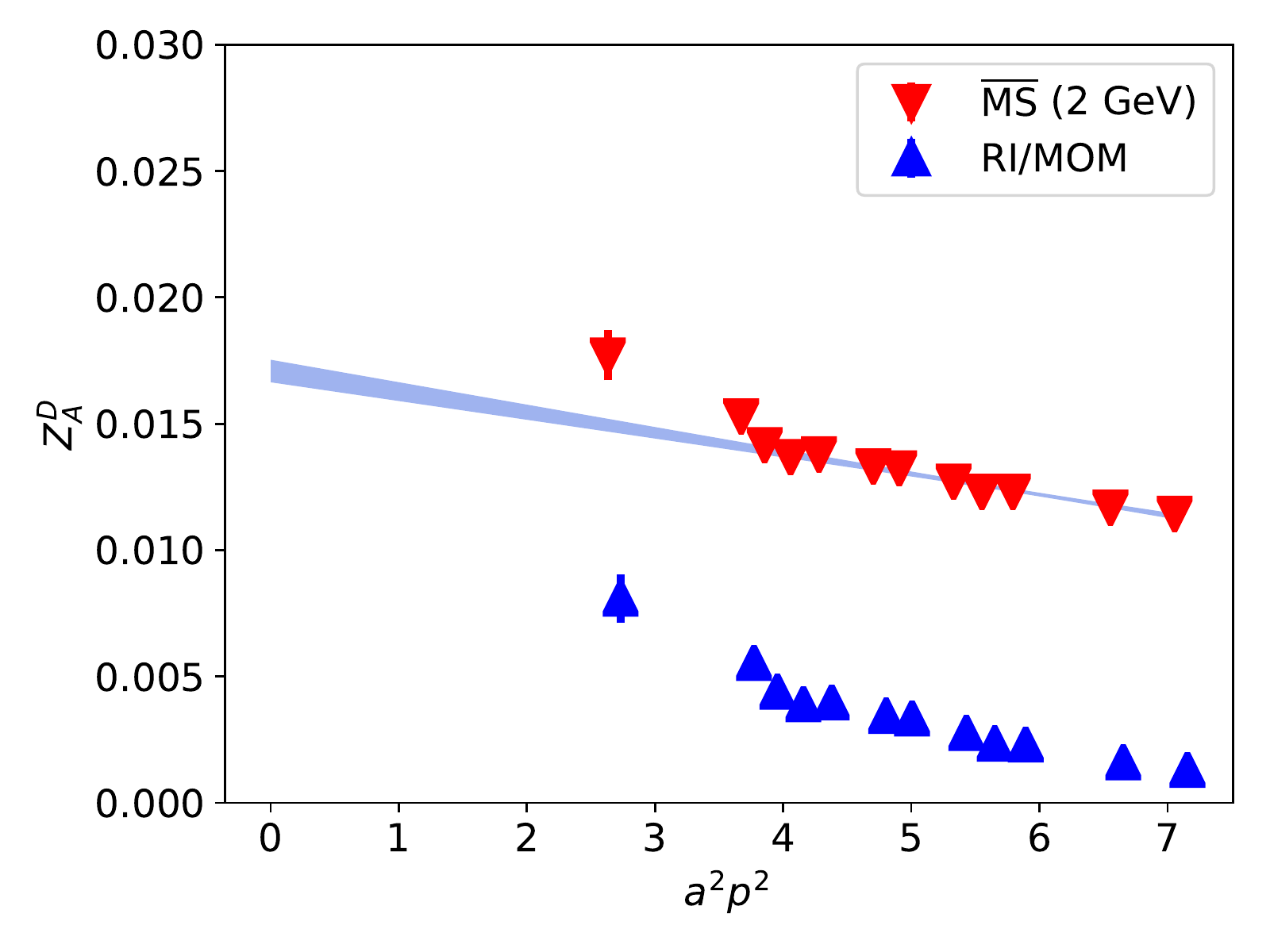}\includegraphics[scale=0.5,page=1]{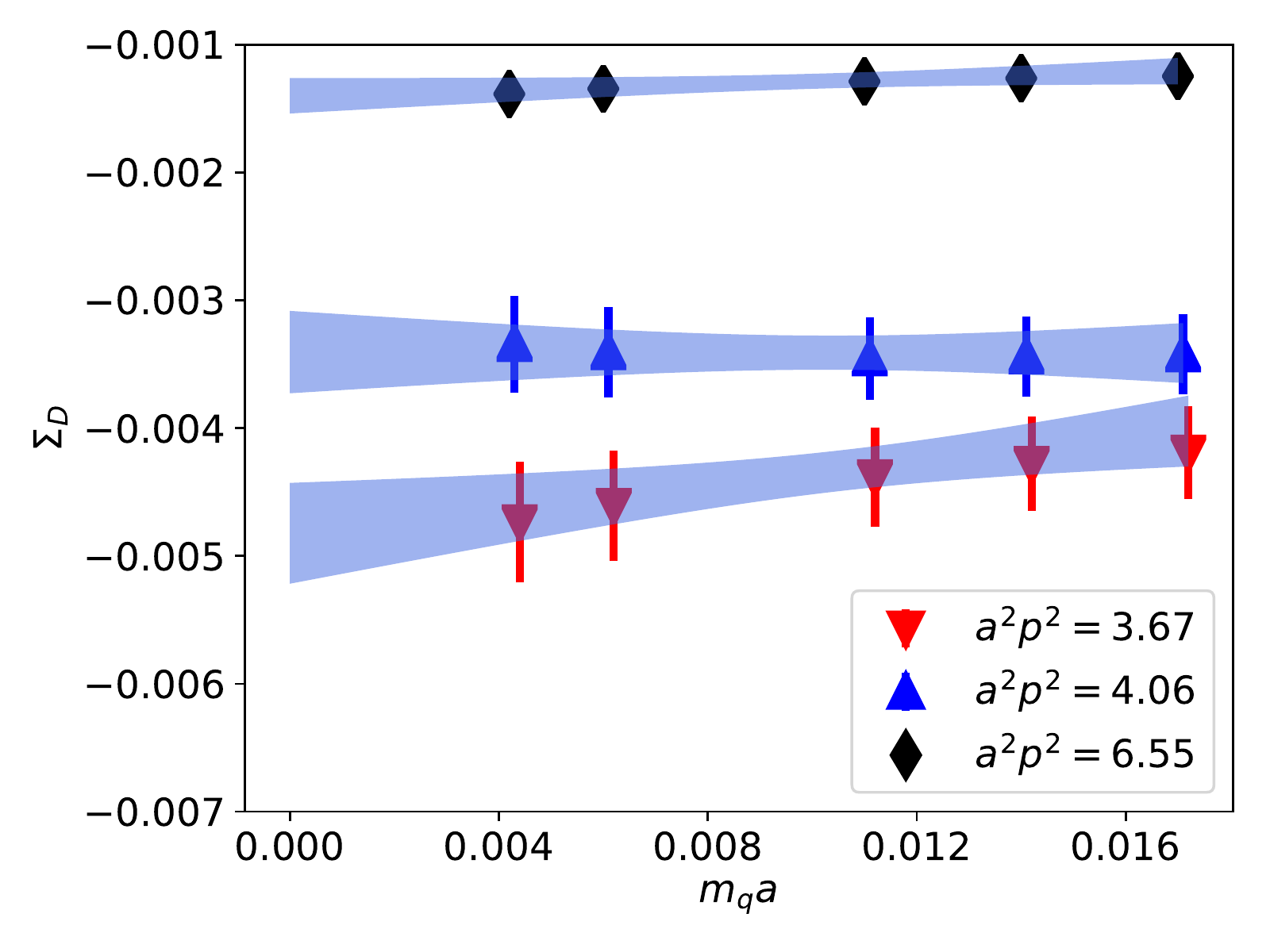}\caption{The same figure as Figure \ref{fg:renorm-32I} but for the 32ID lattice.
\label{fg:renorm-32ID}}
\end{figure}

The results of $Z_{A}^{{\rm D,RI}}$ and $Z_{A}^{{\rm D,\overline{{\rm MS}}}}$
are plotted in Figures \ref{fg:renorm-32I}, \ref{fg:renorm-24I}
and \ref{fg:renorm-32ID} for the three lattices we use respectively.
In each figure, the left panel shows the $a^{2}p^{2}$ dependence
of $Z_{A}^{{\rm D,RI}}$ and also the remaining $a^{2}p^{2}$ effects
of $Z_{A}^{{\rm D,\overline{{\rm MS}}}}$ after we match to the $\overline{{\rm MS}}$
scheme at $\mu=2$ GeV from the RI-MOM results at $p^{2}$ scale.
All the $Z_{A}^{{\rm D,RI}}$
in the figures are already extrapolated to the chiral limit by a linear
fit to $m_{q}a$. The right panel of these figures shows this linear
extrapolation for three typical values of $a^{2}p^{2}$. The blue bands
show the linear fit results of either the $a^{2}p^{2}$ dependence
or the $m_{a}a$ dependence; all the $\chi^{2}/d.o.f.$ of the fits
are less than 1. 
For the fitting of the $a^{2}p^{2}$ dependence, small $a^{2}p^{2}$ values are excluded since the renormalization scale of these points is not large enough such that the
two-loop matching factor can have large truncation error.
The final values we achieve at 2 GeV are $Z_{A}^{{\rm D,\overline{{\rm MS}}}}({\rm 32I})=0.01148(16)$,
$Z_{A}^{{\rm D,\overline{{\rm MS}}}}({\rm 24I})=0.01517(88)$ and $Z_{A}^{{\rm D,\overline{{\rm MS}}}}({\rm 32ID})=0.01709(45)$,
respectively.

The CDER technique is also used in the calculation of all the disconnected
parts of the vertex functions. Since the overall correction of this
part is small, we do not need very precise results, so no aggressive
cutoff is applied. In practice, the cutoffs are chosen to be $22a$,
$38a$ and $15a$ for the 24I, 32I and 32ID lattices respectively.
The improvement of the signal-to-noise ratio is $\sim50\%$ or
less. The criterion of choosing the cutoff is based on the $\chi^{2}$
of the linear fit with respect to $a^{2}p^{2}$ which is described
in detail in Ref. \citep{Yang:2018bft}.

\section{Global fitting and results\label{sec:Global-fitting-and}}

Having the bare MEs and the renormalization constants we obtained
in the above sections, we can now carry out the global fitting to
push our results to the physical pion point, the continuum limit and the
infinite volume limit. The functional form used is
\begin{equation}
g_{A}=c_{0}+c_{1}^{\rm I}/c_{1}^{\rm ID}a^{2}+c_{2}\left(m_{\pi,v}^{2}-m_{\pi,p}^{2}\right)+c_{3}\left(m_{\pi,s}^{2}-m_{\pi,p}^{2}\right)+c_{4}e^{-m_{\pi,v}L}
\end{equation}
where $m_{\pi,v}^{2}$ means the valence pion mass, $m_{\pi,s}^{2}$
means the sea pion mass and $L$ is the size of the lattice. We have
two $m_{\pi}^{2}$ terms in the fitting since we are using partially-quenched
valence quark masses. 
We use two coefficients $c_{1}^{\rm I}/c_{1}^{\rm ID}$ for the lattice spacing dependence term which reflects the fact that 
the ensembles we are using are generated with two slightly different gauge actions (Iwasaki for 24I and 32I and Iwasaki plus DSDR for 32ID).
We use the form like $\left(m_{\pi,v}^{2}-m_{\pi,p}^{2}\right)$
where $m_{\pi,p}^{2}$ is the physical pion mass in order to let $c_{0}=g_{A}^{{\rm phy}}$
to be the value in the physical limit. However, not all the coefficients
in the fitting function have statistical significance during the fit,
meaning that the lattice data has no constraint on the corresponding
term, or in other words, the effect of the corresponding term is
weak enough to be ignored with the current statistical uncertainty.
To be specific, the coefficient $c_{3}$ has no statistical significance in all the cases, 
so we exclude this term in all fittings to avoid overfitting. The difference
between the results with and without the $c_{3}$ term is included in the systematic uncertainties.
The other four terms (although not all of them have signals) are all kept in the fitting such that our final predictions are in the physical limit.
Since we use the improved axial-vector current and the finite lattice spacing effects are very weak,
additional prior values for the coefficients 
of the $a^2$ terms are used to guarantee stable results. 
We use the fitting results without separating 
the lattice spacing dependence into two groups as the central value of the prior and the widths are set to be 100\% of the central value.
The final results of the coefficients of the DI case are listed
in Table \ref{tab:global_fit_DI}; corresponding results are also collected in Table \ref{tab:global_fit_CI} for the CI case.

\begin{table}[ht]
\caption{The results of the coefficients and the corresponding $\chi^{2}/d.o.f.$ in the global fitting for the DI case.
Results for both the light quark and strange quark are listed.
\label{tab:global_fit_DI}}
\centering{}%
\begin{tabular}{cccccc}
 & $c_{0}$ & $c_{1}^{\rm I}/c_{1}^{\rm ID}$ & $c_{2}$  & $c_{4}$  & $\chi^{2}/d.o.f.$\tabularnewline
\hline 
l & -0.070(12) & 0.64(79)/0.97(44) & 0.131(51) & 0.11(31) & 0.39\tabularnewline
\hline 
s & -0.035(06) & 0.20(29)/0.35(21) & 0.024(29) & -0.06(23) & 0.41 \tabularnewline
\hline 
\end{tabular}
\end{table}

\begin{table}[ht]
\caption{The results of the coefficients and the corresponding $\chi^{2}/d.o.f.$ in the global fitting for the CI case for both $d$ and $u$ quarks.
\label{tab:global_fit_CI}}
\centering{}%
\begin{tabular}{cccccc}
 & $c_{0}$ & $c_{1}^{\rm I}/c_{1}^{\rm ID}$ & $c_{2}$ & $c_{4}$  & $\chi^{2}/d.o.f.$\tabularnewline
\hline 
d & -0.337(10) & -0.087(90)/-0.006(90) & 0.25(10) & -0.17(48) & 0.15\tabularnewline
\hline 
u & 0.917(13)  & 0.060(60)/0.061(60) & -0.01(11) & -0.56(51) & 0.04\tabularnewline
\hline 
\end{tabular}
\end{table}

\begin{figure}[tbp]
\centering{}\includegraphics[scale=0.5,page=1]{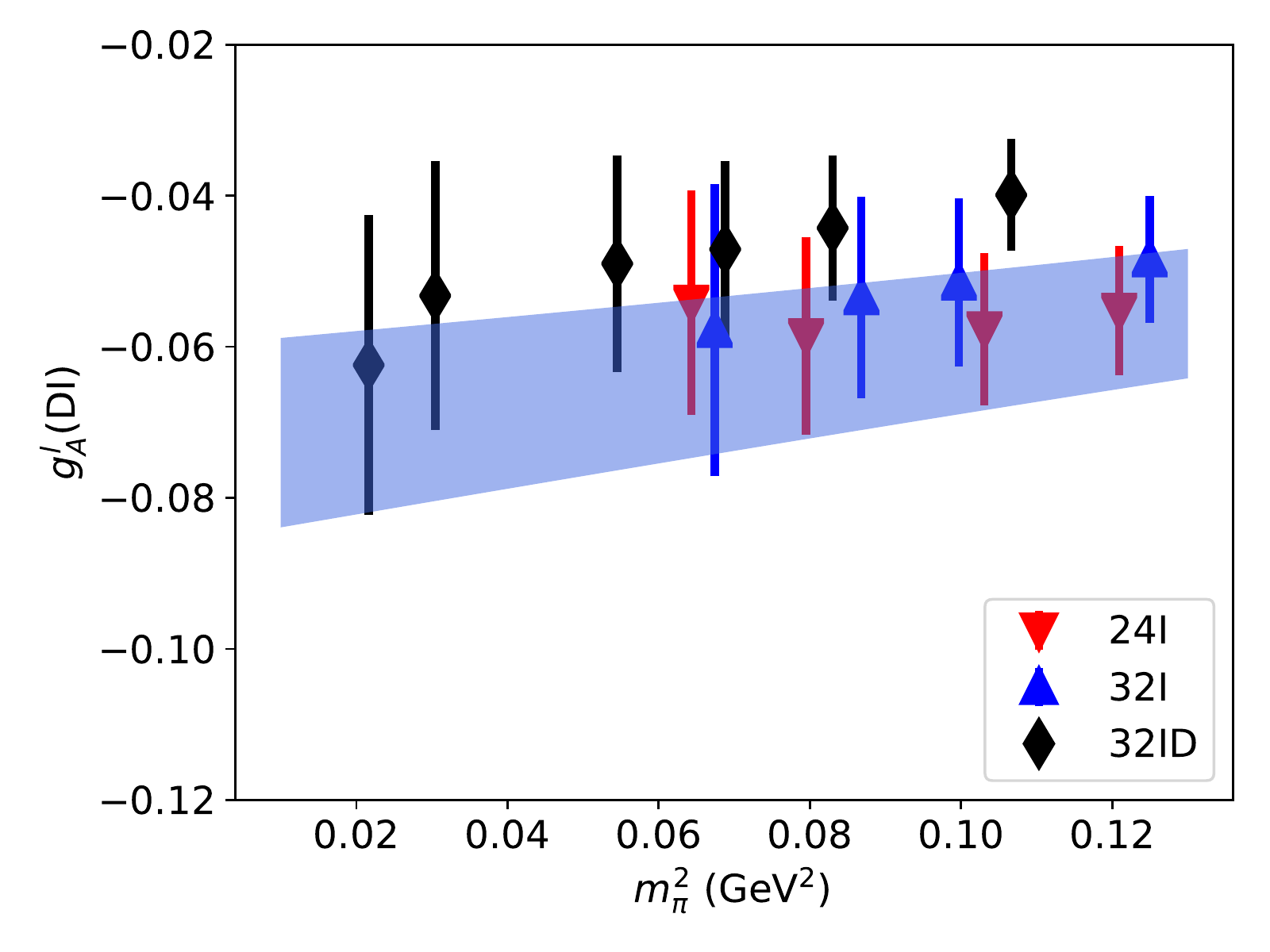}\includegraphics[scale=0.5,page=1]{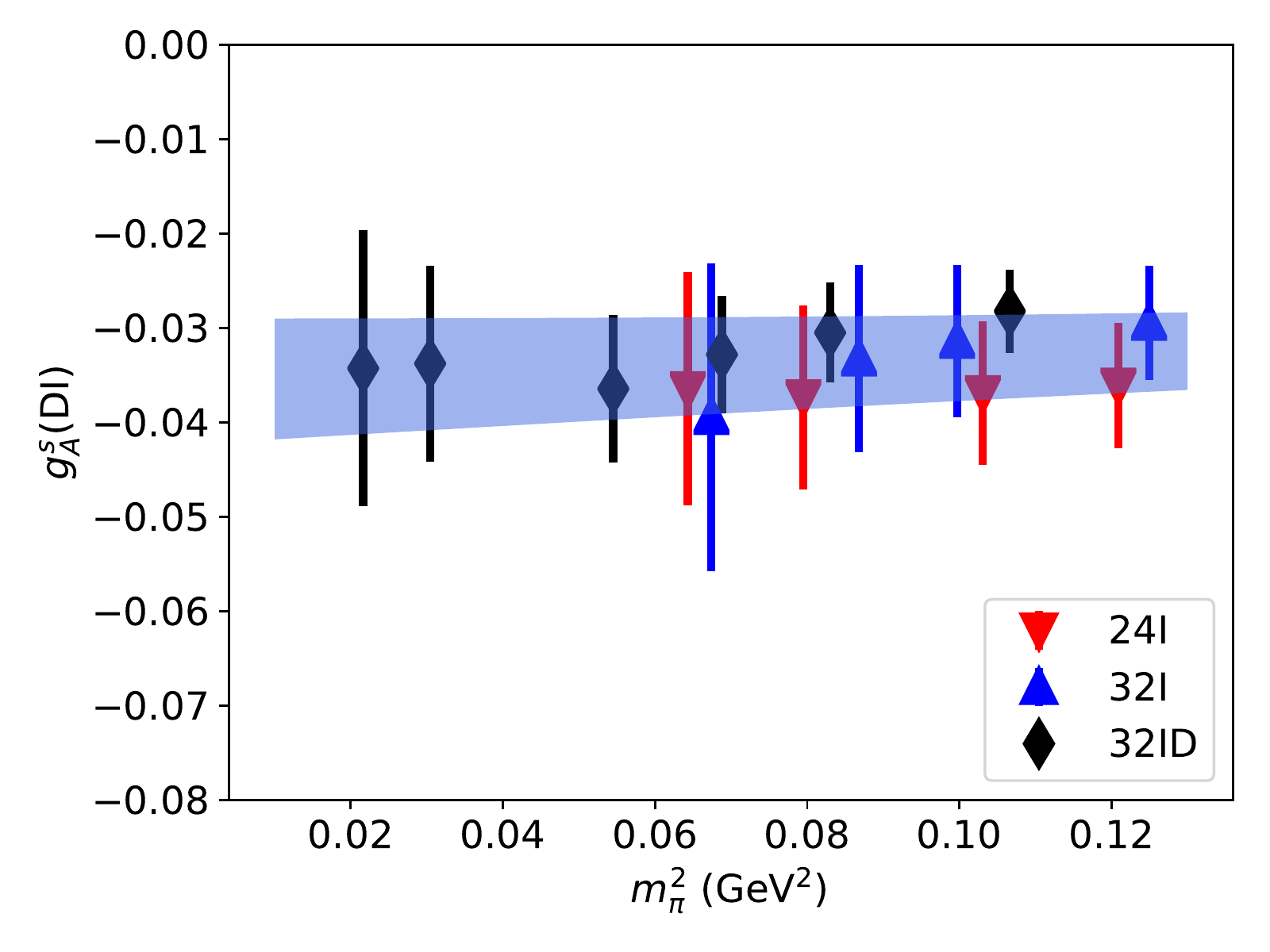}\caption{The global fit of the DI case for both light and strange quark. \label{fig:global-fit_DI}}
\end{figure}

\begin{figure}[tbp]
\centering{}\includegraphics[scale=0.5,page=1]{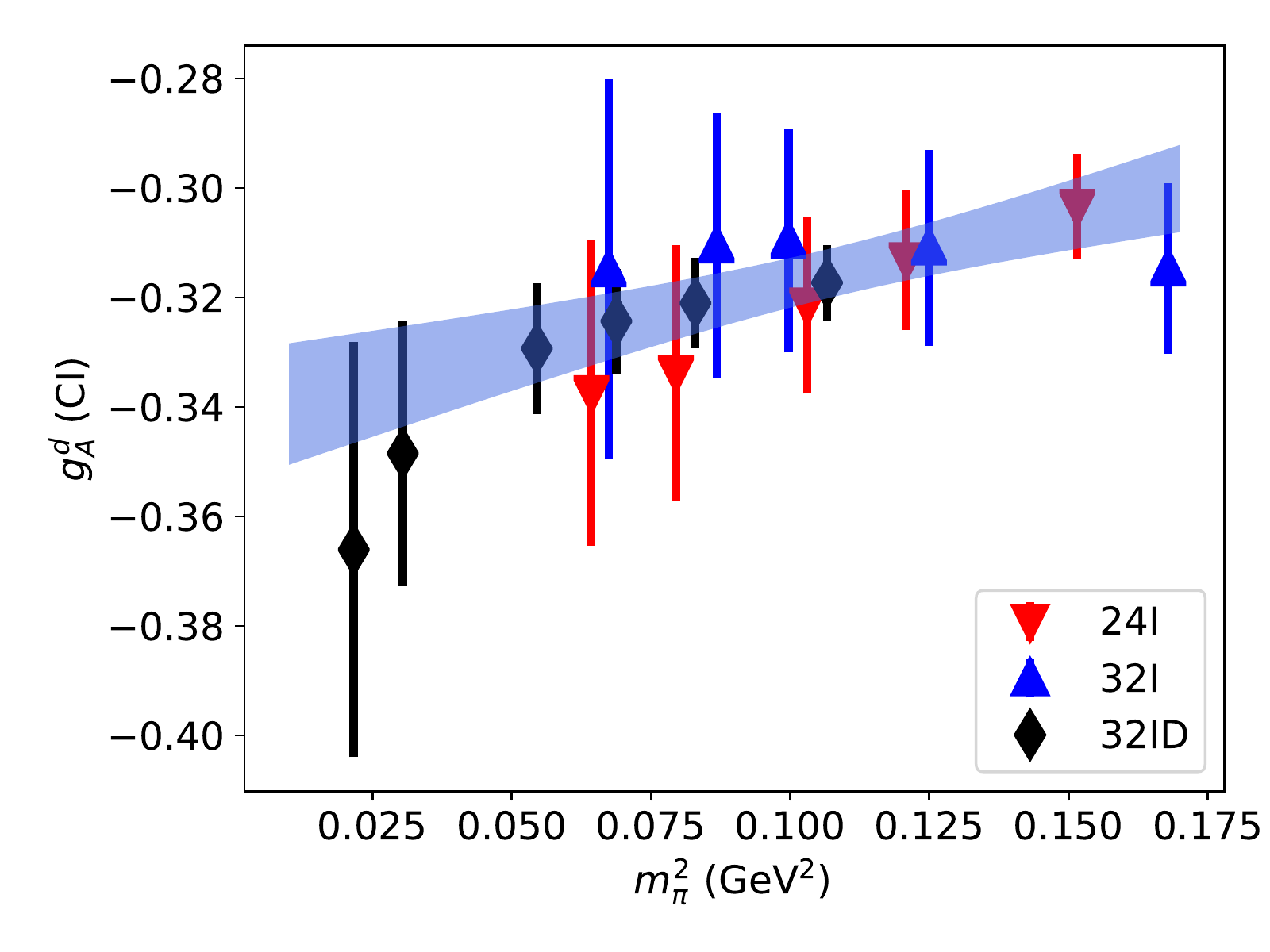}\includegraphics[scale=0.5,page=1]{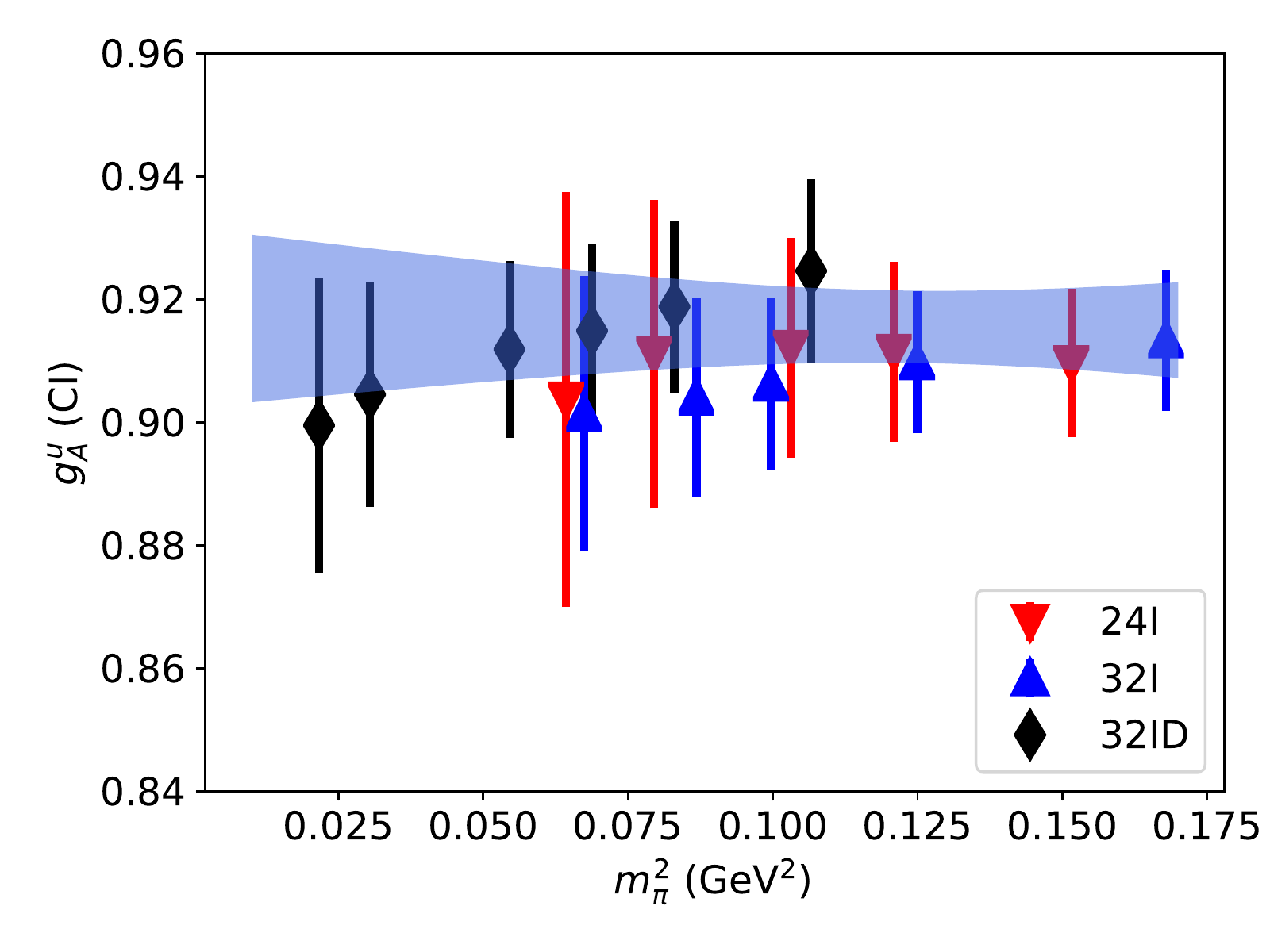}\caption{The global fit of the CI case for both $d$ and $u$ quark.\label{fig:global-fit_CI}}
\end{figure}

The final results of global fitting are shown in Figure \ref{fig:global-fit_DI}
and Figure \ref{fig:global-fit_CI} respectively for DI and CI. The
blue bands show the fitting results with only valence pion mass dependence.
A table listing the $\overline{{\rm MS}}$ numbers at 2 GeV with both
statistical and systematic errors is presented below (Table \ref{tab:final_results}).
The systematic errors are estimated by combining the systematic uncertainties
coming from the CDER technique, the fitting windows and function forms,
the extrapolations and the excited-states contamination. To be specific,
for the CI case, the systematic error coming from varying fitting
windows and function forms is estimated to be $3\%$. For the DI case,
the total systematic error is estimated to be $20\%$. 
The final errors of $g_A^3$ and $\Delta\Sigma$ are combined from the errors of $\Delta u$, $\Delta d$ and $\Delta s$ in quadrature.
Two sets of results
from recent lattice calculations and
three sets of results from recent global fittings of experiments are also listed in that table for comparison.
The results from D. de Florian {\it et al.} \cite{deFlorian:2009vb} and NNPDFpol1.1 \cite{Nocera:2014gqa} are at $Q^2=10$ GeV$^2$ 
and the integration range over the momentum fraction is from $10^{-3}$ to $1$.
The COMPASS results \cite{Adolph:2015saz} are at scale $Q^2=3$ GeV$^2$.
All the lattice results are calculated in the $\overline{\rm MS}$ scheme at $\mu=2$ GeV. Since the evolution of $\Delta\Sigma$ involves two-loop 
anomalous dimension (Eq.~\ref{evolve}), it does not vary much from $\mu=2$ GeV to $\mu=3$ GeV.
The calculation by C. Alexandrou {\it et al.} \citep{Alexandrou:2017oeh} is carried out on one
ensemble at the physical point with 2-flavor clover-improved twisted
mass fermions and the calculation by J. Green {\it et al.} \cite{Green:2017keo} is on one
ensemble with $2+1$-flavor clover fermions at $m_\pi=$ 317 MeV.
The total quark spin contribution of our present calculation is $\Sigma=0.405(25)(37)$ which agrees with that of C. Alexandrou {\it et al.} ($\sim0.402$)
and is consistent with the experimental results. The isovector $g_A^3=1.254(16)(30)$ agrees with the PDG value of 1.2723(23) within one sigma. It has a combined statistical and
systematic error of $\sim3\%$. This is consistent with the recent percent-level lattice calculation \cite{Chang:2018uxx} at 1.271(13), but
in contrast with the other two lattice calculations in the table which are lower than
the experimental value. There is another lattice calculation from the JLQCD Collaboration \cite{Yamanaka:2018uud} that uses dynamical overlap fermions at a single lattice spacing with four pion masses in the range 290--540 MeV. Their results $g_A=1.123(28)(95)$, $\Delta s=-0.046(26)(9)$, and $\Delta\Sigma=0.398(86)(99)$ are all consistent with ours.

\begin{table}[ht]
\centering{}%
\begin{tabular}{ccccccc}
 & \textit{$\Delta u$}{ } & \textit{$\Delta d$}{ } & \textit{$\Delta s$}{ } & \textit{$g_A^3$} & \textit{$g_A^8$}  & \textit{$\Delta\Sigma$}{ }\tabularnewline
\hline 
\{\makecell{D. de Florian {\it et al.}\\($Q^2$=10 GeV$^2$)} & $0.793^{+0.028}_{-0.034}$ & $-0.416^{+0.035}_{-0.025}$ & $-0.012^{+0.056}_{-0.062}$ &  &  & $0.366^{+0.042}_{-0.062}$\tabularnewline
\hline 
{\makecell{NNPDFpol1.1\\($Q^2$=10 GeV$^2$)}} & {0.76(4)} & {-0.41(4)} & {-0.10(8)} &  && {0.25(10)}\tabularnewline
\hline 
{\makecell{COMPASS\\($Q^2$=3 GeV$^2$)}} & {$[0.82,0.85]$} & {$[-0.45,-0.42]$} & {$[-0.11,-0.08]$} & 1.22(5)(10) &&{$[0.26, 0.36]$} \tabularnewline
\hline
{J. Green {\it et al.}} &  {0.863(7)(14)} & {-0.345(6)(9)} & {-0.0240(21)(11)} &  {1.206(7)(21)} & 0.565(11)(13) & {0.494(11)(15)}\tabularnewline
\hline 
{C. Alexandrou {\it et al.}} & {0.830(26)(4)} & {-0.386(16)(6)} & {-0.042(10)(2)} &  {1.216(31)(7)} & 0.526(39)(10) & {0.402(34)(10)}\tabularnewline
\hline 
{$\chi$QCD (this work)} & 0.847(18)(32) & -0.407(16)(18) & -0.035(6)(7) & 1.254(16)(30) & 0.510(27)(39) & 0.405(25)(37)\tabularnewline
\hline 
\end{tabular}\caption{The final results of quark spin matched to the $\overline{{\rm MS}}$
scheme at 2 GeV. The errors of $g_A^3$ and $\Delta\Sigma$ are combined from the errors of $\Delta u$, $\Delta d$ and $\Delta s$ in quadrature.
Results from two recent lattice calculations by J. Green {\it et al.} \cite{Green:2017keo} and the
Cyprus group \citep{Alexandrou:2017oeh} 
and three experimental results from D. de Florian {\it et al.} \cite{deFlorian:2009vb}, the NNPDF collaboration \cite{Nocera:2014gqa} and the COMPASS collaboration \cite{Adolph:2015saz}
are also listed for comparison.
\label{tab:final_results}}
\end{table}

\section{Summary\label{sec:Summary}}

In this work, we calculate the quark spin using overlap valences on
3 RBC/UKQCD domain-wall ensembles 24I, 32I and 32ID. The pion mass
of 32ID is around 171 MeV which is close to the physical point. The
anomalous Ward identity is checked carefully and we find that the identity
holds very well in our calculation with normalized axial-vector current
if the divergence of the axial-vector current is inserted as an operator between nucleon states. This is an
important check indicating that the lattice artifacts are under control.
For the disconnected-insertion part, the CDER technique is used for
the 32ID lattice when constructing 3-point functions and the statistical
error can be reduced by $10\%\sim40\%$. The DI contributions to the light and strange quark $\Delta l({\rm DI})$ and $\Delta s({\rm DI})$ are determined 
to be $-0.070(12)(15)$ and -0.035(6)(7), respectively.
For the connected-insertion
part, we use the improved axial-vector current aiming to reduce the
finite lattice spacing effects. 
The results of the CI contribution to $u$ and $d$ quarks $\Delta u({\rm CI})$ and $\Delta d({\rm CI})$ are $0.917(13)(28)$ and $-0.337(10)(10)$ respectively.
As we mentioned in Sec.~\ref{sec:Renormalization}, they are scale independent due to the
chiral Ward identity and can be compared to other lattice calculations. They can be extracted from deep inelastic scattering, provided the connected-sea and
disconnected-sea partons are separated in the global fit~\cite{Liu:2017lpe}.
Nonperturbative renormalization is
carried out so the reported results are all in the $\overline{{\rm MS}}$
scheme at 2 GeV scale. The numerical results are collected in Table
\ref{tab:final_results}; the total intrinsic quark spin contribution
is $\Delta\Sigma=0.405(25)(37)$, which is consistent with
the recent global fitting results of experimental data \cite{deFlorian:2009vb,Nocera:2014gqa,Adolph:2015saz}.
The isovector $g_A^3=1.254(16)(30)$ with $\sim3\%$ combined statistical and systematic error is within one sigma of that of the experimental value at 1.2723(23).

When checking the axial Ward identity, we find that the effects of
the excited states are crucial to understand the violation of the
extended Goldberger-Treiman relation and even a two-term fit cannot
always extract the MEs unbiasedly, so our estimations of the systematic
uncertainties are relatively large. Our results can be further improved
by carrying out the same calculation at the physical point directly
and by using larger source-sink separations to reduce the excited-states
contamination.

\begin{acknowledgments}
We thank the RBC and UKQCD Collaborations for providing their DWF
gauge configurations. This work is supported in part by the U.S. DOE
Grant No. DE-SC0013065. 
YY is
supported by the US National Science Foundation under
grant PHY 1653405 ``CAREER: Constraining Parton
Distribution Functions for New-Physics Searches.''
This research used resources of the Oak Ridge
Leadership Computing Facility at the Oak Ridge National Laboratory,
which is supported by the Office of Science of the U.S. Department
of Energy under Contract No. DE-AC05-00OR22725. This work used Stampede
time under the Extreme Science and Engineering Discovery Environment
(XSEDE), which is supported by National Science Foundation Grant No.
ACI-1053575. We also thank the National Energy Research Scientific
Computing Center (NERSC) for providing HPC resources that have contributed
to the research results reported within this paper. We acknowledge
the facilities of the USQCD Collaboration used for this research in
part, which are funded by the Office of Science of the U.S. Department
of Energy.
\end{acknowledgments}

\bibliography{library}

\end{document}